\documentclass[prb,twocolumn,showpacs,floatfix]{revtex4}
\usepackage[final]{graphicx}
\usepackage{amssymb}
\usepackage{amsmath}

\usepackage{graphicx}
\usepackage[breaklinks=true,colorlinks=true,linkcolor=blue,urlcolor=blue,
citecolor=blue]{hyperref}

\usepackage[utf8]{inputenc}
\usepackage[T1]{fontenc}
\usepackage{times}
\usepackage[american]{babel}

\usepackage{braket}
\usepackage{siunitx}

\DeclareMathOperator{\sgn}{sgn}
\newcommand{\phdag}{{\phantom{\dag}}}
\newcommand{\HeliumGroundState}{\mathrm{He^0(1s^2,1^1S_0)}}
\newcommand{\HeliumPositiveIon}{\mathrm{He^+(1s,1^2S_{1/2})}}
\newcommand{\HeliumNegativeIon}{\mathrm{He^{\ast-}(1s2s^2,2^2S_{1/2})}}
\newcommand{\HeliumMetaSing}{\mathrm{He^\ast(1s2s,2^1S_0)}}
\newcommand{\HeliumMetaTrip}{\mathrm{He^\ast(1s2s,2^3S_1)}}

\begin{document}
\bibliographystyle{apsrev}
\preprint{APS/123-QED}

\title{Towards an integrated modeling of the 
plasma-solid interface}

\author{M Bonitz$^1$, A Filinov$^{1,2,3}$, J W Abraham$^1$, K Balzer$^4$, H Kählert$^1$, E Pehlke$^1$,  FX Bronold$^5$, M Pamperin$^5$, M M Becker$^2$, D Loffhagen$^2$, H Fehske$^5$}
\address{$^1$ Institut f\"ur Theoretische Physik und Astrophysik,
Christian-Albrechts-Universit\"at, Leibnizstr. 15, D-24098 Kiel, Germany}
\address{$^2$ Leibniz-Institut für Plasmaforschung und Technologie e.V. (INP), Felix-Hausdorff-Str.~2, D-17489 Greifswald, 
Germany}
\address{$^3$ Joint Institute for High Temperatures RAS, Izhorskaya Str.~13, 
125412 Moscow, Russia}
\address{$^4$ Rechenzentrum der Christian-Albrechts-Universit\"at Kiel}
\address{$^5$ Institut für Physik, Universit\"at Greifswald}


\begin{abstract}
Solids facing a plasma are a common situation in many astrophysical systems and laboratory setups. Moreover, many plasma technology applications rely on the control of the plasma-surface interaction, i.e. of the particle, momentum and energy fluxes across the plasma-solid interface. However,  presently often a fundamental understanding of them is missing, so most technological applications are being developed via trial and error. 
%
The reason is that the physical processes at the interface of a low-temperature plasma and a solid are extremely complex, involving a large number of elementary processes in the plasma, in the solid as well as fluxes across the interface. An accurate theoretical treatment of these processes is very difficult due to the vastly different system properties on both sides of the interface: quantum versus classical behavior of electrons in the solid and plasma, respectively; as well as the dramatically differing electron densities, length and time scales. Moreover, often the system is far from equilibrium.
In the majority of plasma simulations surface processes are either neglected or treated via phenomenological parameters such as sticking coefficients, sputter rates or secondary electron emission coefficients. However, those parameters are known only in some cases and with very limited accuracy. Similarly, while surface physics simulations have often studied the impact of single ions or neutrals, so far, the influence of a plasma medium and correlations between successive impacts have not been taken into account. Such an approach, necessarily neglects the mutual influences between plasma and solid surface and cannot have predictive power.

In this paper we discuss in some detail the physical processes a the plasma-solid interface which brings us to the necessity of coupled plasma-solid simulations.
We briefly summarize relevant theoretical methods from solid state and surface physics that are suitable to contribute to such an approach and identify four methods. The first are mesoscopic simulations such as kinetic Monte Carlo (KMC) and molecular dynamics (MD) that are able to treat complex processes on large scales but neglect electronic effects. The second are quantum kinetic methods based on the quantum Boltzmann equation that give access to a more accurate treatment of  surface processes using simplifying models for the solid. The third approach are \textit{ab initio} simulations of surface process that are based on density functional theory (DFT) and time-dependent DFT. The fourths are nonequilibrium Green functions that able to treat correlation effects in the material and at the interface. The price for the increased quality is a dramatic increase of computational effort and a restriction to short time and length scales. We conclude that, presently, none of the four methods is capable of providing a complete picture of the processes at the interface. Instead, each of them  provides complementary information, and we discuss possible combinations.

\end{abstract}



\maketitle


\section{Introduction}\label{s:intro}
\begin{figure}[h]
  \begin{center} 
  \hspace{-0.cm}\includegraphics[width=0.5\textwidth]{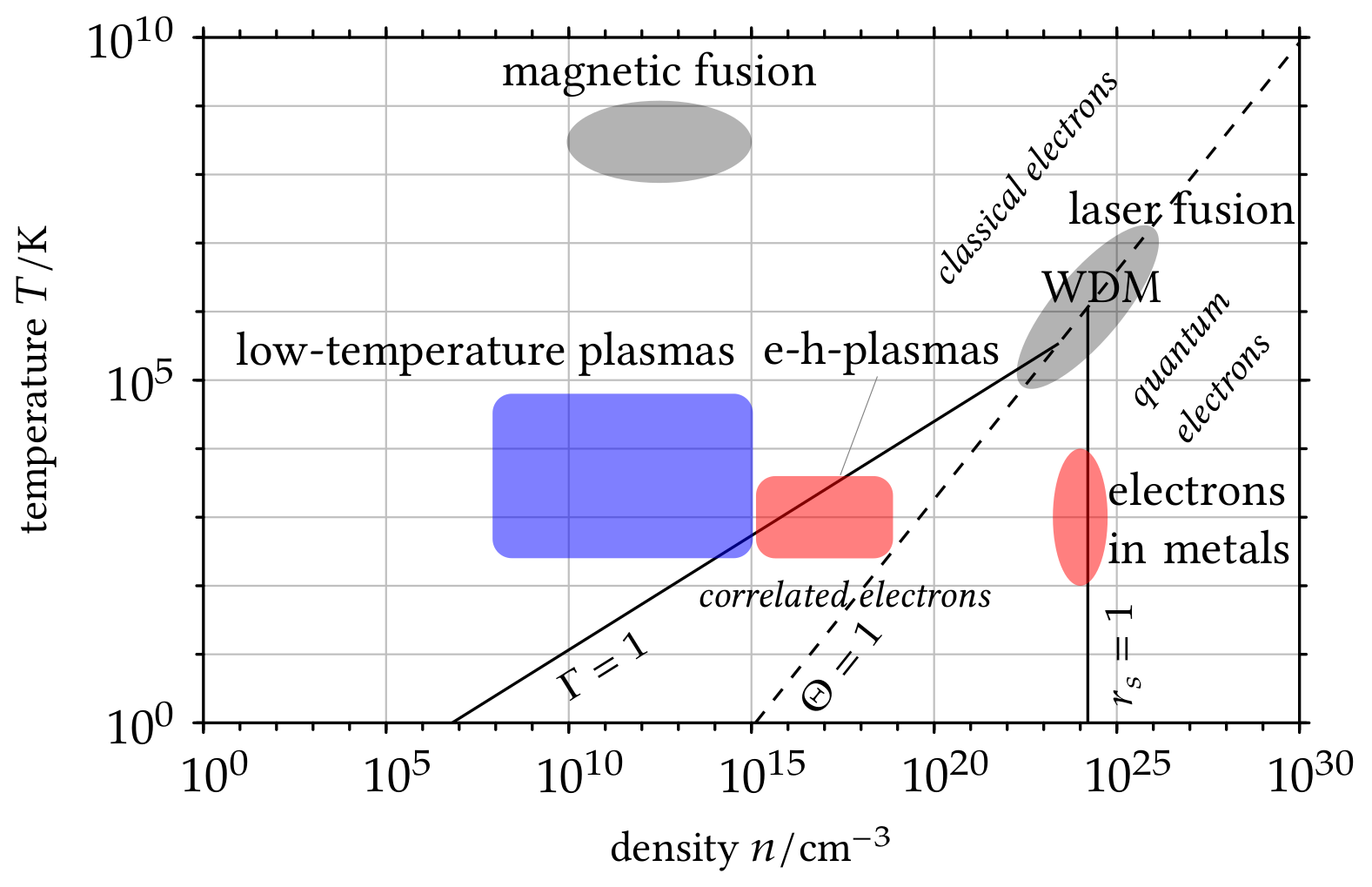}
  \end{center}
  \caption{Low-temperature plasmas (LTP) are one of three main current research topics in plasma physics, aside from magnetic fusion and dense plasma (warm dense matter, WDM) \cite{doe-report-17}. These systems cover a huge parameter range in the density-temperature plane. LTP (the blue box) range from low (electron) density to atmospheric pressure (right edge). Representatives of solids facing the plasma are metals and semiconductors (electron-hole plasmas) sketched by the red areas. Relevant dimensionless parameters are the classical coupling parameter, $\Gamma=e^2/{\bar r} k_BT$, the quantum coupling parameter, $r_s={\bar r}/a_B$ and the degeneracy parameter of the electrons, $\Theta=k_BT/E_F$, with ${\bar r}, a_B$ and $E_F$ denoting the mean interparticle distance, the Bohr radius and the Fermi energy, respectively.} 
  \label{fig:nt-plane}
  \end{figure}

%
Modern plasma physics has three main research topics [\onlinecite{doe-report-17}]: high-temperature plasmas, in particular magnetic fusion; high-density plasmas (``warm dense matter'', laser plasmas, inertial confinement fusion); and low-temperature plasmas. The location of these areas in the density-temperature plain is sketched in Fig.~\ref{fig:nt-plane}. In each of these fields, the processes at the plasma wall play a crucial role, both, for fundamental understanding and for technological applications. Therefore, progress in the simulation of plasma-solid interaction is of crucial importance in each of these fields.
Here we concentrate on plasma-solid processes in low-temperature plasma, although some of our results are expected to be of interest also for the high-temperature plasmas. This  field has experienced impressive progress over the last two decades, both, in experiments and 
applications. 

Aside from traditional applications also new materials, in particular \textit{nanomaterials}, are coming into the focus [\onlinecite{meyyappan_jpd_11,ostrikov_aip_13}]. This includes carbon based materials such as carbon nanotubes and graphene nanoribbons that have a size-dependent bandgap and promise exciting electronic and optical properties, e.g. [\onlinecite{son_prl_06, prezzi_prb_08}]. The role of plasmas, in the context of these novel materials, is only poorly explored yet. There are impressive first results on plasma synthesis of these systems. On the other hand, it will also be interesting to use such materials inside a plasma and to utilize their properties in a discharge environment.

\begin{figure}[h]
  \begin{center} 
  \hspace{-0.cm}\includegraphics[width=0.49\textwidth]{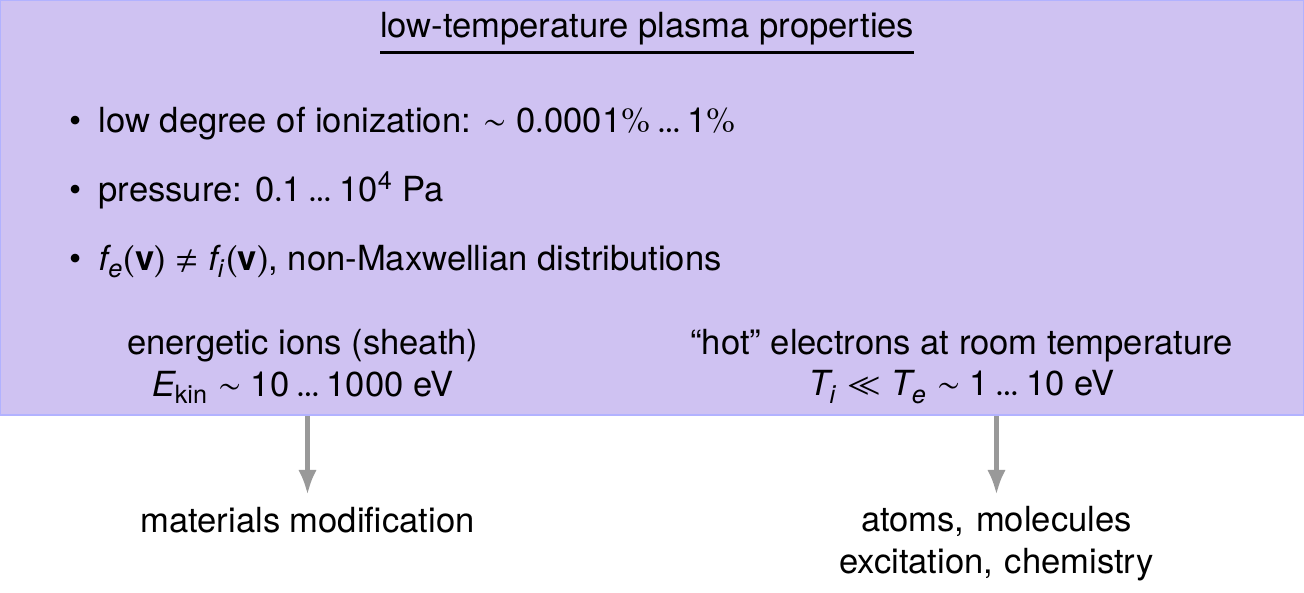}
  \end{center}
  \caption{Low-temperature plasmas being  composed of neutrals (atoms, molecules), ions and electrons comprise a number of very unusual properties: they are non-isothermal ($T_i\ne T_e$), far from thermal equilibrium (non-Maxwellian velocity distributions), and they contain electrons and ions of a very broad range of kinetic energies. In the plasma boundary region (``sheath'') ions may reach high energies that can be exploited for materials modification, sputtering and ion implantation. At the same time, electrons with energies in the eV range are able to excite and ionize neutrals and trigger chemical reactions.}
  \label{fig:ltp-properties}
  \end{figure}

These applications are only emerging, and a brief discussion of some aspects will be given in this paper. Yet for their success, and for solids embedded in plasmas in general, it will be crucial to have available accurate simulations of the plasma-solid interface, as has been pointed out in many places, e.g. [\onlinecite{adamovich_2017_plasma, doe-report-17}].
The interest in low-temperature plasmas arises from the peculiar properties of these systems. These plasmas  typically have a low degree of ionization and cover a broad pressure range, from below one Pa to atmospheric pressure, see Fig.~\ref{fig:ltp-properties}.
These plasmas are non-thermal, i.e. the electron and ion temperatures may differ by several orders of magnitude. Moreover, electrons and ions may be far from thermal equilibrium, being described by non-Maxwellian velocity distributions. The existence of energetic electrons with temperatures in the eV-range, which is sufficient to excite or ionize atoms and molecules, is of high interest for applications in surface chemistry, biology and even medicine. On the other hand, these plasmas may contain highly energetic ions that are accelerated by strong electric fields, in particular in the surface near region (the ``plasma sheath''). These energies are sufficient for mechanical modification of the solid surface  such as defect creation, ion implantation or sputtering.

\begin{figure}[h]
  \begin{center} 
  \hspace{-0.cm}\includegraphics[width=0.5\textwidth]{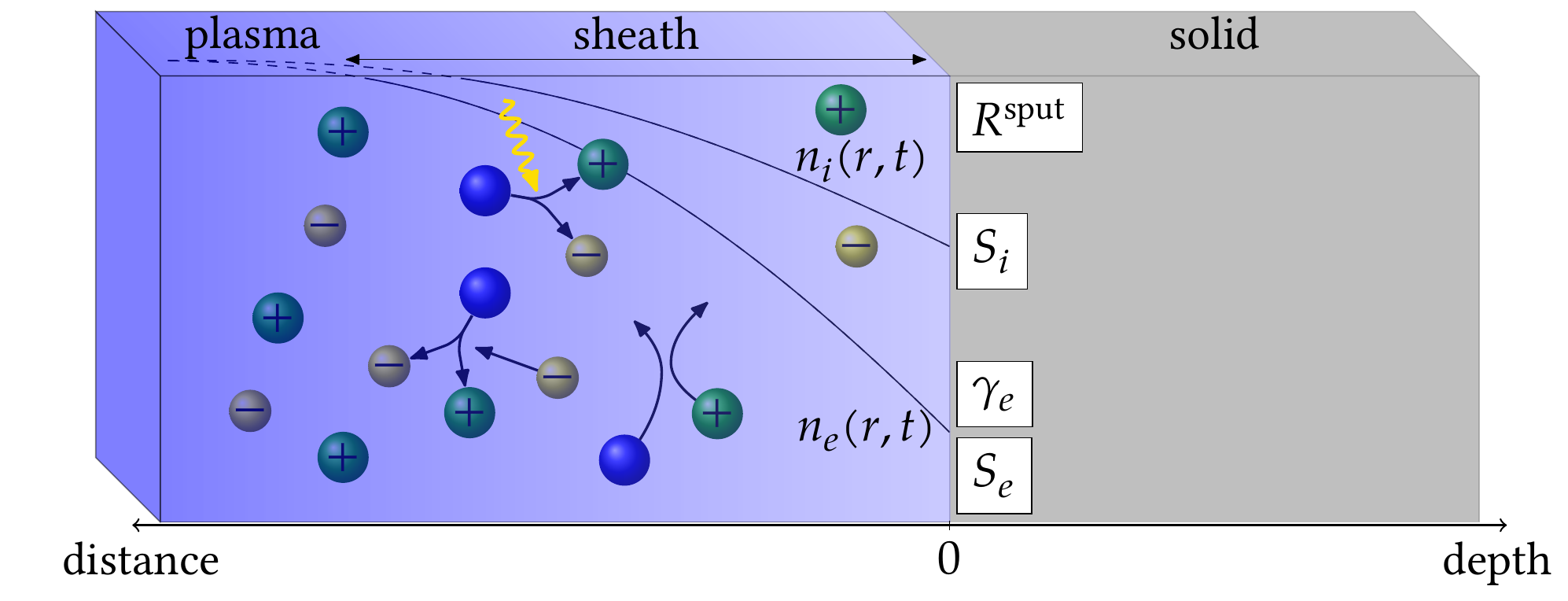}\\ 
    \hspace{-0.cm}\includegraphics[width=0.5\textwidth]{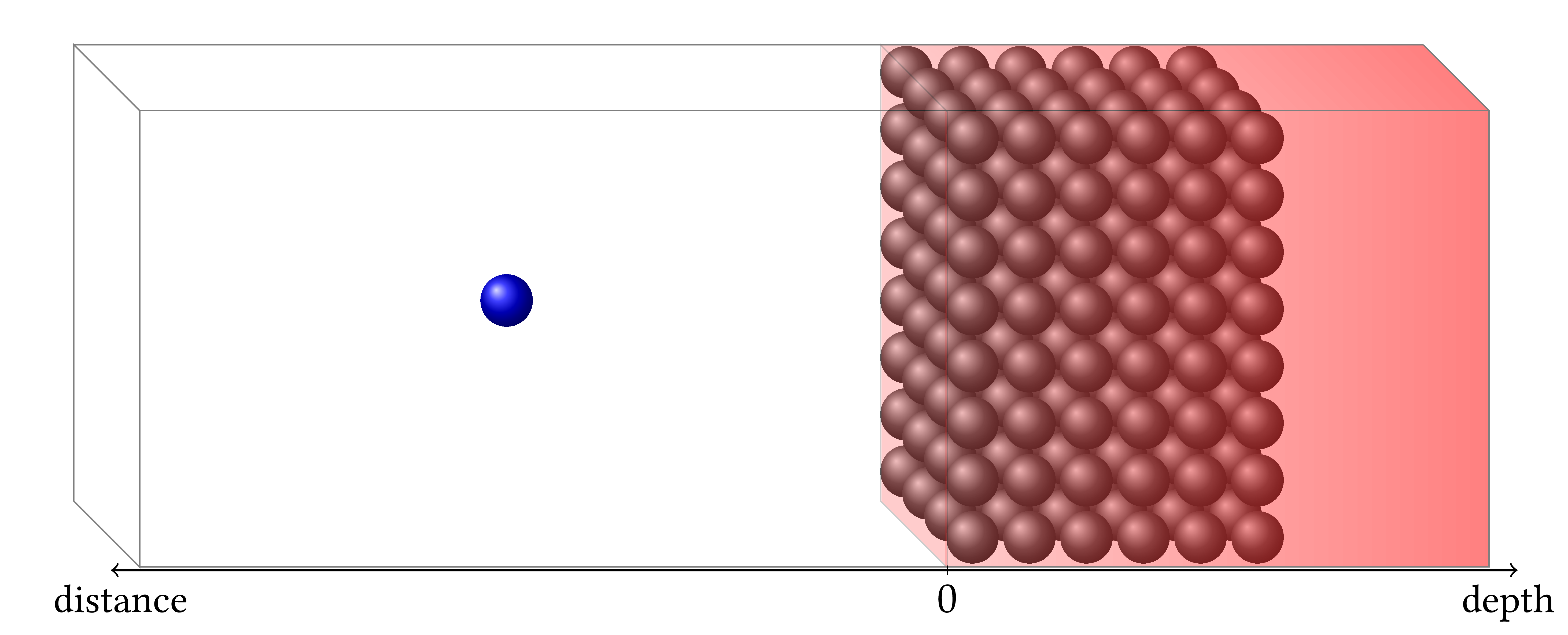}
  \end{center}
  \caption{\textbf{Top}: Sketch of the current approach to include surface properties in plasma simulations via phenomenological parameters such as sputter rates $R^{\rm sput}$, electron and ion sticking coefficients, $S_e, S_i$ and the secondary electron emission coefficient, $\gamma_e$. While the plasma is treated by advanced approaches, the atomic structure of the solid and the surface is not resolved. \textbf{Bottom:} In contrast, in surface science, atomic level information of the surface is taken into account, whereas plasma effects are approximated via independent impacts of neutrals or ions.}
  \label{fig:psi_today}
  \end{figure}

Even though there have been remarkable recent advances, both, in plasma modeling 
and surface science simulations, the combination of the two is still at a very early 
stage. 
Current simulations in low-temperature plasma physics that are based on kinetic theory or fluid theory have achieved a high quality description of the dynamics of electrons, ions and neutrals in the plasma bulk, including elastic and inelastic scattering processes, cf. top part of Fig.~\ref{fig:psi_today}. At the same time, these simulations often omit plasma-surface processes or treat them phenomenologically. 
For example, in many advanced kinetic simulations based on the Boltzmann equation, e.g. [\onlinecite{hagelaar_2005,donko_2016}] or particle in cell (PIC-MCC) simulations, e.g. [\onlinecite{ebert_2016,becker_psst17}] neutrals are treated as a homogeneous background, and their interaction with surfaces is not included in the description. However, the effect of energetic neutrals maybe crucial for secondary electron emission (SEE), as was demonstrated in PIC simulations of Derszi \textit{et al.} [\onlinecite{derzi_2015_effects}]. 
Another effect that can be important for the behavior of the plasma are the properties of the surface, such as surface roughness, oxidation or coverage by an adsorbate. For example, Phelps and Petrovic [\onlinecite{phelps_1999_cold}] convincingly demonstrated that the plasma modification of a metal surface may change SEE by several orders of magnitude, see Sec.~\ref{s:influences}. The conclusion is that an accurate theoretical treatment of the solid surface may be very important for low-temperature plasma simulations. This suggests to resort to surface science methods where a very accurate atomic level treatment of solid surfaces has been achieved by \textit{ab initio} methods such as density functional theory (DFT). Surface simulations have also incorporated the impact of energetic projectiles to simulate sputtering, e.g. [\onlinecite{
brault_frontiers_18}] or energy loss (stopping power), e.g. [\onlinecite{zhao_2015_comparison,balzer_prb16}]. However, these are typically simulations where single ions or neutrals are treated, but the effect of a plasma and of its nonequilibrium properties [cf. Fig.~\ref{fig:ltp-properties}] and the plasma-induced modifications of the surface have not been taken into account so far.

From the examples presented above that demonstrated the mutual influences between plasma and surface, it is clear that further progress in an accurate modeling of the plasma-solid interface requires to go beyond an independent treatment of both sides. Instead it is necessary to develop a combined theory and simulation of the entire system. This is a challenging project that requires strong input from plasma physics and surface science, simultaneously. In fact, such a research effort is under way at Kiel University, in collaboration with Greifswald University and the INP Greifswald. The concept of this project was first presented in 2015 [\onlinecite{interface}] and continuously developed since then. It is the goal of this article to present these ideas and first simulation results and to outline further directions and perspectives of development. We note that similar concepts have been developed by Graves, Brault and Neyts and others in the frame of MD simulations, e.g. [\onlinecite{graves_2009_molecular, neyts2017molecular}], see Sec.~\ref{s:mesoscopic}. The main difference is that those simulations usually neglect the electronic degrees of freedom, in particular, quantum effects and internal relaxation processes in the solid.

This paper is organized as follows. In Sec.~\ref{s:influences} we discuss the mutual influences between plasma and solid that motivates the development of a novel approach combining plasma and surface science methods.
In Sec.~\ref{s:concepts} we discuss more in detail the relevant physical processes and effects at the plasma-solid interface. 
This sets the basis for the required theoretical approaches that are capable to accurately simulating plasma-surface processes and discuss their respective advantages and problems. We identify four different methods that are discussed in detail with increasing degree of complexity.
In Sec.~\ref{s:mesoscopic} we discuss the first one -- mesoscopic approaches to the plasma-solid interface -- in particular molecular dynamics and discuss  acceleration approaches that are of potential relevance for plasma-surface interaction. In Sec.~\ref{s:qkin} we discuss the second class of methods that is based on the quantum Boltzmann equation.  In Secs.~\ref{s:abinit} we consider the third approach that is based  DFT and TDDFT. Finally, in Sec.~\ref{s:negf} the fourth approach is discussed that is based on nonequilibrium Green functions (NEGF) and leads to generalized quantum kinetic equations. The analysis concludes with Sec.~\ref{s:integrated} where we outline first steps towards an integrated plasma-surface modeling, 
%
and we present our conclusions in Sec.~\ref{s:conclusion1}.
\section{Mutual influences of plasma and  solid}\label{s:influences}
  \begin{figure}
  \begin{center} 
  \hspace{-0.cm}\includegraphics[width=0.4\textwidth]{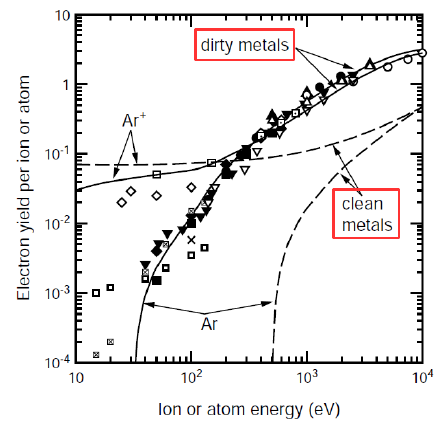}\\ 
    \hspace{-0.cm}\includegraphics[width=0.4\textwidth]{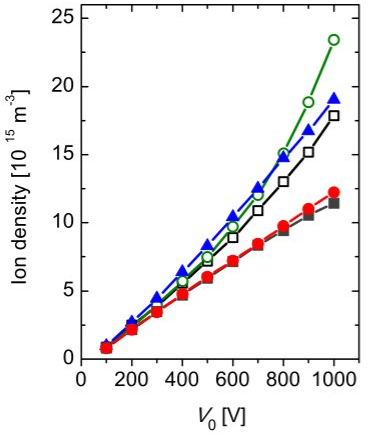}
  \end{center}
  \caption{\textbf{Top}: Secondary electron emission yield per argon atom or ion for a broad variety of metals (symbols). ``Clean metals'' refers to beam experiments where the surface was cleaned via ion sputtering. ``Dirty metals'' denotes measurements following varying degrees of surface exposure to oxygen, to water or ambient gas. Figure from Ref.~\onlinecite{phelps_1999_cold} where additional details are given. \textbf{Bottom:} Effect of SEE on ion density in the bulk of a  AC discharge (13.56 MHz, p=5Pa, for varying electrode voltage), for different SEE models: no SEE (black) to a full treatment of ion and atom induced SEE (green). From Ref.~\onlinecite{derzi_2015_effects}.}
  \label{fig:see}
  \end{figure} 
As we discussed in the Introduction there are many processes that couple the plasma and the surface. Here we discuss a few examples. The first is the effect of energetic neutrals impacting the surface. These neutrals are efficiently produced in the case of a strong sheath electric field, by means of charge exchange collisions. Energetic neutrals maybe crucial for secondary electron emission from metal surfaces, as was shown in Ref.~\onlinecite{phelps_1999_cold}, see also Fig.~\ref{fig:see}. This was  confirmed by PIC simulations where neutrals above a threshold energy of 23eV were traced individually [\onlinecite{derzi_2015_effects}]. 

An example where the surface properties affect the plasma is related to plasma electrons hitting a solid surface. The standard assumption in plasma simulations is that these electrons are lost without reflection, e.g. [\onlinecite{sheehan_2013_kinetic}]. Only recently a microscopic calculation of the electron sticking coefficient was performed by Bronold and Fehske [\onlinecite{bronold_2015_absorption}] that demonstrated that this picture has to be revised. Recent PIC simulations by Sun \textit{et al.} confirmed that finite electron reflectivity, in combination with SEE, may have a significant influence on the plasma parameters of a CCRF discharge for pressures of several tens of Pa [\onlinecite{sun_psst_2018}].

An even more striking example of the effect of the  properties of the surface on the plasma properties is secondary electron emission. Phelps and Petrovic compiled extensive experimental data for the SEE yield from different metals over a broad range of impact energies of argon atoms and ions [\onlinecite{phelps_1999_cold}]. They compared SEE from surfaces that were cleaned by ion sputtering (cf. curves labeled ``clean metals'' in the top part of Fig.~\ref{fig:see}) to the SEE yield obtained from surfaces that 
have been in contact with a plasma (or to oxygen or ambient gas, for the original references see Ref.~\onlinecite{phelps_1999_cold}). The authors note a dramatic difference of the SEE yield from a ``dirty'' surface compared to a ``clean'' one which may exceed two orders of magnitude for SEE due to neutral atoms. For energetic atoms (energies above 200 eV) this difference is much larger than the difference between different metals. 

The crucial importance of SEE has been confirmed in many simulations. For example, Derszi \textit{et al.} performed PIC simulations where they included SEE according to various models via modified cross sections ~[\onlinecite{derzi_2015_effects}]. In that work it was found that a realistic (``dirty'' [\onlinecite{phelps_1999_cold}]) surface gives rise to a significant increase (up to a factor of two) of the ion density, even far away from the electrode compared to simulations where SEE is neglected. This trend is seen in the bottom part of Fig.~\ref{fig:see}, compare the green and black curves. At the same time, the experimental data shown in the top part of Fig.~\ref{fig:see}  suggest that there remain substantial uncertainties in the values of the SEE coefficient. It is also not clear how long the surface was treated. In a real plasma treatment experiment a ``clean'' surface may correspond to the initial state of an electrode which, ultimately, turns into a ``dirty'' metal that is covered by adsorbates or an oxide layer.
Thus, more accurate experimental and theoretical knowledge of the SEE for different materials and varying degrees of surface coverage will be very important for applications.

Strong surface effects have also been observed in the field emission. For example, Li \textit{et al.} studied the effect of surface roughness on the field emission by including a phenomenological geometric enhancement factor [\onlinecite{li_2013}].


These examples show that the simple concept of fixed phenomenological surface parameters, such as the SEE coefficient $\gamma_e$, for a given surface material, that was sketched in Fig.~\ref{fig:psi_today} has to be questioned. Instead the SEE coefficient has to take into account the properties of the surface and also the exposure to the plasma.
  \begin{figure}[h]
  \begin{center} 
  \hspace{-0.cm}\includegraphics[width=0.5\textwidth]{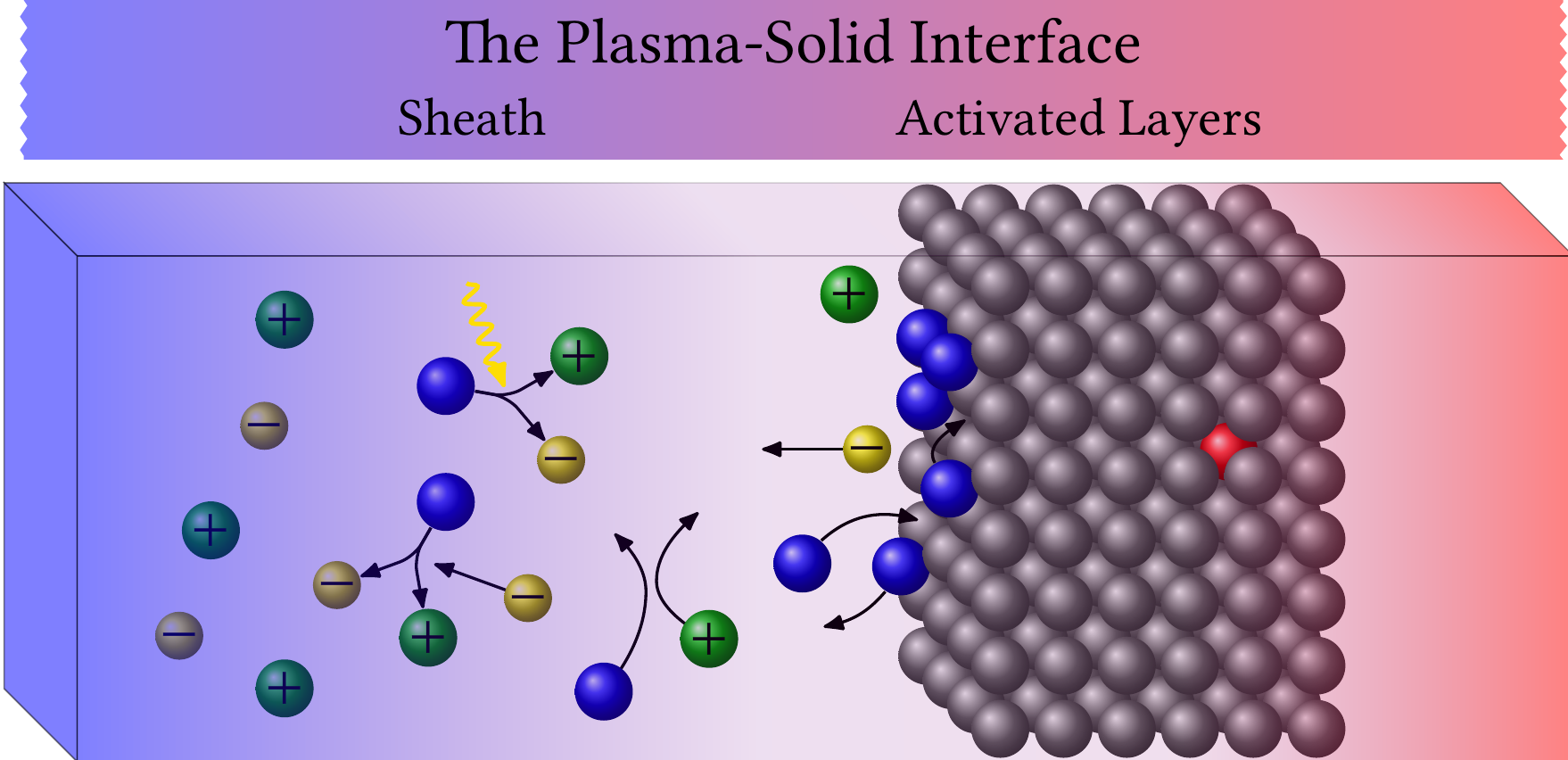} 
  \end{center}
  \caption{Sketch of the plasma-solid interface which comprises the plasma sheath and the plasma facing activated layers of the solid [\onlinecite{interface}]. Among the relevant processes are diffusion, adsorption (``sticking'') and desorption of neutrals, penetration (stopping) of ions and electron transfer between solid and plasma. 
  The influence of the plasma on the solid and vice versa is a major challenge for a predictive theoretical treatment and require a combination of various theoretical approaches, see Fig.~\ref{fig:theory}.}
  \label{fig:psi}
  \end{figure} 
In order to make reliable predictions, novel theoretical concepts are needed that include the whole scope of complex physical and chemical processes that 
occur at the plasma-solid interface
which include secondary electron emission, sputtering, neutralization and stopping of ions, adsorption and desorption of neutral particles as well as chemical reactions.
Therefore, the plasma-solid system should be treated as single entity which we call ``plasma-solid interface'' [\onlinecite{interface}]. It comprises the plasma sheath and the plasma facing atomic layers of the surface that are influenced  (excited or ``activated'') by the plasma. 
This new theoretical concept is sketched in Fig.~\ref{fig:psi}.

\section{The plasma-solid interface: Physical processes and theoretical approaches}\label{s:concepts}
\subsection{Overview on the physical processes at the plasma-solid interface}\label{ss:psi-processes}
Let us now look more in detail at the properties of the plasma-solid interface and the relevant processes. We already discussed in the beginning, cf. Fig.~\ref{fig:nt-plane}, that plasma and solid are characterized by a huge density gap leading, in many cases, to an enormous difference in length and time scales.
First of all, the plasma-solid system is in a stationary state. This state differs from thermal equilibrium due to the nonequilibrium character of the plasma [cf. Fig.~\ref{fig:ltp-properties}] and due to fluxes of particles (electrons, ions, neutrals) that cross the boundary in both directions.

\begin{figure*}[h]
  \begin{center} 
  \hspace{-0.cm}\includegraphics[width=0.88\textwidth]{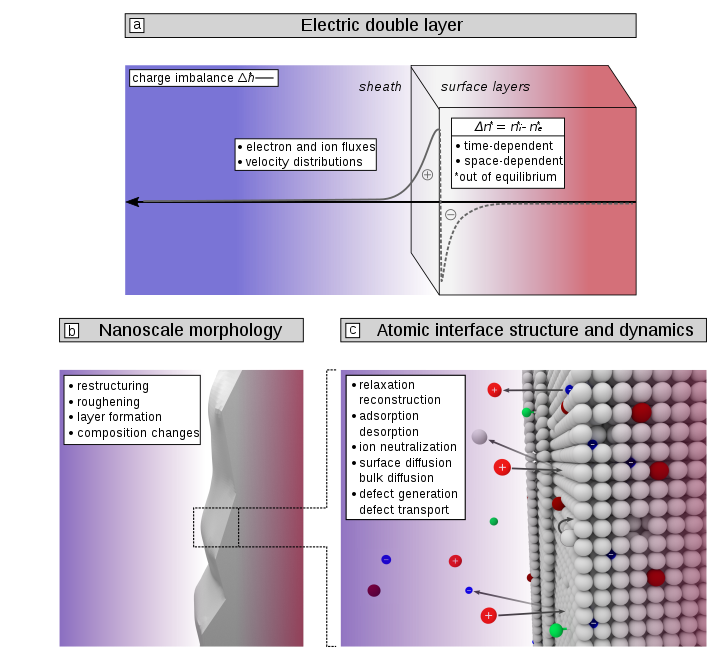} 
  \end{center}
  \caption{Sketch of the physical processes at the plasma-solid interface--from the largest to the smallest length scale. \textbf{Top}: The electric double layer (on the scale of the Debye length, on the plasma side, and a few nanometers, in the solid) resulting from electron depletion in the plasma sheath [cf. Fig.~\ref{fig:psi_today}] is characterized by the local  difference of the nonequilibrium ion and electron densities and is accompanied by electron accumulation in the solid which is influenced by the processes in figure parts b) and c). \textbf{Bottom left}: on the scale of the surface roughness (typically nanometers) the surface exhibits local variations of the morphology and chemical composition. \textbf{Bottom right}: atomic scale modification of the surface and the plasma sheath caused by individual particle impacts, charge transfer, chemical reactions etc. 
  The relevant processes are indicated inside the figure parts.}
  \label{fig:interface-physics}
  \end{figure*}

On the largest scale (the scale of the Debye length, Fig.~\ref{fig:interface-physics}.a) the interface is characterized by the density profiles of electrons and ions.
Electron depletion in the plasma sheath near a surface [cf. Fig.~\ref{fig:psi_today}] gives rise to an excess positive charge in front of the surface. The missing negative charge has to accumulate inside the surface giving rise to an \textit{electric  double layer}. Charged double layers are a common phenomenon in liquids and were originally studied in electrolytes by Helmholtz~[\onlinecite{helmholtz}]. The corresponding effect for the plasma-solid interface was predicted by Bronold and co-workers [\onlinecite{heinisch_2012_electron}] and turns out to be very different, by the different composition of the system, and more complex, due to its nonequilibrium nature.
The peculiarities of the electric double layer that are caused by the plasma properties--the non-Maxwellian velocity distributions and the time variation, in case of an rf field, on the scale of
nanoseconds--are qualitatively understood. On the other hand, what is far less understood, is the
impact of the solid on this
charge distribution: the
influence of the nanoscale surface structure [Fig.~\ref{fig:psi_today}.b] as well as of the
atomic scale structure [Fig.~\ref{fig:psi_today}.c] and processes
between solid and plasma. 

Thus, an analysis of the
surface and  near-surface structure of  the solid in the presence of the plasma and of surface processes under these conditions is required.  This includes the
structural and chemical response of the material. Surface science experiments are required to investigate
the atomic-scale structure [Fig.~\ref{fig:psi_today}.c], e.g., surface relaxation and reconstruction, adsorbate species,
and surface defects, as well as the evolution of the nanoscale morphology, e.g., the formation of
islands, pits, steps, and ultrathin films on the surface [Fig.~\ref{fig:psi_today}.b]. Also the modification of the material
in the near-surface region (e.g., crystallinity, porosity, composition, defect density) has to be studied as
function, e.g., of ion and neutral particle energy and plasma density. It would be highly desirable that surface science and plasma physics
experiments obtain the concentration of charged and neutral species at and near the surface as
well as chemical binding energies of various species. The information that should be obtained from theory includes
the relevant cross sections and rate coefficients, neutral sticking coefficients, SEE coefficients,
diffusion coefficients, information on energy dissipation channels and time scales etc.

Another key topic is the charge transfer dynamics across the interface in the presence of the plasma. 
Here, plasma physics experiments are needed that provide key information, including the electron and ion fluxes to and from the surface [Fig.~\ref{fig:psi_today}.a]. This should be complemented by surface science experiments measuring the secondary electron current from the surface and the plasma-induced modification of the band structure of the surface material. These quantities are the combined result of a multitude of physical processes. 
Theory and simulations should attempt to resolve the individual contributions due to the neutralization of ions in front of the surface or electron impact ionization and ion stopping inside the solid.
Furthermore, it would be desirable if  surface science experiments could provide insight into the complex energy landscape of the solid surface, its modification due to the plasma environment, as well as the plasma-induced space charge region inside the solid. The experimental information should be complemented by  novel theoretical approaches that will lead to accurate results for electron and ion velocity distributions in the plasma sheath, on secondary electron emission, on the sticking coefficients of electrons, ions, and neutrals and on the charge of nanoparticles embedded in the plasma. 

All these processes evolve in time (on
the scale of seconds), as a result of the surface modification by the plasma.
Thus, there is a direct coupling between effects on the atomic scale and the macroscopic plasma behavior that needs to be explored. 
\subsection{Theoretical approaches for the plasma-solid interface}\label{ss:psi-theory}
\begin{figure*}
  \begin{center} 
  \hspace{-0.cm}\includegraphics[width=0.78\textwidth]{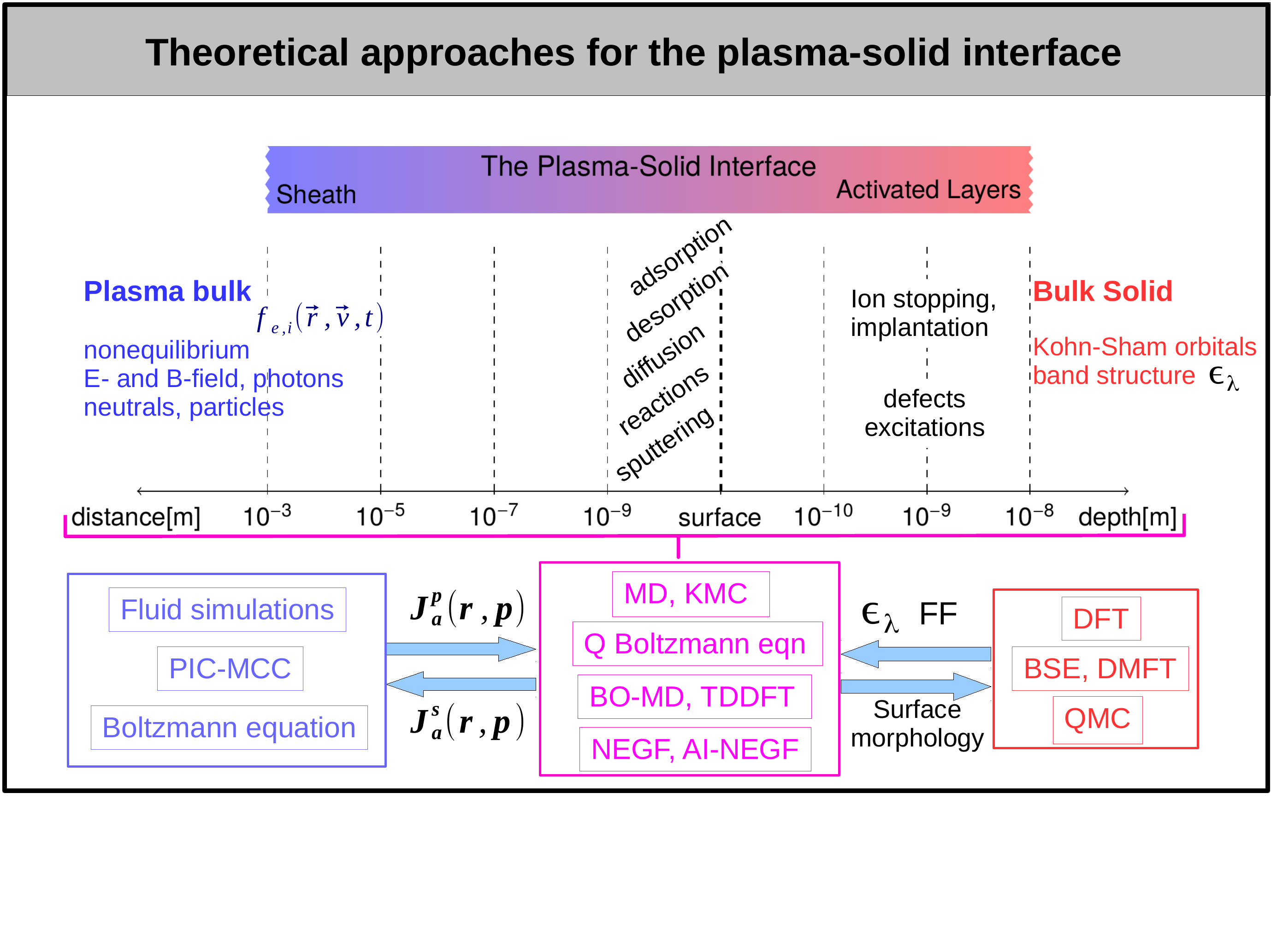} 
  \end{center}
  \vspace{-2.0cm}
  \caption{Theoretical methods for the description of the plasma-solid interface \cite{interface}, as sketched in  Fig.~\ref{fig:psi}. Some of the processes of interest are listed in the figure. Note the dramatically different length scales and the very different properties of the plasma and the solid, requiring fundamentally different methods to be applied on the plasma and the solid side. Standard methods for the bulk solid are Density functional theory (DFT),  Bethe-Salpeter equation (BSE), Dynamical Mean Field Theory (DMFT), and Quantum Monte Carlo (QMC). To simulate  surface processes (central box), additional non-adiabatic (time-dependent) approaches are required: molecular dynamics (MD), Kinetic Monte Carlo (KMC), Quantum Boltzmann eqution, Born-Oppenheimer MD (BO-MD), time-dependent DFT (TDDFT), Nonequilibrium Green functions (NEGF) and \textit{ab initio} NEGF (AI-NEGF). To account for the complex interactions between plasma and solid, the corresponding methods have to be properly linked: plasma simulations should provide the momentum dependent fluxes $\textbf{J}^p_a$ of all species ``a'' to the surface whereas surface simulations deliver the corresponding fluxes $\textbf{J}^s_a$ that leave the surface. Bulk solid simulations provide the band structure $\epsilon_\lambda$ and reactive force fields (FF), whereas surface simulations return the updated surface morphology, chemical modifications etc. For details see text.}
  \label{fig:theory}
  \end{figure*}

Let us discuss, in the following the  theoretical strategy needed to tackle these problems. Aside from the different pressure, length and time scales, the main difficulty is that both sides of the plasma-solid interface are governed by completely different physics: low-density gas-like behavior, in the plasma, versus quantum dynamics of electrons, in the solid, coupled to the lattice dynamics; this situation, the relevant processes and scales are sketched in ~Fig.\ref{fig:theory}.

An accurate simulation of plasma-surface processes, first of all, requires a reliable description of the plasma and the solid. Standard tools in plasma simulations are fluid simulations and kinetic theory (cf. blue box in Fig.~\ref{fig:theory}). Here two main approaches are in use: direct solution of the Boltzmann equation or particle in cell simulations with Monte Carlo collisions (PIC-MCC). The particle dynamics have to be coupled to the dynamics of the electromagnetic field on the basis of Maxwell's equations or the Poisson equation for the electrostatic potential. Finally, these simulations require surface parameters as an input---the fluxes $\textbf{J}^s$---and deliver the corresponding fluxes $\textbf{J}^p$, as an output, cf. Fig.~\ref{fig:theory}. 

To obtain the necessary surface information, one first of all needs to obtain the ground state properties of the solid--the energy spectrum (band structure) and the Kohn-Sham orbitals--which is done by density functional theory (DFT) simulations [Sec.~\ref{s:abinit}], cf. right part of Fig.~\ref{fig:theory}. However, DFT is known to have problems, in particular, in treating materials with strong electronic correlations including various oxides. Here, many-body approaches are being used that include the Bethe-Salpeter equation (BSE), e.g. [\onlinecite{onida_rmp_02}], dynamical mean field theory (DMFT), e.g. [\onlinecite{kotliar_rmp_06}], or quantum Monte Carlo (QMC) for the ground state or finite temperature [\onlinecite{foulkes_rmp_01,DORNHEIM_physrep18}].
  \begin{figure*}
  \begin{center} 
  \hspace{-2.cm}\includegraphics[width=1.1\textwidth]{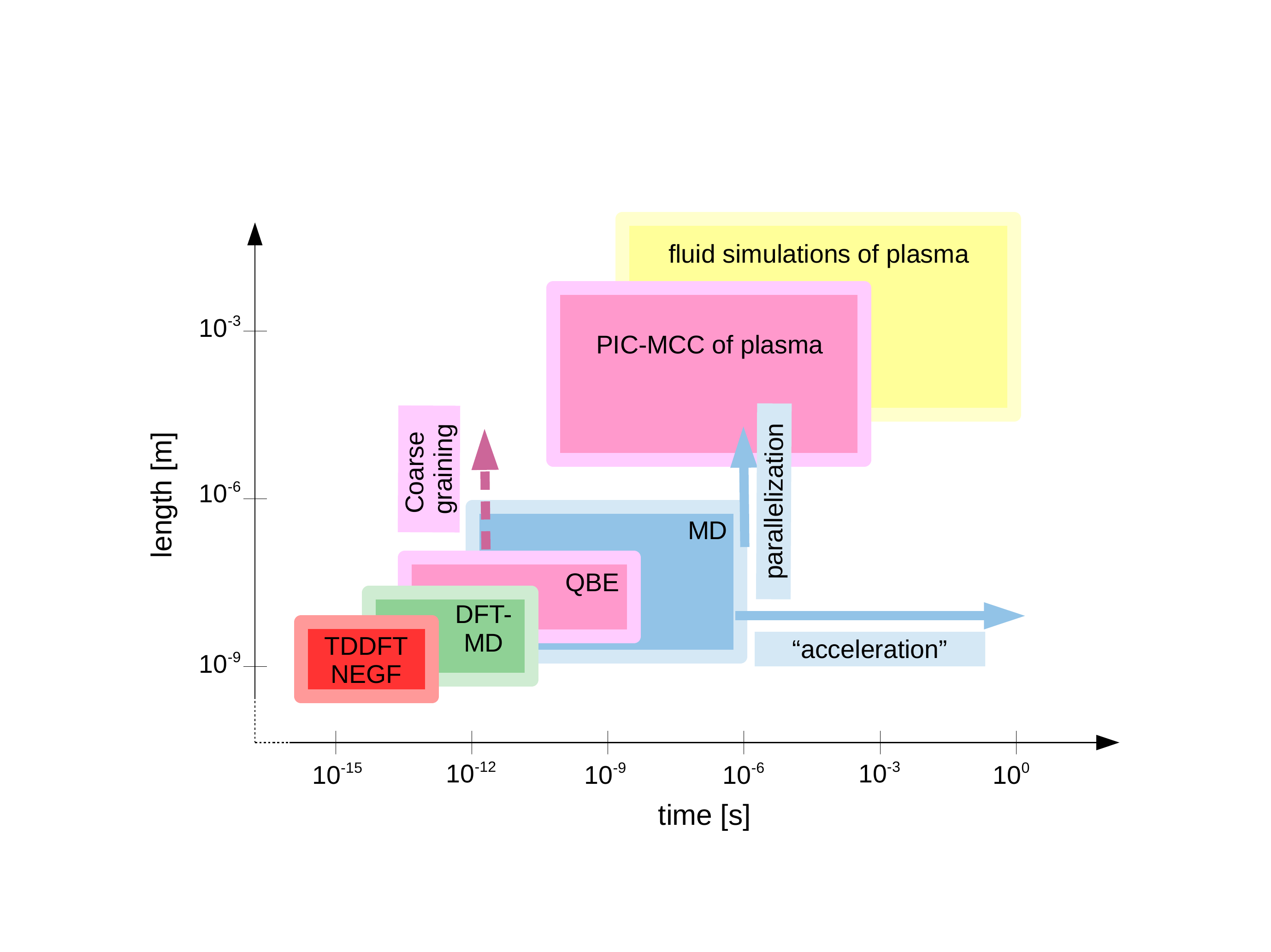} 
  \end{center}
  \vspace{-2.0cm}
  \caption{
  Approximate length and time scales accessible with different  simulation methods for plasma-surface applications that are listed in Fig.~\ref{fig:theory}. The shortcuts are the same as in Fig.~\ref{fig:theory} except for DFT-MD, which is equivalent to BO-MD, and QBE which stands for Quantum Boltzmann equation. Note that the comparison is only qualitative as different methods may apply to different processes. Also, the physical time resolution is often much less (larger minimum scale) than the required time step that is dictated by numerical stability.
  The \textit{ab initio} approaches \textbf{TDDFT} and \textbf{NEGF} resolve the electronic length and time scales and apply to ultrafast processes. \textbf{DFT-MD} has a more limited resolution of electronic relaxation processes. The upper limits of DFT, TDDFT and NEGF are set by the required basis dimension and accessible number of time steps. \textbf{QBE} resolves spatial details on the level of 100 inter-atomic distances and the relaxation time of the electrons in the solid. The upper length limit is determined by the imposed accuracy (level of coarse graining, pink arrow). \textbf{MD} propagates only the heavy particles neglecting electronic degrees of freedom. The accessible simulation dimensions can be increased via parallelization. Simulation times are restricted by the number of time steps and can, in some cases, be extended by additional ``acceleration'' methods (blue arrows).
  On the plasma side, \textbf{PIC-MCC} simulations resolve approximately one tenth of the electron Debye length and one plasma period and may extend to centimeters and milliseconds. \textbf{Fluid simulations} contain an additional coarse graining with respect to the particle velocities which limits their lower scale limits, compared to PIC, but extends their upper limits.
  KMC may, in principle, extend plasma-surface simulations to minutes and millimeters but is not considered here due to its largely uncontrolled character for the present applications, see text. Figure adapted from Ref.~\onlinecite{abraham_phd_18}.   }
  \label{fig:scales}
  \end{figure*} 

Next, if the solid comes in contact with a low-temperature plasma, energetic electrons,  ions or neutrals may excite the electrons of the solid and the lattice. This is already not captured by ground state DFT but requires time-dependent extensions, cf. the approaches listed in the central box of Fig.~\ref{fig:theory}.
\textit{The top row in the central box} lists mesoscopic approaches--molecular dynamics (MD) and kinetic Monte Carlo (KMC)--that do not treat the quantum dynamics of the electrons explicitly. This simplification allows one to access comparatively large time and length scales, see also Fig.~\ref{fig:scales}. In particular with KMC, in principle, one can reach minutes of simulation time and length scales of centimeters. 
KMC has been successfully used in many plasma simulations, e.g. [\onlinecite{marinov_2016, guerra, abraham_jap_15, fujioka_phd_15, polonsky_epjd18}], and it is an integral part of many multiscale simulation concepts because it is very flexible with respect to the inclusion of new processes. 
However, all these processes are treated in a very simplified manner using process rates that often do not include all relevant parameter dependencies. In cases where the complete set of relevant processes is well known, KMC may, nevertheless, be a powerful tool. In contrast, for new problems--such as the plasma-solid interface--where the complete set of events is not known \textit{a priori}, the accuracy and predictive power of KMC is rather limited.
We will, therefore, not discuss KMC in detail here--the interested reader is referred to Refs.~\onlinecite{guerra, marinov_2016,rosenthal_phd_13, abraham_phd_18}.
Instead, here we will concentrate on MD because it has a much stronger foundation and can  retain a first-principle character once the information on the interaction potentials is derived from \textit{ab initio} approaches in cases when electronic and quantum effects are not important. We will give a more detailed discussion in Sec.~\ref{s:mesoscopic}. For an overview on the length and time scales that can be reached by MD and the other methods, see Fig.~\ref{fig:scales}.  

The \textit{second row in the central box} of Fig.~\ref{fig:theory} lists models based on the quantum Boltzmann equation -- the quantum generalization of classical kinetic equations. Here extensive recent work is due to Bronold and co-workers who developed simulations of the charge transfer, electron sticking and other processes. The corresponding approach and some results will be summarized in Sec.~\ref{s:qkin}.

The \textit{third row in the central box} of Fig.~\ref{fig:theory} is devoted to time-dependent simulations that are based on DFT. The first approach is Born-Oppenheimer MD where the ions are moved with classical molecular dynamics whereas the electrons are assumed to follow the ion dynamics adiabatically, thereby remaining in the (time evolving) ground state that is obtained quantum-mechanically, by ground state DFT.
 However, for strong excitation and/or fast processes the adiabatic approximation fails. Even though in plasma-surface interaction the mean excitation of the surface (per atom) may be small, local excitations may be strong, e.g. due to the impact of plasma particles.
 The corresponding non-adiabatic extension of DFT is time-dependent DFT (TDDFT) [\onlinecite{runge_gross}]. TDDFT is successfully being used for many surface processes. Applications to ion stopping have been performed e.g. in Ref. \onlinecite{zhao_2015_comparison}. An overview on this method is presented in Sec.~\ref{s:abinit}.

Finally, the \textit{fourth row in the central box} contains another \textit{ab initio} method: nonequilibrium Green functions (NEGF), e.g. [\onlinecite{balzer-book,schluenzen_cpp16}]. This method generalizes the quantum Boltzmann equation (second row, Sec.~\ref{s:qkin}) to fast processes and is, in particular, well suited to  accurately treat electronic correlation effects in the surface material. First applications to ion stopping were performed recently [\onlinecite{balzer_prb16}]. Finally, we also list 
 \textit{ab initio}  NEGF simulations -- a recently developed combination 
of ground state DFT and NEGF [\onlinecite{marini_2009_yambo}]. An overview on NEGF methods will be presented in Sec.~\ref{s:negf}.

Even though the \textit{ab initio} methods are, obviously, the most accurate theoretical approaches,
they are extremely CPU-time demanding which strongly limits the accessible length and time scales. 
 For example, Born-Oppenheimer  MD simulation require a 
time step around $0.1\dots 1$ femtoseconds, which allows one to treat on the order $100\dots 
1000$ atoms for $1\dots 100$ picoseconds, during a week of simulations on 
massively parallel hardware, e.g. Ref.~\onlinecite{hutter_wires_12}. The demand for TDDFT and NEGF is several orders of magnitude larger.
Typical length and time scales are summarized in Fig.~\ref{fig:scales}. 
Thus, despite their accuracy, it is prohibitive to apply \textit{ab initio} methods to all problems of the plasma-solid interface. They should be applied to those processes where such an involved treatment is without an alternative, in particular, when important processes would be lost otherwise, e.g. via an averaging or coarse graining procedure.

Even though there is an impressive list of applications of all four plasma-surface simulation approaches, until now these have been developed essentially in isolation from each other.
At the same time there is a high need for smart combinations of the different methods to cover the length and time scales of the plasma-solid interface at a sufficient accuracy and to capture all relevant physical and chemical processes. We hope that the present analysis of each of theses methods will highlight their strengths and weaknesses and stimulate such combinations.

\section{Mesoscopic simulation approaches: molecular dynamics. Acceleration and extension concepts}\label{s:mesoscopic}
We start with the method that 
extends to the largest time and length scales of those that are listed in the center of Fig.~\ref{fig:theory}: molecular dynamics [for the discussion of KMC, see Sec.~\ref{s:concepts}]. This method is based on empirical interaction potentials and does not include quantum effects in the dynamics of the particles, in particular no explicit electronic effects. This approach has to be clearly distinguished from \textit{ab initio} MD (or Born-Oppenheimer) MD where the electrons are time-propagated as well using density functional theory simulations (see Sec.~\ref{s:abinit}). Nevertheless, such a semiclassical modeling is often sufficient for the treatment of the dynamics of neutral particles on a surface: diffusion, adsorption and desorption or many chemical reactions--a technique that is well developed in surface science and in theoretical chemistry, e.g. Ref.~ \onlinecite{gross-book}. 
Similarly, MD simulations are well established in low-temperature plasmas, e.g. to compute first principle structural properties of dust particles [\onlinecite{com-plasma_springer_14}] or the diffusion coefficient in a strongly correlated magnetized plasma [\onlinecite{ott_prl_11}].
MD simulations are also actively used in plasma-surface simulations, e.g. [\onlinecite{graves_2009_molecular,neyts2017molecular}], and recent applications include cluster growth [\onlinecite{abraham_jap16}] and sputtering [\onlinecite{brault_frontiers_18}].

The semiclassical MD simulations solve Newton's equations for all particles exactly. The quality of the results, obviously, crucially depends on the accuracy of the input data, most importantly, the effective pair potentials or force fields. These quantities are usually derived from microscopic quantum simulations or are adjusted to reproduce experimental data.   Typically MD simulations for atoms require a time step of the order of $1$ fs and can, in principle, treat huge systems by using massively parallel hardware. For example, Ref.~\onlinecite{nakano_08} reported simulations of a system containing $10^{11}$  atoms that reach times of the order of several milliseconds. However, this is 
presently only possible on the largest supercomputers or on dedicated hardware, 
e.g.~[\onlinecite{piana_13}]. At the same time, even though parallelization allows to reach larger system sizes, it does not help to extend the simulation duration.

Despite these impressive records, it is clear that in the near future MD 
simulations for plasma-surface processes will remain many orders of magnitude short of system 
sizes and length scales needed to compare with experiments. 
In plasma physics, these are minutes and (at least) micrometers, respectively. Therefore,  additional strategies are needed. One way is  of course the use 
of additional approximations leading to simplified models at the expense of 
accuracy and reliability. Here, we are concentrating on other methods that avoid simplifications of the equations of motion and to retain the 
first principles character
of the MD simulations. 
The idea is to 
invoke additional information on the system properties that allow one to 
effectively \textit{accelerate} the simulations and/or to extend them to larger scales 
\textit{without loosing accuracy}.

There exists a variety of acceleration strategies including hyperdynamics 
[\onlinecite{voter_97}], metadynamics [\onlinecite{LaioParrinello2002}] or temperature 
accelerated dynamics [\onlinecite{Sorensen2000}], for additional comments see Sec.~\ref{s:abinit}. A more recent concept is collective 
variable driven hyperdynamics [\onlinecite{Bal2015}] that was reported to achieve, for 
some applications, speed-ups of about nine orders of magnitude. These methods have been successfully applied in surface physics and chemistry, and a more detailed discussion of these very diverse acceleration/extension developments, of their respective strengths and limitations was presented recently in Ref.~\onlinecite{bonitz_psst18}. 

The above methods are not easily applied to the heterogeneous plasma-solid system. Here recently a multi-scale simulation concept has been proposed that overcomes the low-density problem of the gas phase, for details see Ref.~[\onlinecite{brault_frontiers_18}].
Another 
approach developed by two of the present authors [\onlinecite{abraham_jap16, abraham_cpp18}] 
uses a \textit{selective acceleration of some relevant processes} and, thereby, 
achieved speed-ups exceeding a factor $10^9$, see Sec.~\ref{ss:spa}.
The third direction of developments we underline here does not aim at accelerating the \textit{ab 
initio} simulations but to extend them to longer times by a suitable combination 
with analytical models [\onlinecite{franke_2010_diffusion, filinov_psst18_1}]. These methods will be called below 
\textit{Dynamical freeze out of dominant modes} (DFDM) and are briefly discussed in Sec.~\ref{ss:freezout}

%
\subsection{MD simulations employing selective process acceleration (SPA)}\label{ss:spa}
%
%
The authors of reference \onlinecite{abraham_jap16} considered, as an example for plasma-surface interaction, the deposition of gold atoms onto a polymer surface. The MD simulations tracked each individual atom, its diffusion on the surface, the emergence and growth of clusters and, eventually, the coalescence of the latter.  
The influence of the plasma environment was treated statistically by taking into account the impact of energetic ions that leads to the formation of surface defects that trap incoming atoms and prevent their diffusion. Varying the fraction of trapped atoms allows to mimick the flux of ions to the surface. 
In the MD simulations the  isotropic Langevin equation of motion for all gold particles
with the mass $m$ and spatial coordinates $\mathbf{r}=(\mathbf{r}_1, 
\mathbf{r}_2, \ldots)$
were solved:
\begin{equation}
m \ddot{\mathbf{r}} = - \nabla U(\mathbf{r})
- \frac{m}{t_\mathrm{damp}} \dot{\mathbf{r}}
+  \sqrt{\frac{2m k_\mathrm{B}T}{t_\mathrm{damp}}} \,\mathbf{R}(t)\,,
\label{eq:eqmotion_langevin}
\end{equation}
where the potential $U$ describes the interaction between gold particles. For this potential \textit{ab initio} force field data are being used (the MD simulations used the LAMMPS package).
Further, 
$t_\mathrm{damp}$ has the role of a damping parameter,
and
$\mathbf{R}$
is a delta-correlated Gaussian random process.
This random force and the viscous damping simulate the effect of the polymer on the heavy gold particles. This is motivated by the fact that the interaction between gold atoms by far exceeds the one between gold and polymer, so details of the latter are of minor importance.
While the first term on the r.h.s. of
Eq.~(\ref{eq:eqmotion_langevin}) favors cluster formation (gold atoms settle in the minima of the total potential), the last two terms 
induce a diffusive motion with the diffusion coefficient
\begin{equation}
\label{eq:diffusioncoeff}
D=\frac{1}{m} {k_\mathrm{B}T t_\mathrm{damp}}\,.
\end{equation}
Thus, it is  clear that the utilization of the Langevin dynamics allows one to control
the speed of the surface diffusion and bulk by choosing a specific combination of the temperature $T$
and the damping parameter $t_\mathrm{damp}$.
Beyond that, it is possible to add a spatial (or directional) dependence to the diffusion 
coefficient
if one lets the damping parameters depend on the position of the particle. 

Based on the above considerations,
Abraham \textit{et al}.~[\onlinecite{abraham_jap16}] developed a procedure to
simulate the growth of nanogranular gold structures on a thin polymer film.
By choosing appropriate ratios of the damping parameters,
one can make sure that the atoms spend most of the time in the surface layer 
 where they
perform a random walk. 
The use of Langevin dynamics is restricted to the polymer surface 
whereas in the plasma, the dynamics of the gold atoms are purely microscopic. 
This allows one to add particles to the system by creating particles
at the top of the simulation box and assigning them an initial velocity towards the substrate.
Therefore, it is  possible to perform the simulation with values of the 
deposition rates $J_\mathrm{sim}$
and diffusion coefficients $D_\mathrm{sim}$ that are much higher than the values 
in typical
experiments.

In Ref.~[\onlinecite{abraham_jap16}], it was argued that the simulations
yield an adequate description of a real experimental deposition process
if the ratio $J_\mathrm{sim}/D_\mathrm{sim}$ is equal
to the ratio 
$J_\mathrm{exp}/D_\mathrm{exp}$ 
of the corresponding quantities in the experiment.
The idea behind that is that -- at least at the early stage of the
deposition process -- the growth should be essentially
determined by the average distance an atom travels on the surface between 
successive
depositions of atoms. 
Hence, the absolute time of the process is  assumed to be irrelevant.
The results presented in Ref.~[\onlinecite{abraham_jap16}]
were obtained with a time step of $\SI{1}{fs}$,
and a damping parameter for the diffusion in $x$- and $y$-directions of 
$\SI{1}{ps}$. 
The temperature and the deposition rate were set to match the conditions of the 
experimental results in Ref.~\onlinecite{schwartzkopf_2015_real} for the sputter deposition 
of gold on polystyrene.
Using these parameters, the direct MD simulation time for the growth of a thin 
gold film
is roughly $10^9$ times shorter than the corresponding time in the experiment. Or in other words, the duration of the MD simulations could be extended by nine orders of magnitude.

To verify the validity of such a dramatic shift of the time scales and to obtain the applicability limits,
comprehensive tests of the method against experimental results were performed, see also 
Refs.~\onlinecite{abraham_phd_18,abraham_epjd18}
for a discussion.
In particular, as one  accelerates only selected processes, i.e., the deposition 
of atoms 
and the diffusion of atoms on the surface,
one has to make sure that the neglect of other processes,
e.\,g., the relaxation of a cluster structure,
does not lead to artifacts in the simulation results. 
In Ref.~\onlinecite{abraham_jap16}, 
the method was tested by comparing
several quantities describing the evolution of the gold film morphology
with the results of time-resolved in situ grazing incidence x-ray scattering (GISAXS) experiments of Schwartzkopf \textit{et al.} [\onlinecite{schwartzkopf_2015_real}]. Indeed, many of the observed features could be reproduced by the simulations, for film thicknesses
up to \SI{3}{nm}. This thickness corresponds to an impressive effective 
simulation
time of \SI{367}{s} which is directly suited for comparison with measurements. 

The present approach of selective acceleration of dominant processes can be 
generalized to other systems as well. 
A recent application concerned 
the deposition and growth of bi-metallic clusters on a polymer surface 
[\onlinecite{abraham_cpp18}] where the acceleration allowed one to study the very slow 
process of demixing of the two metals. Applications of this approach to various plasma processes should be possible as well. 
In addition to the deposition of neutral atoms, 
the method also allows one to describe the impact of ions and the growth of 
charged clusters.

%

\subsection{Dynamical freeze out of dominant modes (DFDM)}\label{ss:freezout}
We now discuss an approach that does not accelarate MD, as the ones mentioned above, but takes advantage of the intrinsic hierarchy of relaxation processes existing in any many-body system. When a system is excited, typically, small scale processes and correlation effects will tend to equilibrate fast whereas large scale effects such as particle transport will occur on longer time scales. In between these scales one expects the establishment of equilibrium (or stationary) velocity distributions, for details and more examples, see Refs.~[\onlinecite{bonitz_qkt,bonitz_psst18}]. This means that during the course of the evolution the information and detail required to describe the entire process is systematically reduced. Finally, in thermodynamic equilibrium the system would be completely described by a few macroscopic variables such as temperature and density. 

This hierarchical character of the evolution is known for a long time and also called ``coarse graining''. It means that a full microscopic description is only needed for the early period of the evolution whereas, for later times, it is sufficient to capture the dynamics of the relevant ``modes'' or degrees of freedom. This has lead to simplified models such as kinetic equations, fluid equations or rate equations where the full N-particle information is mapped onto a limited number of quantities.
Even though each of these equations is an approximation to the full many-body 
equations, these equations are accurate within their respective relaxation stages and time scales. 

Of course, kinetic equations, fluid models or rate equations contain input parameters (such as collision cross sections or collision integrals, transport coefficients or reaction rates, respectively) that are usually derived by applying suitable approximations to the many-body problem. It is the accuracy of these approximations that governs the accuracy of the model. 
Consider, as an example, a master (or rate) equation
\begin{eqnarray}
\frac{dp_i(t)}{dt} &=& \sum\limits_{j\ne i}
\left\{ \Gamma_{j\to i}\, p_j(t) - \Gamma_{i\to j}\, p_i(t)
\right\},\; 
\label{eq:master}   
\\
& &0 \le  p_i(t) \le  1, \quad \sum_i p_i(t)=1\,, \nonumber
\end{eqnarray}
where $i$ is a multi-index numbering the configurations the system can have at some stage of the relaxation which occur with a probability $p_i(t)$. $\Gamma_{i\to j}$ are the transition rates 
(probability per unit time) from state $i$ to $j$. The first term on the right hand
side of Eq.~(\ref{eq:master}) describes processes which increase the probability 
to realize state $i$ (``gain''), whereas the second term describes the analogous 
loss processes. 

Let us now return to our general question about the accuracy of the model (\ref{eq:master}). Such an equation can often be rigorously derived from the underlying classical or quantum equations of motion, and its accuracy is only limited by the accuracy of the transition rates $\Gamma_{i\to j}$ that are typically calculated using various approximation schemes. If, on the other hand, no approximations would be made and a general form be permitted where $\Gamma_{i\to j}\to \Gamma_{i\to j}(t;\{p_k(t)\})$ may depend on time as well as on all $p_i$ for the present and earlier times (memory) this would, in general, result in an exact equation (\ref{eq:master}). 

In fact, in Ref.~[\onlinecite{bonitz_psst18}] it was proposed to obtain the exact expressions for $\Gamma_{a\to b}$ from first principle MD simulations. Imagine that in the course of the evolution the system reaches a state where only a small number of configurations  can be realized the description would much simpler compare to the full N-body dynamics, even if the involved rates are complicated. Moreover, it can be expected that, in many cases, the functional form of the rates $\Gamma_{i\to j}(t;\{p_k(t)\})$ will simplify in the course of the evolution, and they even may converge to stationary values.

This procedure was demonstrated in Refs.~[\onlinecite{filinov_psst18_1, filinov_psst18_2}] for a simple example: the adsorption kinetics of argon atoms on a platinum (111) surface. There it was shown that the sticking probability of the atoms can be derived from an equation of the type (\ref{eq:master}) with just three different states. The transition probabilities between these states were computed by first principle MD, and their time dependence was analyzed. It turned out that, in fact, these probabilities converge to constant values, $\Gamma_{i\to j}(t;\{p_k(t)\}) \to \Gamma_{i\to j}^{\rm EQ}$, within approximately $t=t^{\rm EQ}\approx 20$ ps, i.e. as soon as the adsorbate atoms have equilibrated with the surface. Thus, using the first-principle transition probabilities obtained from MD in the master equation (or rate equation), its solution will be essentially exact, for times $t\ge t^{\rm EQ}$. This allows one to extend first principle-type simulations to times  long enough to compare with experiments [\onlinecite{bonitz_psst18}] without actually performing MD simulations. The reason is that one was able to identify the dominant collective modes that fully describe the system at long times and emerge dynamically during the evolution.

A similar approach was developed by Franke and Pehlke [\onlinecite{franke_2010_diffusion}]. They performed DFT simulations of the diffusion of a 1,4-butanedithiol molecule on a gold surface and also mapped this on a master equation. Even though they did not consider in detail \textit{ab initio} results for the transition probabilities, their results confirm that the type of extension of accurate simulations can also be applied with quantum simulations, instead of semiclassical MD.
The idea of mapping the microscopic dynamics onto a small set of relevant degrees of freedom is also utilized in the computation of the secondary electron emission coefficient. This also leads to a system of coupled rate equations as is explained in  section~\ref{s:qkin}.

\section{Quantum Boltzmann equation}\label{s:qkin}


In the previous section the focus was on semiclassical or even classical methods 
for describing the interaction of atoms with surfaces, based on Newton'a equations 
of motion for the constituents involved and their solution by molecular dynamics 
techniques. Not all surface scattering processes can be treated in that manner. 
In particular, charge-transferring processes require a quantum-mechanical approach.
The most prominent scattering process (cf. Sec.~\ref{s:influences}), which has to be treated quantum-mechanically
and, at the same time, has great relevance for plasma modeling, is secondary 
electron emission from surfaces/plasma walls due to impacting heavy species.

\subsection{Quantum Kinetic Approach to secondary electron emission. Generalized Newns-Anderson model}\label{ss:qke-see}
It has been known for a long time that secondary electron emission is an important process 
in bounded plasmas, affecting the structure of the plasma sheath, the overall charge balance, 
and the operation modii of basically all types of low-temperature discharges~[\onlinecite{LL05}]. To 
quantify the process is, however, a rather challenging task. Experimentally it requires sophisticated 
instrumentation and theoretically it asks for the solution of a scattering problem involving
many-body targets and projectiles~[\onlinecite{Rabalais94,Winter02,Winter07}]. It is thus not
surprising that little is known quantitatively about secondary electron emission from 
plasma-exposed surfaces which, moreover, are usually also insufficiently well characterized. 
The collection and discussion of secondary electron emission data by Petrovi\'c and 
Phelps~[\onlinecite{PP99}] is still the main reference, cf. Sec.~\ref{s:influences}. Only recently the plasma physics 
community initiated a number of new investigations devoted to the 
issue~[\onlinecite{DBS16,MCA15}]. 
\begin{figure*}[t]
  \includegraphics[width=\linewidth]{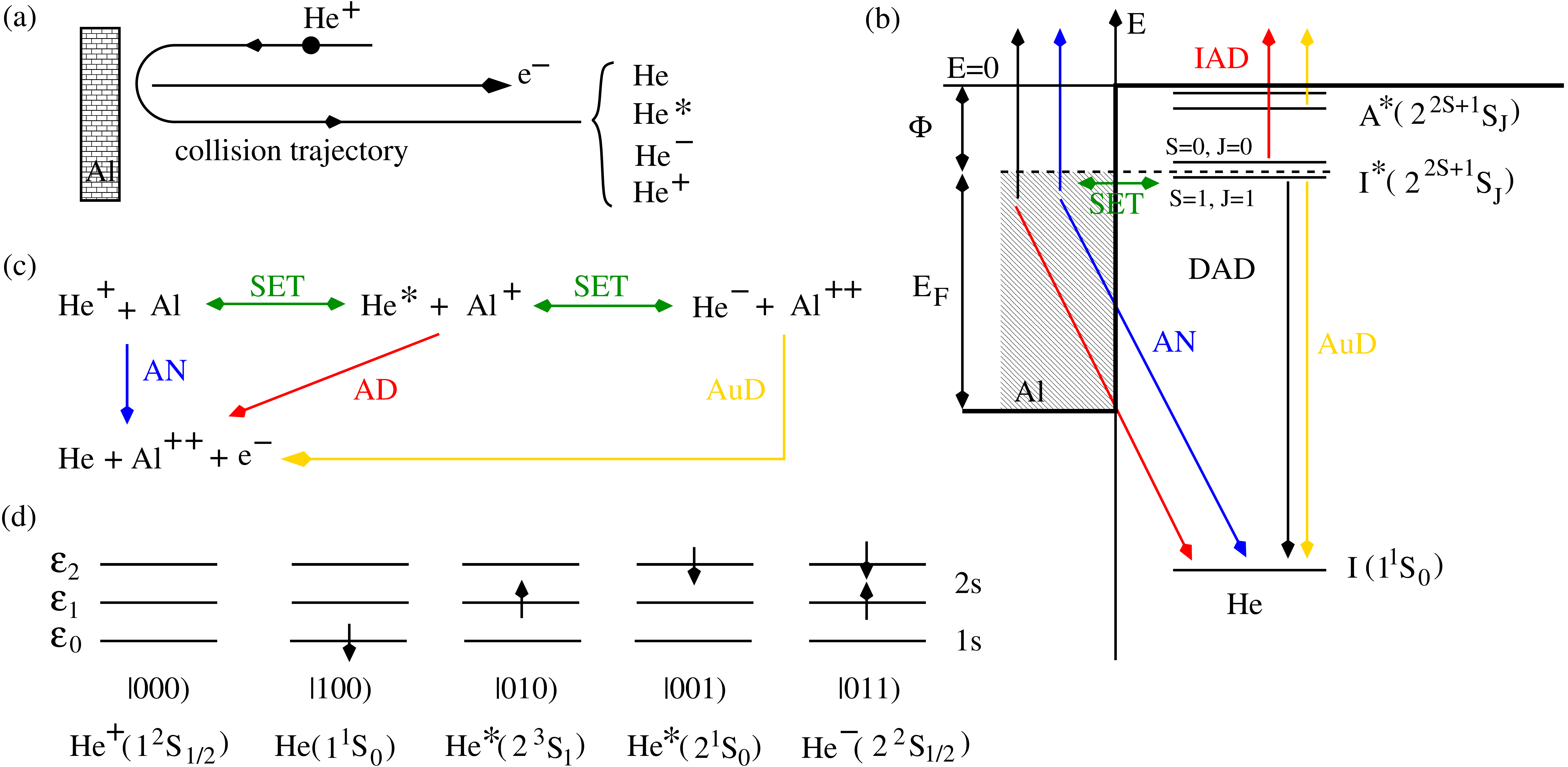}
  \caption{(Color online) Illustration of the main ingredients of an Anderson-News model
   based quantum-kinetic analysis of the neutralization of a helium ion on an aluminum 
   surface characterized by a step potential with depth $E_{\rm F}+\Phi$, where 
   $E_{\rm F}$ is the Fermi energy and $\Phi$ the work function
   of aluminum. The collision trajectory leading to time-dependent matrix elements 
   enforcing a quantum-kinetic analysis is shown in (a) and the 
   reaction channels included in the modeling are indicated in the (on scale) 
   energy diagram (b) and the reaction diagram (c). For simplicity, the projectile 
   levels shown are the ones far away from the surface, level shifts and broadening 
   due to the interaction with the surface are not visualized. As indicated in (c), 
   there are three routes to the projectile groundstate, each one leading to the 
   emission of an electron. The helium ion may capture an electron from the metal by a 
   single-electron transfer (SET), changing its configuration from $\HeliumPositiveIon$ to
   either $\HeliumMetaTrip$ or $\HeliumMetaSing$, which may either Auger de-excite (AD) 
   to $\HeliumGroundState$ or attract another electron from the metal to form a 
   $\HeliumNegativeIon$ ion releasing an electron then by auto-detachment (AuD). In 
   addition to these two routes the $\HeliumGroundState$ configuration may be also reached 
   by Auger neutralization (AN) of the $\HeliumPositiveIon$ ion setting also free an 
   electron. Panel (d) finally depicts the three-level system which can be employed to 
   represent the helium configurations taking part in the collision. Depending on the 
   process the levels act as ionization or affinity levels. Auxiliary bosons are 
   used to assign the functionality needed to the levels. Not included in the three 
   level modeling, since it is unaffected by the collision, is the spin-up electron in 
   the 1s-shell common to all configurations listed in (d). For details see 
   Ref.~\onlinecite{PBF18}.}
  \label{HeNeutralProcesses}
\end{figure*}

In the following we focus on electron emission at low impact energies where the atomic projectile remains outside the surface and emission is driven by the transfer of internal 
potential energy from the projectile to the target. 
In principle there are two 
theoretical approaches to the problem. The first approach attempts to describe the 
processes from first principles, using density functional theory or quantum-chemical 
techniques. There exist various facets of this approach (see, for instance, the work 
of More and coworkers~[\onlinecite{MMM98}] and the monograph~[\onlinecite{Winter07}]) depending on how 
many electronic and lattice degrees of freedom are kept at an ab-initio level. Ideally, 
one keeps all. But this sophistication can be hardly maintained in a realistic description of plasma-surface interaction. The second approach, 
more modest in theoretical detail and to be followed in this subsection, is to keep only the most important degrees
of freedom taking part in the scattering process and to construct effective models 
for them. Typically these are Anderson-Newns models~[\onlinecite{NMB83,YM86,LG90}]
where the collision trajectory of a projectile is prescribed externally, leading to a time-dependent Hamiltonian for the participating electrons. 
The matrix elements of the Hamiltonian can be, of course, calculated from first principles 
but it is more in the spirit of an effective modeling to parametrize the matrix elements 
by physical considerations. The resulting time-dependent Hamiltonian is then fed into a  
quantum-kinetic treatment, based on nonequilibrium Green functions (NEGF), which, if 
combined with pseudo-operator techniques, has the advantage that all collision channels,
which may open-up when the projectile approaches the target surface, can be treated simultaneously. The NEGF approach is the same as listed in the fourth line of the pink box in Fig.~\ref{fig:theory} and will be discussed in some detail in Sec.~\ref{s:negf}. Here we will concentrate on the main steps that are involved in deriving the quantum Boltzmann equation from the NEGF equations and on the derivation of the rate equations model for the relevant degrees of freedom.

The use of Anderson-Newns models for the description of charge-transferring atom-surface 
collisions is well established and has a long history, recent 
applications~[\onlinecite{WGM01,VGB05,MBF12a,MBF12b,IGG13,IGG14,PBF15a,PBF15b,PBF18}] 
differ only in the way the matrix elements are obtained and the number of channels
included. Both of course depend on the scattering process to be modelled. It is beyond 
the scope of this paper to review it here. Instead, we describe the Anderson-News type 
modeling developed at the University Greifswald~[\onlinecite{MBF12a,MBF12b,PBF15a,PBF15b}] 
taking a helium ion hitting an aluminum surface as an example. The description 
is necessarily sparse, details and data for other metals can be found in the work of 
Pamperin, Bronold, and Fehske~[\onlinecite{PBF18}] to which we refer as PBF.  

Figure~\ref{HeNeutralProcesses} summarizes the approach as applied to the collision of 
a positive helium ion with an aluminum surface. In accordance with the general spirit 
of the Anderson-Newns model~[\onlinecite{NMB83,YM86,LG90}] the center-of-mass motion of the ion, 
for simplicity assumed to be normal to surface, is put on a prescribed trajectory. Due 
to the large mass of the projectile, the motion is indeed to a good approximation 
classical. It leads however to time-dependent single- and two-particle matrix elements 
and in turn to the necessity of using quantum kinetics to extract from the model 
the secondary electron emission coefficient $\gamma(\varepsilon_{\vec{q}})$, where 
$\varepsilon_{\vec{q}}$ is the energy of the emitted electron. The electronic degrees 
of freedom of the projectile and the target are treated quantum mechanically, using a 
step potential of depth $E_F+\Phi$ to mimic the aluminum surface, with $E_F$ the Fermi
energy and $\Phi$ the work function, and a few-level system to emulate the electronic
states of the projectile affected by the collision, that is, the ionization and affinity 
levels which may take part in an electron transfer. As indicated, there will be 
typically more than one ionization and affinity levels involved. As a result, electron
emission can occur via many channels. In the case depicted there are three main channels: 
Auger neutralization (AN), direct (DAD) and indirect (IAD) Auger de-excitation, and 
auto-detachment (AuD). All three can be included in the present NEGF approach and the quantum Boltzmann equation derived from it. It 
is sensible to parametrize the model Hamiltonian such that, far away from the surface, 
the energy levels of the projectile coincide with the experimental ionization energies 
and electron affinities of the isolated helium projectile. As it comes closer to the  surface it starts to interact with it leading to level shifts and broadening (not shown 
in the figure), as well as to electron transfer due to Auger- and single-electron processes. 
For the results presented below the matrix elements of the model Hamiltonian were 
obtained from mean-field wave functions and physical considerations, taking image 
charges~[\onlinecite{Gadzuk67a,Gadzuk67b}] as well as tunneling through potential 
barriers~[\onlinecite{Propst63,PA90}] into account in case they arise. Details are given 
in PBF~[\onlinecite{PBF18}]. If instead of the physics-guided manner the matrix elements are obtained from first-principles, the quantum-kinetic treatment of the Hamiltonian  described below remains the same. 

The Hamiltonian describing the physics sketched in Fig.~\ref{HeNeutralProcesses}
is best written down in the notation of second quantization, that is, in terms of
annihilation and creation operators  [cf. Sec.~\ref{s:negf}]. Distinguishing between electron states 
belonging to the step potential, the few level system, and the unbound continuum, 
three types of Fermi operators are introduced: $c_{\vec{k}\sigma}^{(\dagger)}$, 
$c_{\vec{q}\sigma}^{(\dagger)}$, $c_n^{(\dagger)}$ annihilating (creating) an
electron with spin $\sigma$ in, respectively, a surface state $|\vec{k}\rangle$, 
a continuum state $|\vec{q}\rangle$, and a projectile state $|n\rangle$. From a 
calculational point of view it is advantageous to replace the operators 
$c_n^{(\dagger)}$ by pseudo-operators, $e^{(\dagger)}$, $d^{(\dagger)}$, and 
$s_{n\sigma}^{(\dagger)}$ defined by 
\begin{align}
        \label{allPseudoStates}
        \vert000\rangle &= e^\dagger\vert {\rm vac}\rangle \,,\,
        \vert011\rangle = d^\dagger\vert {\rm vac}\rangle \,,
        \vert100\rangle = s^\dagger_{1\downarrow}\vert {\rm vac}\rangle \,,\, \nonumber\\
        \vert010\rangle &= s^\dagger_{2\uparrow}\vert {\rm vac}\rangle \,,\,
        \vert001\rangle = s^\dagger_{2\downarrow}\vert {\rm vac}\rangle \,.
\end{align}
They stand for whole electronic configurations of the projectile and not for single
electron states. The number of electrons required for the electronic configurations 
represented by the pseudo-operators determines their statistics. Is the number 
odd (even) the operators obey Fermi (Bose) statistics. The labeling of the 
pseudo-operator is a reminder that the projectile's configurations involve the 
1s and 2s shell of helium. In spectroscopic terms, the configurations included 
are the $\HeliumPositiveIon$ positive ion, the $\HeliumGroundState$ groundstate, 
the $\HeliumMetaTrip$ triplet and $\HeliumMetaSing$ singlet metastable states, 
and the $\HeliumNegativeIon$ negative ion. Employing the reasoning developed 
in~[\onlinecite{MBF12b,PBF15b,PBF15a}], the Hamiltonian describing the neutralization of 
a $\HeliumPositiveIon$ ion on an aluminum (or any other metal) surface within the 
scenario summarized in Fig.~\ref{HeNeutralProcesses} becomes~[\onlinecite{PBF18}]
\begin{widetext}
\begin{align}
       \label{pseudoHamiltonian}
       H(t) &= \varepsilon^0_{1s\downarrow}(t) s^\dagger_{1\downarrow} s^\phdag_{1\downarrow}
       + \sum_\sigma \varepsilon^*_{2s\sigma}(t) s^\dagger_{2\sigma} s^\phdag_{2\sigma} 
       + \big[\varepsilon^-_{2s\uparrow}(t) + \varepsilon^-_{2s\downarrow}(t)\big]
       d^\dagger d^\phdag 
       + \sum_\sigma \omega_\sigma(t) b^\dagger_\sigma b^\phdag_\sigma
       + \sum_{\vec{k} \sigma} \varepsilon_{\vec{k} \sigma}
         c^\dagger_{\vec{k} \sigma} c^\phdag_{\vec{k} \sigma} \nonumber\\
       &+ \sum_{\vec{q} \sigma} \varepsilon_{\vec{q}\sigma}(t)
        c^\dagger_{\vec{q} \sigma} c^\phdag_{\vec{q} \sigma}
       + \sum_{\vec{k} \sigma} \big[ V^{\rm SET}_{\vec{k}\sigma}(t)
        c^\dagger_{\vec{k} \sigma} e^\dagger s^\phdag_{2\sigma} + \text{H.c.} \big] 
       - \sum_{\vec{k} \sigma} \big[ \sgn(\sigma) V^{\rm SET}_{\vec{k}\sigma}(t) c^\dagger_{\vec{k} \sigma}
        b^\dagger_\sigma s^\dagger_{2-\sigma} d^\phdag + \text{H.c.} \big] \nonumber\\
       &+ \sum_{\vec{k}_1\vec{k}_2\vec{k}^\prime \sigma} \big[ V^{\rm AN}_{\vec{k}_1\vec{k}_2\vec{k}^\prime \sigma}(t)
        c^\dagger_{\vec{k}^\prime \sigma} s^\dagger_{1\downarrow} e^\phdag c^\phdag_{\vec{k}_1 \downarrow}
        c^\phdag_{\vec{k}_2 \sigma}  + \text{H.c.} \big] 
       + \sum_{\vec{k}\vec{k}^\prime \sigma} \big[ V^{\rm DAD}_{\vec{k}\vec{k}^\prime \sigma}(t)
         c^\dagger_{\vec{k}^\prime \sigma} s^\dagger_{1\downarrow}
         c^\phdag_{\vec{k} \sigma} s^\phdag_{2\downarrow} + \text{H.c.} \big] \nonumber\\
        &+ \sum_{\vec{k}\vec{q}\sigma} \big[ V^{\rm IAD}_{\vec{k}\vec{q}\sigma}(t) c^\dagger_{\vec{q} \sigma}
         s^\dagger_{1\downarrow} c^\phdag_{\vec{k} \downarrow} s^\phdag_{2\sigma}  + \text{H.c.} \big] 
       + \sum_{\vec{q}} \big[ V^{\rm AuD}_{\vec{q}}\, c^\dagger_{\vec{q} \uparrow} s^\dagger_{1\downarrow} d^\phdag 
        + \text{H.c.} \big] \,.
\end{align}
\end{widetext}
where we included auxiliary Bose operators $b_\sigma^{(\dagger)}$ enabling us to switch
between projectile states with defects in their internal energy~[\onlinecite{PBF18}]. The physical 
meaning of the various terms is easy to grasp. For instance, the second last term stands for 
indirect Auger de-excitation (IAD) of the projectile (red arrows in Fig.~\ref{HeNeutralProcesses}b),
that is, the creation of the groundstate of the projectile ($s^\dagger_{1\downarrow}$) by 
creating an electron in an unbound continuum state ($c^\dagger_{\vec{q}\sigma}$) and 
annihilating either a singlet ($s_{2\downarrow}$) or a triplet ($s_{s\uparrow}$) metastable
state and an electron bound in the surface ($c^\dagger_{\vec{k}\downarrow}$). The
electron in the continuum is the secondary electron released in the course of Auger 
de-excitation. 

The time-dependence of the Hamiltonian~(\ref{pseudoHamiltonian}) reflects the dependence 
of most of its matrix elements on the actual position $z(t)$ of the projectile in front 
of the surface. The calculation of the matrix elements is analytically and numerically 
rather demanding. In fact, most of the computation time required for the quantum-kinetic 
modeling of secondary electron emission is allocated to the numerical evaluation of the matrix 
elements (and the selfenergies they give rise to). Explicit expressions for the matrix 
elements worked out along the lines developed in Refs.~\onlinecite{MBF12b,PBF15a} are given in 
PBF~[\onlinecite{PBF18}]. The time-dependencies of the energies $\varepsilon_{1s\downarrow}^0(t)$, 
$\varepsilon^*_{2s\sigma}(t)$, $\varepsilon^-_{2s\sigma}(t)$, and 
$\varepsilon_{\vec{q}\sigma}(t)$ are caused by long-range polarization effects and 
short-range non-orthogonality corrections. Assuming the projectile staying a few Bohr 
radii in front of the surface, the latter can be neglected while the former can be 
approximated by image shifts. The time-dependencies of the Coulomb matrix elements 
$V^{\rm AN}_{{\vec{k}_1}{\vec{k}_2}{\vec{k}^\prime\downarrow}}(t)$ 
for Auger neutralization, $V^{\rm DAD}_{\vec{k}\vec{k}\prime\sigma}(t)$ for direct Auger 
de-excitation, and $V^{\rm IAD}_{\vec{k}\vec{q}\sigma}(t)$ for indirect Auger de-excitation, 
as well as the time-dependence of the single-electron transfer matrix element
$V_{\vec{k}\sigma}^{\rm SET}(t)$ arise from the overlap of projectile and target wave 
functions which of course also depends on the separation $z(t)$ of the projectile and target. 
\subsection{Derivation of the quantum Boltzmann equation for SEE}\label{ss:qbe-see}
Once the model is constructed it is analyzed within the quantum kinetic approach pioneered  
by Langreth and co-workers~[\onlinecite{LN91,SLN94a,SLN94b}]. As result one obtains a linear set of 
ordinary first order differential equations for the probabilities with which the projectile's
electronic configurations occur in the course of the collision. In the case under discussion, 
one obtains equations for the occurrence probabilities of the ion, the groundstate, the two 
metastable states and the negative ion of the helium projectile. Three steps are required 
to obtain the rate equations. First, one has to set up two-time Dyson equations for the 
projectile Green functions $G$, cf. Eq.~(\ref{eq.kbe}). Second, the selfenergies $\Sigma$ on the right hand side contain all interaction effects and have to be calculated. In the absence 
of strong electron-electron correlations on the projectile, the non-crossing approximation suffices for that 
purpose. Finally, the time integrals entering the selfenergies and the Dyson equations (\ref{eq.kbe})
are evaluated within a saddle-point approximation utilizing the fact that various functions 
are peaked around the time-diagonal. For more details, see~Refs.~\onlinecite{LN91,SLN94a,SLN94b} as well
as Refs.~\onlinecite{MBF12b,PBF15a,PBF15b,PBF18}. Following this reasoning one obtains the following system of rate equations
\begin{widetext}
\begin{align*}
\label{rateEq}
\frac{d}{dt}
\begin{pmatrix}
n_+ \vspace{1.25mm}\\
n_\uparrow \,\vspace{1.25mm}\\
n_\downarrow \,\vspace{1.25mm}\\
n_- \vspace{1.25mm}\\
n_g\,
\end{pmatrix}
=
\begin{pmatrix}
-[\Gamma_\uparrow^<\! +\! \Gamma_\downarrow^<\! +\! \Gamma_{\rm AN}^<] & \hspace*{-5mm}\Gamma_\uparrow^> & \hspace*{-5mm}\Gamma_\downarrow^> & \hspace*{-5mm} 0 & \,\,0 \hspace*{1mm}\vspace{1.25mm}\\
\Gamma_\uparrow^<  & \hspace*{-5mm}-[\Gamma_\uparrow^>\! + \!\Gamma^<_{-,\downarrow} \!+\! \Gamma_{\rm IAD\uparrow}^<] & \hspace*{-5mm} 0 & \hspace*{-5mm}\Gamma^>_{-,\downarrow} & \,\,0 \hspace*{1mm}\vspace{1.25mm}\\
\Gamma_\downarrow^< &  \hspace*{-5mm}0 & \hspace*{-5mm}-[\Gamma_\downarrow^> \!+\! \Gamma^<_{-,\uparrow} \!+\! \Gamma_{\rm IAD\downarrow}^< \!+\! \Gamma_{\rm DAD\downarrow}^<]  & \hspace*{-5mm} \Gamma^>_{-,\uparrow} & \,\,0 \hspace*{1mm}\vspace{1.25mm}\\
0 & \hspace*{-5mm}\Gamma^<_{-,\downarrow} &  \hspace*{-5mm}\Gamma^<_{-,\uparrow} & \hspace*{-5mm} -[\Gamma^>_{-,\uparrow} \!+\! \Gamma^>_{-,\downarrow} \!+\! \Gamma^<_{\rm AuD}] & \,\,0 \hspace*{1mm}\vspace{1.25mm}\\
\Gamma_{\rm AN}^<  & \hspace*{-5mm}\Gamma_{\rm IAD\uparrow}^< & \hspace*{-5mm}\Gamma_{\rm IAD\downarrow}^< \!+\! \Gamma_{\rm DAD\downarrow}^< & \hspace*{-5mm} \Gamma^<_{\rm AuD} & \,\,0 \hspace*{1mm}
\end{pmatrix}
\cdot
\begin{pmatrix}
n_+ \vspace{1.25mm}\\
n_\uparrow \,\vspace{1.25mm}\\
n_\downarrow \,\vspace{1.25mm}\\
n_- \vspace{1.25mm}\\
n_g \,
\end{pmatrix}  \,
\end{align*}
\end{widetext}
where $n_+(t)$, $n_\uparrow(t)$, $n_\downarrow(t)$, $n_-(t)$, and $n_g(t)$ denote, respectively, 
the occurrence probabilities at time $t$ for the positive ion, the triplet and singlet 
metastable state, the negative ion, and the groundstate of the projectile. Expressions for the 
time-dependent rates $\Gamma_{...}^\gtrless(t)$, a discussion of the physical content of the rate 
equations as well as a route for obtaining the secondary electron emission coefficient $\gamma_e$ 
and the energy spectrum $\gamma_e(\varepsilon_{\vec{q}})$ of the emitted electron from the solution 
of the rate equations are given in PBF~[\onlinecite{PBF18}].

\begin{figure*}[t]
  \begin{minipage}{0.5\linewidth}
  \includegraphics[width=0.99\linewidth]{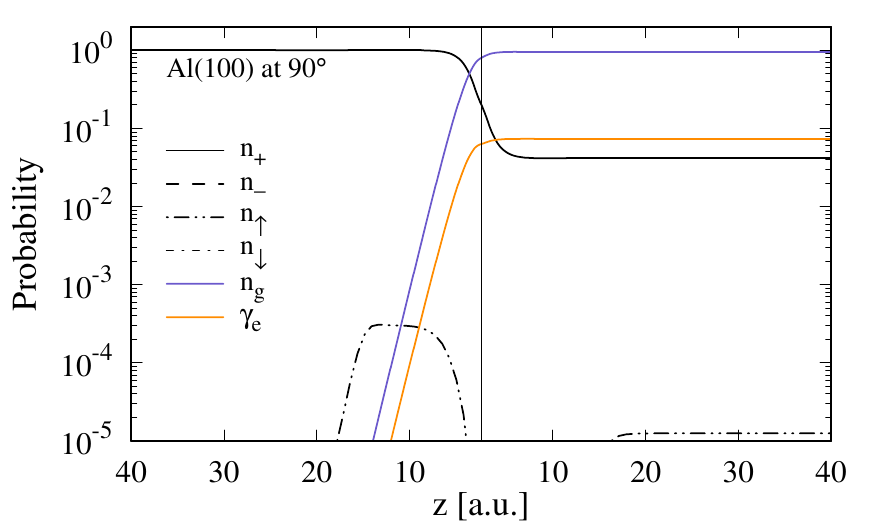}
  \end{minipage}\begin{minipage}{0.5\linewidth}
  \includegraphics[width=0.99\linewidth]{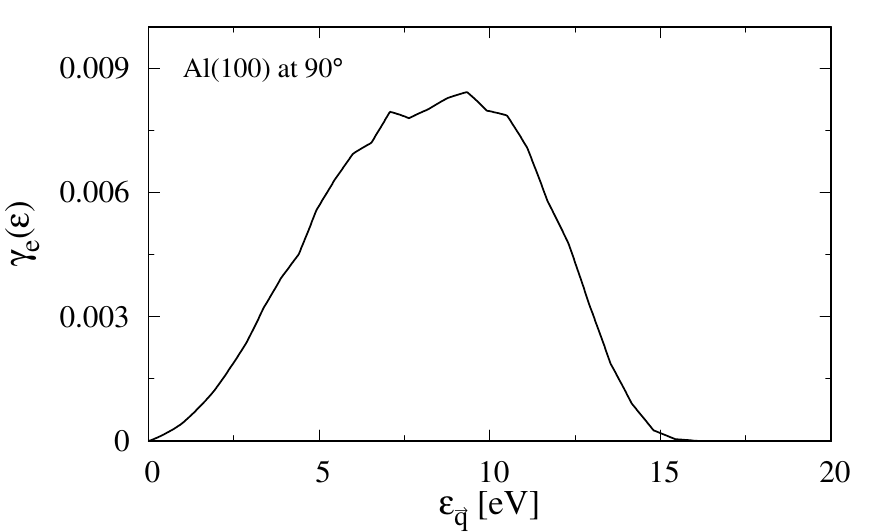}
  \end{minipage}
  \caption{(Color online) Left panel: Instantaneous probabilities $n_+(t)$, 
  $n_-(t), n_\uparrow(t), n_\downarrow(t)$, and $n_g(t)$ for the projectile to be at 
  time $t$ in the $\HeliumPositiveIon$, $\HeliumNegativeIon$, $\HeliumMetaTrip$, 
  $\HeliumMetaSing$, and the $\HeliumGroundState$ configuration together with 
  the instantaneous probability $\gamma_e(t)$ for emitting an electron.
  The projectile hits the aluminum surface as a positive ion (solid black line) 
  with a kinetic energy $E_{\rm kin}=60$\, eV and an angle of incidence with 
  respect to the surface of $\varphi=90^\circ$. The thin vertical line denotes 
  the turning point $z_{\rm TP}=2.27$, separating the incoming (left) form the 
  outgoing (right) branch of the collision trajectory. The final probabilities,
  after the collision is completed, which are also the numbers relevant for plasma 
  modeling, are the values at $z=40~a_{\rm B}$ on the outgoing branch. Right panel: 
  Energy spectrum $\gamma_e(\varepsilon_{\vec{q}})$ of the emitted electron after the 
  collision is completed.
  }
  \label{HeNeutralData}
\end{figure*}

To indicate the type of data which can be produced by the quantum-kinetic modeling of 
secondary electron emission, Fig.~\ref{HeNeutralData} shows numerical results for a 
positive helium ion hitting perpendicularly an aluminum surface with $E_{\rm kin}=60\,{\rm eV}$
using the material parameters listed in PBF~[\onlinecite{PBF18}]. The left panel 
depicts the instantaneous occurrence probabilities $n_+(t), n_-(t), n_\uparrow(t)$,
$n_\downarrow(t)$, and $n_g(t)$ as well as the instantaneous probability $\gamma_e(t)$ for 
emitting a secondary electron. The energy spectrum of the emitted electron is plotted in 
the right panel. The projectile starts in the $\HeliumPositiveIon$ configuration at a distance
$z=40\,a_{\rm B}$, where $a_{\rm B}$ is the Bohr radius, moves along the trajectory shown
in Fig.~\ref{HeNeutralProcesses}a towards the turning point $z_{\rm tp}=2.27\, a_{\rm B}$,
where it is specularly reflected to move back to the distance $z=40\,a_{\rm B}$ to complete
the collision. The probabilities for finding the projectile at the end of the collision in 
any one of the configurations included in the modeling can be read off from the left panel. 
For instance, the probability for returning in the groundstate configuration 
$\HeliumGroundState$ is $n_g(t_{\rm max})\approx 0.96$ whereas the probability for surviving
the collision in the $\HeliumPositiveIon$ configuration is $n_+(t_{\rm max})\approx 0.04$. 
There is also a very small probability $n_\uparrow(t_{\rm max})\approx 10^{-5}$ to come back 
in the $\HeliumMetaTrip$ triplet configuration. The probability for electron emission, 
the secondary electron emission coefficient $\gamma_e=\gamma_e(t_{\rm max})\approx 0.07$. 
It has the correct order of magnitude suggesting that by careful testing, benchmarking, and 
comparison with experimental data the semi-empirical approach can be advanced to a level 
where it can produce very reliable data. The energy spectrum in the right
panel indicates that the secondary electron can have rather high energies. It can thus 
become chemically very active in the gas discharge. A detailed analysis shows that for the 
aluminum surface secondary electron emission is dominated by Auger neutralization. For other 
metals, Auger de-excitation may become also important. The relative importance of the two 
channels depends on the line-up of the projectile's affinity and ionization levels with the 
Fermi level of the metal. It can thus be controlled by a judicious choice of the metal and 
the projectile, that is, in the plasma context, of the wall material and the composition of 
the background gas.

The presentation of the quantum Boltzmann equation approach for charge-transferring 
atom-surface collisions was couched in a particular application: secondary electron 
emission. It can be, however, also applied to other atom-surface scattering processes
affecting the electronic structure of the projectile-target system. The 
pseudo-particle representation of the projectile states opens the door for handling, within a single Hamiltonian, complex collisions, involving more than one channel. 
Numerically the approach is rather involved, preventing thus a quick production of 
surface parameters needed in plasma modeling. The main obstacle is the calculation 
of the matrix elements. So far they have been mostly obtained in a semi-empirical 
manner, guided by physical considerations and experimental data. Using \textit{ab initio}
techniques, such as DFT [cf. Sec.~\ref{s:abinit}], instead may eliminate some of the uncertainties of the matrix elements 
but the numerical effort will remain the same or even increase. Making the 
quantum Boltzmann equation-based modeling of charge-transferring atom-surface collisions numerically 
more efficient is one of the challenges for the future. 

Finally, we notice that the quantum Boltzmann equation cannot resolve ultra-short time and length scales of the solid surface that are related electron correlations (cf. Fig.~\ref{fig:scales}). The extension of quantum kinetic theory into these ranges can be achieved using the method of Nonequilibrium Green functions which is discussed in detail in Sec.~\ref{s:negf}. There we will also present a more accurate treatment of the solid surface that is complementary to the present approach.



%
%
\section{DFT-based {\it ab initio} simulations}\label{s:abinit}

\subsection{Born-Oppenheimer MD}\label{ss:bomd}

Density functional theory (DFT) based methods constitute the most 
widely applied approach to calculate the electronic structure in 
physics and chemistry. It can be combined with Born-Oppenheimer Molecular 
Dynamics (BOMD) to simulate the dynamics of atoms within molecules and 
solids [\onlinecite{Marx2009}]. {\it Ab initio} BOMD has been applied successfully 
by many groups, see  e.g. Refs.\ \onlinecite{hafner_2008_initio,Gross_O2Pt_2016,Kuehne2014}. 
The BOMD approach starts from the Born-Oppenheimer (BO) approximation for 
the coupled system 
of nuclei and electrons [\onlinecite{Baer2006}], where all non-adiabatic coupling 
terms between the different BO surfaces of the electronically excited states 
are neglected. Furthermore, the classical limit for the motion of the nuclei 
is assumed, ignoring geometrical phases occurring around conical 
intersections [\onlinecite{Baer2006}]. The equation of motion (EOM) reduces 
to a set of coupled Newton's equations. Like in semi-classical MD described 
in Sect.\ IV, in case of an ensemble of $N$ atoms the micro-canonical MD 
simulation ``boils down'' to the integration of
\begin{equation}
 M_i \ddot {\bf R}_i = {\bf F}_i = -\nabla_{{\bf R}_i} U_0({\bf R}_1, ..., {\bf R}_N)\,
\label{eq:BOMD_Newton_EOM}
\end{equation}
starting from some suitable initial conditions.
$U_0({\bf R}_1, ..., {\bf R}_N)$ denotes the ground state potential energy surface 
(PES) of the system. Canonical ensembles can be simulated by coupling to a Nos\'{e}-Hoover 
thermostat [\onlinecite{Nose1984}]
or Nos\'{e}-Hoover chain thermostat [\onlinecite{Marx2009}].
Alternatively, the Langevin scheme can be used as detailed in Sect.\ IV.

As already pointed out, rare events pose a serious 
problem to any MD scheme, and even more so to the very CPU-time intensive 
{\it ab initio} based MD programs. Thus, for the determination of reaction and 
diffusion energy barriers, algorithms like the nudged elastic band method 
(NEB, CI-NEB) [\onlinecite{JCP:H00_TAN,JCP:H00_CI}] are in use. Together with 
calculated vibrational frequencies, reaction and diffusion rates can be 
estimated within transition state theory (TST) [\onlinecite{JPCS:VIN57}]. 
The rates can be entered into a Master equation 
approach 
or kinetic Monte Carlo simulations as described in Sec.\ref{s:mesoscopic}. In addition, 
Meta-dynamics approaches, as described by Laio and Parrinello [\onlinecite{LaioParrinello2002,Laio2008}] can help to escape local or 
global potential energy minima and explore further regions of the PES.
A general difficulty when applying Meta-dynamics is the judicious choice of
few collective variables for the simulation.

While not without open basic problems, total energy calculations based 
on density functional theory (DFT) for the electronic ground state -- together 
with a suitable approximation to the exchange-correlation energy 
functional -- constitutes today's standard approach in electronic structure 
theory -- both in physics and in 
chemistry [\onlinecite{Martin2004,Marx2009,Burke2012,Becke_Perspective_2014,Yu_Perspective_2016}]. This 
is in general ascribed to, in case of many applications, a favorable balance between 
the accuracy of the calculated total energy differences and forces 
acting on the ions as required for MD on the one hand and the (nonetheless 
still very large) computational costs on the other hand. DFT describes the exact 
mapping of the electronic ground state of the interacting many-particle Hamiltonian 
onto the ground state of a system of fictitious spin-1/2 particles not 
interacting with each other, which are moving in an effective potential 
that itself depends on the electron density [\onlinecite{Hohenberg1964, Kohn1965}]. 
The crucial point for practical applications is the required approximation 
to the exchange-correlation (XC) energy functional. A careful choice of 
this approximation for the problem in question is essential for the accuracy 
of the obtained results. Various approximations are available.  
A universally applicable approximation to the XC energy functional giving 
chemical accuracy is lacking -- this constitutes a field of active current 
research [\onlinecite{Burke2012,Becke_Perspective_2014,Yu_Perspective_2016,Klimes2012}]. 
In many practical computations one of the generalized gradient approximations 
GGA-PW91 [\onlinecite{Perdew1992_PW91}] or GGA-PBE [\onlinecite{Perdew1996_PBE}] are 
applied. Other approximations comprise meta-GGAs [\onlinecite{Tao2003}], hybrid functionals, 
and many other approaches, see {\it e.g.} Refs. \onlinecite{Burke2012,Becke_Perspective_2014,Yu_Perspective_2016}
for reviews. In view of MD simulations relevant for plasma physics of, 
{\it e.g.}, noble gas atoms interacting with a metal surface, 
van der Waals forces (which cannot be represented
by semi-local XC functionals like the local density approximation, 
LDA, or the GGAs) have to be considered [\onlinecite{Klimes2012,filinov_psst18_1}].
Furthermore, as soon as open-shell atoms or other spin-polarized systems 
come into play, Spin-DFT has to be used [\onlinecite{Martin2004}].  

Widely used DFT total-energy computer programs ({\it e.g.} VASP 
[\onlinecite{CMS:VASP96,PRB:VASP93,PRB:VASP94,PRB:VASP96}], the Quantum Espresso package [\onlinecite{Giannozzi_QE_2009, Giannozzi_QE_2017}] and various other programs) 
employ super-cells, which are repeated periodically 
in all three spatial directions, to simulate surfaces, clusters, etc., 
a plane-wave expansion of the Kohn-Sham orbitals and special {\bf k}-points 
for approximate Brillouin zone integration [\onlinecite{Mattsson2005}]. 
The electron--ion interaction is represented 
by pseudopotentials [\onlinecite{Martin2004}] ({\it e.g.} norm-conserving 
[\onlinecite{Hamann1989,Trouiller1991,Fuchs1999}], ultrasoft [\onlinecite{Vanderbilt1990}], 
projector augmented wave (PAW) [\onlinecite{PRB:PAW94,PRB:PAW99}] pseudopotentials). 
In this way only the valence band states have to be computed, while
the effect of the (frozen) core electrons is accounted for by the 
pseudopotentials. Again, special attention is required in case of MD 
simulations for plasma--surface interactions: If the kinetic energy 
of a projectile is so large that the inter-atomic separations become 
small during a collision, the pseudopotential approximation may 
become invalid and the interaction between the core electrons may 
have to be accounted for explicitly.

\subsection{Time-dependent DFT}\label{ss:tddft}

An ion from a plasma interacting with a solid 
surface [\onlinecite{Modinos1987,Brako1989,Kimmel_1993_alkali_ion_Cu_resonant_charge_transfer,Winter_PhysRep_2002,Race_RepProgPhys_2010,wucher_2011_kinetic}] 
results in inherently electronically non-adiabatic dynamics  -- which 
is beyond the realm of electronic ground state theory. 
Already the initial state, 
corresponding to an ion far away from the surface, is electronically strongly 
excited, although the individual parts of the system, {\it i.e.} the solid 
and the projectile, may initially be in their respective electronic ground 
state. 
Also the scattering of faster than thermal atoms 
at a metal surface can result in electronic energy dissipation, the description of which requires a simulation that accounts for electronically
nonadiabatic effects (for an example see Ref.\ \onlinecite{Lindenblatt2006b} for 2 -- 10 eV H atoms impinging on an Al(111) surface). However, the 
deviation from the BO-surface is distinctly smaller in this case.
In surface chemistry, electronically non-adiabatic effects have been 
observed experimentally (see., 
{\it e.g.} Refs.~\onlinecite{Nienhaus_2002_chemicurrents,Diesing2016,Buenermann1346_2015,Wodtke2016})
and described theoretically by using various approaches including {\it e.g.} 
electronic friction [\onlinecite{Rittmeyer_2015_LDFA,Rittmeyer2018,Alducin2017,Janke2015,Kroes2014,Monturet2010}],
model Hamiltonians [\onlinecite{Mizielinski2005,Mizielinski2007,Mizielinski2008,Mizielinski2010,Bird2008}], or
time-dependent density functional theory (TDDFT) 
[\onlinecite{Lindenblatt2006,Lindenblatt2006a,Lindenblatt2006PRL,Grotemeyer2014,Timmer_2009_H_Al111_perturbation}].
Furthermore, also ion-atom collisions [\onlinecite{Zhao_2015_H+Be}] and 
ions impinging onto a metal cluster [\onlinecite{Moss_2009_Li+_AlCluster,Castro2012_H+Li4}], 
colliding with  carbon nanostructures [\onlinecite{Krasheninnikov2007}], 
graphene fragments [\onlinecite{Bubin2012}], 
graphene or boron nitride (BN) [\onlinecite{zhao_2015_comparison,ojanpera_prb_14}],
the collision of Cl, Cl$^-$ with a MoSe$_2$ monolayer [\onlinecite{Wang2015}] as well as the 
electronic stopping of atoms or ions moving through a 
solid [\onlinecite{Yost2017,Schleife_2015_H_He_Al,Correa_2012_H_in_Al,Zeb2012,Ullah2015}] 
have been simulated using TDDFT.

TDDFT [\onlinecite{Marques_TDDFT_I,Marques_TDDFT_II,Ullrich_TDDFT,Ullrich2014,Maitra_Perspective_2016}] 
is based on the Runge--Gross theorem [\onlinecite{Runge1984}]. Under certain 
restrictions to the single particle potential, and given a fixed initial 
state $|\Psi(t_0)\rangle$, the electron density $n({\bf r}',t')$, $t' \in [t_0,t]$ 
of a finite interacting many-particle system determines the time-dependent 
single particle potential $v({\bf r},t)+c(t)$ apart from an arbitrary function 
$c(t)$ of the time [\onlinecite{Marques_TDDFT_I,Maitra_Perspective_2016}]. The interacting system can be 
mapped onto a system of fictitious spin-1/2-fermions with the same density 
not interacting with each other but moving in a time dependent 
effective Kohn-Sham potential [\onlinecite{Gross_Kohn_1985,Marques_TDDFT_I}]
\begin{eqnarray}
 v_{\rm KS}({\bf r},t) &=& v({\bf r},t) + 
 \int d^3{\bf r}' \frac{n({\bf r}',t)}{|{\bf r}-{\bf r}'|}\ + \nonumber \\[0.4cm]
  & &  v_{\rm XC}([n],[\Phi_0],[\Psi_0])({\bf r},t) .
\end{eqnarray}
The XC-potential involves memory, it depends on the charge density history 
and the initial states $\Psi_0$ and $\Phi_0$ of the interacting and the Kohn-Sham system \cite{Marques_TDDFT_I,Marques_TDDFT_II}. 
The crucial step, which is limiting the range of applicability of the respective 
approximate approach [\onlinecite{Provorse2016}], is the approximation applied to the 
time-dependent XC-potential. Today, in most cases the adiabatic approximation is used 
(together with an approximation to the ground-state XC potential), {\it i.e.} the 
instantaneous electron density $n({\bf r},t)$ is inserted into an 
approximate local ({\it e.g.} LDA or GGA) XC-potential 
$v_{\rm XC}^{\rm approx}$ from ground state DFT:
\begin{equation}
 v_{\rm XC}[n]({\bf r},t) \approx v_{\rm XC}^{\rm approx}[n(\cdot,t)]({\bf r}) .
\end{equation}
In the adiabatic approximation memory and initial state dependence are neglected. 
A critical discussion of these issues can be found in the article by N.T.\ Maitra [\onlinecite{Maitra_Perspective_2016}] and the 
references cited therein and the article by Provorse and Isborn [\onlinecite{Provorse2016}].
For the effect of different approximations to the ground-state 
XC-potential see {\it e.g.} Refs.\ \onlinecite{Zhao_2015_H+Be,Nagano2000}.
The effect of the approximate XC-potential  {\it e.g.} on resonant charge transfer 
requires careful consideration, and comparison of time-dependent simulations of the many-body 
system within TDDFT-MD with results obtained from a many-particle NEGF-approach, as 
described in Sec.~\ref{s:negf}, could provide additional insights to judge the accuracy. 
The application of time-dependent current density functional theory (TDCDFT)
to stopping power and the effect of non-locality is discussed by 
V.U.\ Nazarov {\it et al.} in Ref. \onlinecite{Nazarov2007}.

Another approximation is required for the description of the motion of the 
ions. The complete equations of motion for the combined system of nuclei and electrons can be 
written as a coupled system of equations, where the nuclei are moving 
on the ground and excited state Born Oppenheimer surfaces, depending on the electronic 
excitation [\onlinecite{Baer2006}]. Tully's surface hopping algorithm [\onlinecite{Tully1990,Shenvi_2009}] 
allows to simulate such time evolution stochastically. However, as long as the 
trajectories of the nuclei do not differ qualitatively but, instead, closely follow 
some average trajectory, the motion of the nuclei can be approximated 
by means of the much less expensive Ehrenfest dynamics [\onlinecite{Marx2009}],
\begin{eqnarray}
 M_i \frac{d^2 {\bf R}_i} {dt^2} =
 -\langle \nabla_{{\bf R}_i} \hat V_{\rm ion - el}({\bf R}_1, ..., {\bf R}_n ) \rangle 
 \nonumber\\
-  \nabla_{{\bf R}_i} V_{\rm ion - ion}({\bf R}_1, ..., {\bf R}_n ) .
 \end{eqnarray}
The expectation value in above equation has to be calculated with respect 
to the electronic wave function, which follows, in principle, from the 
integration of the time-dependent many-particle Schr\"odinger equation
of the electrons, or, in practical applications, the time-dependent Kohn-Sham 
equations. $M_i$ denotes the ion masses, ${\bf R}_i(t)$ the ion positions, 
and $V_{\rm ion - ion}$  and $V_{\rm ion - el}$ the ion--ion and 
ion--electron interaction potential energies. 

Pseudopotentials [\onlinecite{Martin2004}] are in use in TDDFT-Ehrenfest MD 
simulations [\onlinecite{Marques_octopus_2003}]. 
This enormously reduces the computational effort as compared to 
an all-electron approach [\onlinecite{Foglia2017}]. Galilei invariance is preserved if non-local pseudopotentials
are multiplied by an ion-velocity dependent gauge factor [\onlinecite{Nagano2000}]. 
In case the projectile has some finite initial velocity, an initial boost 
has to be applied to electronic states of the projectile [\onlinecite{AvendanoFranko_Thesis_2013,Nagano2000}] 
so that the electrons have the same initial velocity as the nuclei. 
The accuracy of TDDFT-MD simulations has been critically evaluated recently 
by Yost {\it et al.} [\onlinecite{Yost2017}] for the case of the electronic stopping 
power of a proton in Si. They report deviations between the all-electron 
and the pseudopotential approach and suggest a correction scheme. The
differences arise at higher proton velocity (beyond 1 a.u.), 
and the correction is larger at small impact parameter [\onlinecite{Yost2017}]. 
A pseudoatom method to account for core electron effects has been 
applied to electronic stopping of {\it e.g.} Li$^+$ ions in graphene 
for impact energies beyond 10 keV/u in Ref.\ \onlinecite{ojanpera_prb_14}.
As the deviation from the all-electron case will depend on the binding energy 
of the electronic states assumed as frozen core states in the 
pseudopotential construction, one should care about electron promotion 
effects missing in the pseudopotential calculation [\onlinecite{German1994}] 
when carrying through TDDFT-MD simulations including high-energy head-on 
collisions.

TDDFT-Ehrenfest molecular dynamics is implemented, e.g., in the program 
\textit{octopus} by Rubio {\it et al.} [\onlinecite{Marques_octopus_2003,Castro_octopus_2006,Andrade_octopus_2015}]. 
In their program, the Kohn-Sham wavefunctions are represented on a 
real space grid and systems of different dimensionality can be treated. 

Applications of TDDFT-MD by many authors using different codes can be 
found in the literature. Examples have already been cited above, 
including electron stopping of H in Al, see {\it e.g.} Refs.\ \onlinecite{Correa_2012_H_in_Al,Schleife2015,Shukri_2016_electronic-stopping}, 
the interaction of Li$^+$ with an Al cluster [\onlinecite{Moss_2009_Li+_AlCluster}], the interaction of H/Al in Ref.\ \onlinecite{Lindenblatt2006b}.  
The H/He anomaly in the electronic stopping power in Au at low 
kinetic energies [\onlinecite{Markin2009}] and the role of the Au $d$ electrons has been 
simulated by the authors of Ref.\ \onlinecite{Zeb2012}. 
These applications refer either to electron stopping in bulk materials 
or, in case of the ion-discharge processes at surfaces, to 
resonant charge transfer, which is an inherently single-electron effect --
as opposed to Auger neutralization, which is a two-electron effect [for a discussion, see Sec.~\ref{s:qkin}]).

It has been pointed out in the literature before that 
the energy transfer into electronic excitations can be simulated 
only for a finite time in case of a finite system due to its discrete 
energy spectrum [\onlinecite{Mason2007, Race_RepProgPhys_2010}]. 
Thus, {\it e.g.} in TDDFT-MD simulations of a vibrating molecule 
interacting with a metal surface [\onlinecite{Grotemeyer2014,Grotemeyer_Dissertation}] 
the size of the unit cell (the number of atomic layers forming the metal 
slab or the cluster size) limits the simulation time. 
Within a simplified tight binding model for the case of a vibrationally 
excited HCl molecule incident on an Al(111) slab this has been exemplified 
by M. Grotemeyer [\onlinecite{Grotemeyer_Dissertation}]. A direct simulation 
of open quantum systems, see {\it e.g.} the article by R. D{'}Agosta and 
M. Di Ventra [\onlinecite{Agosta2013}] and the article by H. Appel 
in Ref.~\onlinecite{Marques_TDDFT_II}, would be advantageous. 

For Ar$^{8+}$--Ar charge transfer collisions Nagano {\it et al.} [\onlinecite{Nagano2000}] observed that no Auger excited 
electrons occurred in their TDLDA-simulations.
Furthermore, as shown by C.A. Ullrich for a model system in Ref. \onlinecite{Ullrich2006},
when applying the ALDA for the XC potential in TDDFT doubly excited 
configurations are not accounted for. Based on the calculation of an 
exact XC-potential, V. Kapoor [\onlinecite{Kapoor2016}] has concluded that autoionization requires inclusion of memory effects in
the XC-potential, which displays a rather complicated structure. 
When calculating {\it e.g.} the 
charge transfer between He ions scattered at an Al-surface, the 
two-particle character of the Auger transitions leading to the charge transfer between projectile and surface has to be accounted for, and other techniques 
are therefore used in case of Auger transitions [\onlinecite{Lorente_1994_Auger_neutralization,Monreal_Review_2014}].

To summarize, DFT and TDDFT simulation methods have 
undergone a very strong development since the formulation of 
the underlying theorems.
We have discussed a number of practical applications and important approximations. However, we are not aware of direct applications to plasma-surface interaction yet. Of course, there have been many simulations of particular processes such as ion stopping, ion neutralization or neutral adsorption, but the influence of a nonequilibrium plasma environment is still an open question. DFT and TDDFT will, without doubt, play a key role in plasma-surface simulations. Their strength will be mostly in testing individual processes, rather than simulating the full dynamics. This knowledge can then be used in other simulations such as those using quantum kinetic equations, cf. Secs.~\ref{s:qkin} and \ref{s:negf}, e.g. as input for models describing the dynamics of electrons in ions, atoms or molecules (such as the Newns-Anderson model used in Sec.~\ref{s:qkin}), as well as for models of solids, such as lattice models used in Secs.~\ref{s:negf} and ~\ref{ss:intgrated_ion_impact}.
\section{Nonequilibrium Green functions-based \textit{ab initio} simulations}\label{s:negf}
We now return to the statistical approach to quantum many-body systems. We have already seen in Sec.~\ref{s:qkin} that a quantum generalization of kinetic theory allows for an efficient description of plasma-surface interaction processes such as secondary electron emission. The analysis of that section, however, used a simplified treatment of the solid, assuming spatial homogeneity and effective mass models and did not resolve the electronic dynamics in the material. Here we present a generalized quantum kinetic approach where these effects in the solid can be included straightforwardly. This will be demonstrated on the example of ion stopping focusing on recent simulation results of Balzer, Bonitz and co-workers [\onlinecite{balzer_prb16, balzer_prl_18}]. At the same time, the NEGF approach is computationally extremely expensive and presently does not allow for a full quantum-mechanical treatment of the electron dynamics inside the projectile as well. So here an Ehrenfest-type dynamics will be employed, as was discussed already in Sec.~\ref{ss:tddft}. This means that the present model is--in terms of effects included and neglected--complementary to both, the quantum Boltzmann approach presented in Sec.~\ref{s:qkin} and to the DFT concept of Sec.~\ref{s:abinit}.
\subsection{Definitions and basic concept of NEGF}\label{sss:negf-defs}
The method of nonequilibrium (real-time) Green functions is a successfull approach to quantum many-body systems out of equilibrium, cf. Refs.~\onlinecite{keldysh64,   kadanoff-baym-book}.
The method is a straightforward generalization of classical kinetic theory (e.g. Boltzmann equation) and of the quantum Boltzmann equation approach overcoming the limitations of the latter. These limitations include the restriction to times larger than the correlation time and fundamental problems such as incorrect conservation laws (e.g. conservation of kinetic energy instead of total energy) and relaxation toward an equilibrium state of an ideal gas (Fermi, Bose or Maxwell distribution) instead of the one of an interacting system, for a detailed discussion, see Refs.~\onlinecite{bonitz_qkt, bonitz-etal.96pla, bonitz-etal.96jpb, bonitz96pla}. Generalized quantum kinetic equations that are based on nonequilibrium Green functions (or, alternatively, on density operator theory [\onlinecite{bonitz_qkt, kremp-etal.97ap}]) overcome these problems.

The NEGF approach has been successfully applied to an impressively diverse array of systems, including nuclear matter, by Danielewicz, Köhler and others, e.g. Refs.~\onlinecite{DANIELEWICZ_84_ap2,koehler_prc_95},  to optically excited semiconductors and quantum dots by Schäfer, Haug, Banyai, Bonitz and others, e.g. Refs.~\onlinecite{banyai_prl_95,bonitz-etal.96jpb, kwong-etal.98pss, binder-etal.97prb, bonitz_prb_7, balzer_prb_9}, to dense laser plasmas by Kremp and Bonitz, e.g. Refs.~\onlinecite{kremp_99_pre, bonitz_99_cpp}, to few electron atoms, Refs.~\onlinecite{stefanucci_cambridge_2013, balzer_pra_10, balzer_pra_10_2}, and correlated fermions in lattice systems, Refs.~\onlinecite{verdozzi_jpcs16,schluenzen_prb16} and many other problems. 
In the present context of plasma-surface interaction we expect that the NEGF approach will allow one to study non-adiabatic effects, in particular, relaxation processes in the surface that are initiated by the impact of plasma particles. 

The NEGF-method is formulated in second quantization (for textbook or review discussions, see Refs.~\onlinecite{stefanucci_cambridge_2013,kadanoff-baym-book, schluenzen_cpp16}), in terms of creation (annihilation) operators $c_{i\sigma}$ ($c^{\dagger}_{i\sigma}$) for electrons in a single-particle orbital $|i\rangle$ with spin projection $\sigma$. Below we will consider a spatially inhomogeneous lattice configuration where $i$ labels the spatial coordinates of individual lattice points.
The creation and annihilation operators are time-dependent via the Heisenberg representation of quantum mechanics. The central quantity that determines all time-dependent observables is the one-particle NEGF (we use $\hbar = 1$), 
\begin{equation}
G_{ij\sigma}(t,t') = 
-i
\langle T_{\cal  C}c_{i\sigma}(t)c^{\dagger}_{j\sigma}(t')\rangle\,,
    \label{eq:negf}
\end{equation}
where the expectation value is computed with the equilibrium density operator of the system. For completeness we mention that times are running along the Keldysh contour $\cal C$, and $T_{\cal C}$ denotes ordering of operators on $\cal C$ (this is merely a formal trick for the theory development, all practical calculations are done for real-time quantitites, for details see Ref.~ \onlinecite{balzer-book}). For example, the time-dependent electron density on site $i$ follows from $G$ via 
$n_i(t)= -i G_{ii\sigma}(t, t^+)$, where $t^+ \equiv t+\epsilon$, with $\epsilon>0$ and $\epsilon \to 0$. If the site indices are taken different, $i\ne j$, the Green function describes time-dependent transitions of electrons between two lattice sites. In similar manner one computes the density matrix, currents, mean energies, optical absorption or electrical conductivity from $G$.

The NEGF obeys
the two-time Keldysh-Kadanoff-Baym equations (KBE)~[\onlinecite{kadanoff-baym-book}] 
\begin{eqnarray}
\label{eq.kbe}
 \sum_k&\left[i\partial_t\delta_{ik}
 -h_{ik\sigma}(t)\right]G_{kj\sigma}(t,t')\\
 &=\delta_{\cal C}(t-t')\delta_{ij}+\sum_{k}\int_{\cal C} ds\,\Sigma_{ik\sigma}(t,s)G_{kj\sigma}(s,t')\,,\nonumber
\end{eqnarray}
where $h$ contains kinetic, potential and mean field energy contributions whereas correlation effects are included in the selfenergy $\Sigma$ [here we do not consider spin changes and we omit the second equation which is the adjoint of (\ref{eq.kbe})].

Without the right hand side, Eq.~(\ref{eq.kbe}) would be equivalent to a Vlasov equation or its quantum generalization (time-dependent Hartree-Fock, TDHF). The r.h.s. contains correlation  effects that are responsible for relaxation, dissipation and include scattering of electrons with electrons, ions or lattice vibrations (phonons). Notice the time integral on the r.h.s. which incorporates memory effects that are important to correctly treat correlations. The standard Boltzmann equation [Sec.~\ref{s:qkin}] is recovered by evaluating this time integral approximately via a retardation expansion [\onlinecite{bonitz_qkt, bonitz_cpp18}] or by a saddle point method as done in Sec.~\ref{s:qkin}. In this Section we will not consider this long-time limit but concentrate on the fast ion stopping dynamics in a solid.

The NEGF formalism is formally exact if the selfenergy would be known exactly. The approach is internally consistent, obeys conservation laws and is applicable to arbitrary length and time scales. Its accuracy is determined by the proper choice for a single function -- the selfenergy. For an overview on the treatment of weak and strong correlations in solids an optical lattices, see Ref.~\onlinecite{schluenzen_cpp16}.
In the following we consider, as an example of relevance to plasma-surface interaction, the NEGF approach to the energy loss of energetic ions in a solid.

\subsection{Nonequilibrium Green functions approach to ion stopping in nanomaterials}\label{ss:negf-stopping}
The energy loss of energetic ions in a solid (stopping power) is an old problem that has been studied in great detail. There exist extensive references and successful code packages such as SRIM [\onlinecite{trim}]. However, standard methods assume linear response, i.e. the material is weakly perturbed by the projectile and the response is computed in perturbation theory. While this maybe correct when considering an extended area of the surface where the effect of the projectile is small, on average, locally the excitation maybe strong. In the following we, therefore, attempt to perform a space and time resolved analysis of the projectile-solid interaction including electronic correlation effects in the solid. This will be particularly important for strongly correlated materials.

We consider a surface with strong electronic correlations that is modeled by a Hubbard hamiltonian (\ref{eq:ham1}) with hopping amplitude $J$ [$\langle i,j\rangle$ denotes nearest neighbors] and onsite interaction strength $U$.
\begin{eqnarray}
\label{eq:ham1}
H_\textup{e}&=- J\sum_{\langle i,j\rangle,\sigma}
c_{i\sigma}^\dagger c_{j\sigma} + U\sum_{i}\left(n_{i\uparrow}-\frac{1}{2}\right)\left(n_{i\downarrow}-\frac{1}{2}\right)&\nonumber\\
&
-\frac{Z_\textup{p}e^2}{4\pi\epsilon_0}\sum_{i,\sigma}\frac{c_{i\sigma}^\dagger c_{i\sigma}}{|\vec{r}_\textup{p}(t)-\vec{R}_i|}
+ \sum_{\langle i,j\rangle,\sigma}W_{ij}(t)c_{i\sigma}^\dagger c_{j\sigma}
\,.
\end{eqnarray}
  \begin{figure}[h]
  \begin{center} 
  \hspace{-0.cm}\includegraphics[width=0.485\textwidth]{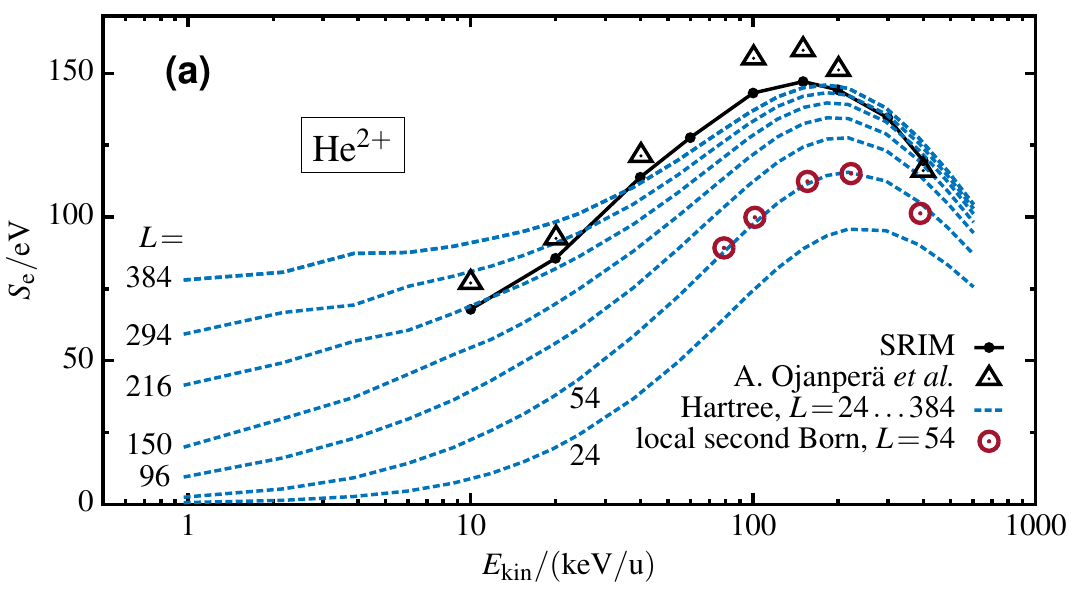}
  \includegraphics[width=0.485\textwidth]{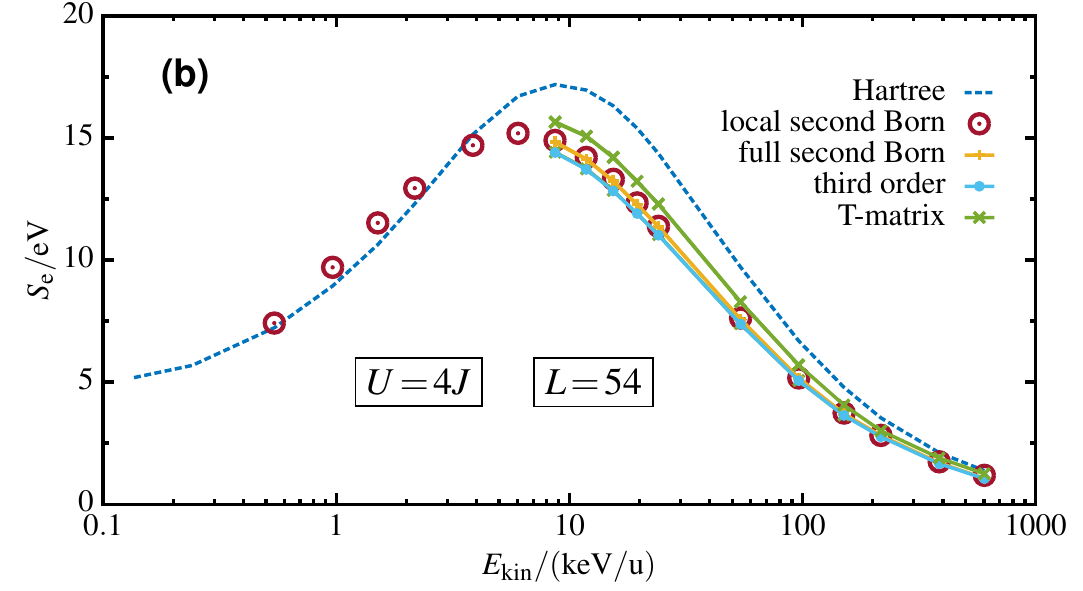}
  \end{center}
  \vspace{-.40cm}
  \caption{Energy loss of doubly charged helium ions in a graphene sheet~[\onlinecite{balzer_prb16}]. \textbf{(a)} Simulation results for $U/J=1.6$, $J=3.15$\,eV and $\gamma=0.55$ and for different cluster size $L$ are compared with SRIM and TDDFT simulations of Ojanper\"a \textit{et al.} [\onlinecite{ojanpera_prb_14}] for an infinite system. The NEGF data (lines: Hartree approximation, red symbols: second Born selfenergy) show good agreement and, in addition, extend to lower projectile energies. 
  \textbf{(b)} A honeycomb cluster with $L=54$ sites, $U/J=4$ $J=2.8$\,eV, $\gamma=0$, is studied with different selfenergy approximations: second Born, third-order and T-matrix approximation showing a clear impact of correlations.}
  \label{fig:stopping_vs_impact}
  \end{figure} 
The strength of correlations is measured by the ratio $U/J$ and is typically in the range from $0$ to $10$.
The second line of Eq.~(\ref{eq:ham1}) contains the coupling of the lattice electrons located at coordinate ${\vec R}_i$ with a positively charged projectile of charge $Z_p$ that is treated classically (Ehrenfest dynamics) by solving Newton's equation for the trajectory $\textbf{r}_p(t)$ under the influence of all Coulomb forces with the lattice electrons. The final term allows to improve the model by accounting for modification of the hopping rates due to the projectile according to $W_{ij}(t)=\gamma [W_{ii}(t)+W_{jj}(t)]/2$, where $W_{ii}$ is the magnitude of the Coulomb potential  of the projectile at lattice site ``i'' and  $\gamma$ is a phenomenological parameter of the order unity [\onlinecite{balzer_prb16}]. 

\begin{figure}[h]
  \begin{center} 
    \hspace{-0.cm}\includegraphics[width=0.2\textwidth]{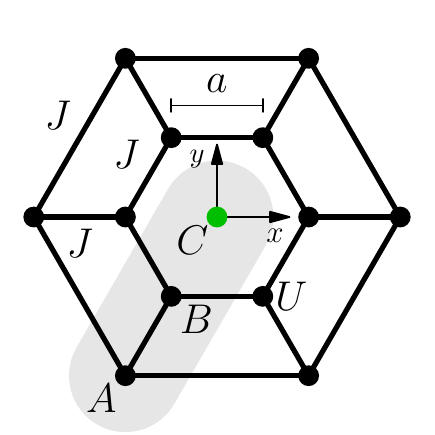}
  \hspace{-0.cm}\includegraphics[width=0.485\textwidth]{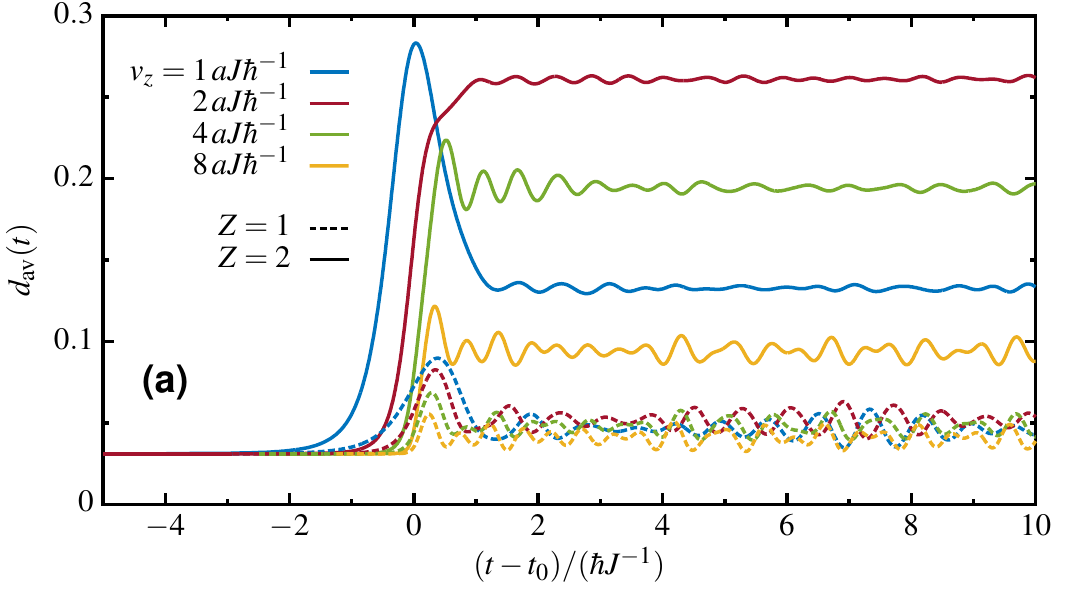}
  \hspace{-0.cm}\includegraphics[width=0.485\textwidth]{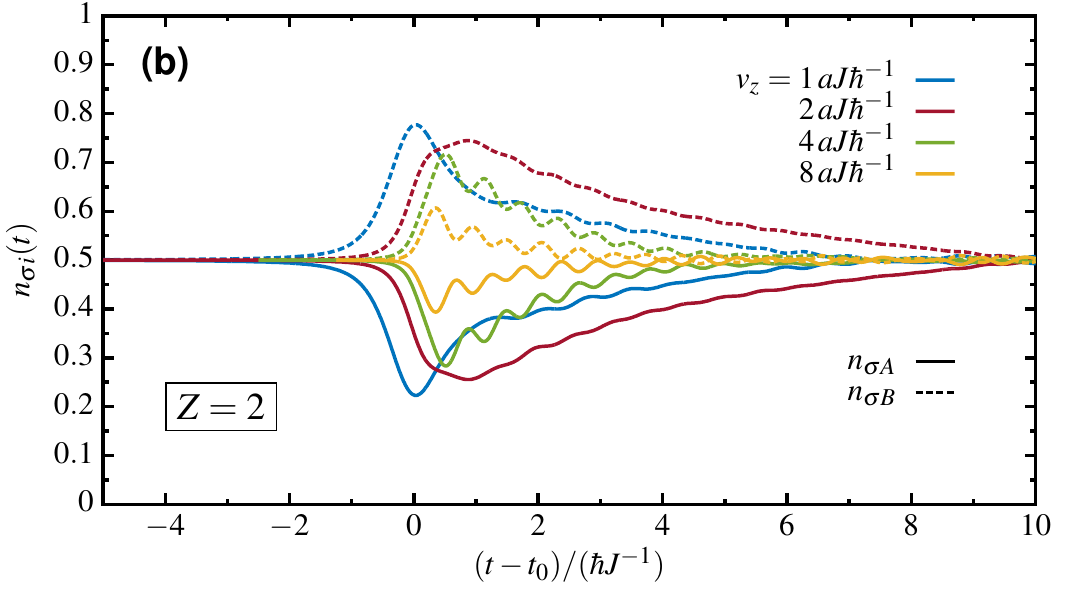}
  \end{center}
  \vspace{-.40cm}
  \caption{Time-dependent response of a strongly correlated finite honeycomb cluster ($L=12$ sites, $U/J=10$) to a charged projectile penetrating through the center (point C in top figure). \textbf{(a)} site-averaged double occupation, $d_\textup{av}(t)=\frac{1}{L}\sum_i d_i(t)$, for  charge $Z=1$ (dashed lines), and $Z=2$ (full lines). \textbf{(b)} The densities on sites A (full line) and B (dashes) closest to the projectile, for the case $Z=2$, after Ref.~[\onlinecite{balzer_prl_18}]. }
  \label{fig:doublons}
  \end{figure}

The KBE (\ref{eq.kbe}) with the hamiltonian (\ref{eq:ham1}) have been solved numerically for a two-dimensional hexagonal lattice as is known e.g. for graphene [cf. top part of Fig.~\ref{fig:doublons}] of finite size (the number of lattice sites $L$ varies from $24$ to $384$)  for a broad range of projectile velocities~[\onlinecite{balzer_prb16}], see Fig.~\ref{fig:stopping_vs_impact}. For the correlation selfenergy $\Sigma$ several approximations were used: no selfenergy (this corresponds to the mean field or Hartree approximation), second Born (second order in the electron-electron interaction), the third order and the T-matrix (strong coupling) approximation. 
The results demonstrate good agreement with available data from SRIM [\onlinecite{trim}] and TDDFT [\onlinecite{ojanpera_prb_14}] simulations for high and intermedeate impact energies, down to 10 keV which confirms the choice of the selfenergy. 
The strength of the NEGF approach is that it also provides data for lower energy, as are typical for low-temperature plasmas. Moreover, the simulations are not restricted to spatially uniform systems but are also directly applicable to finite systems such as graphene nanoclusters, nanostructured surfaces. A particular interesting example are small graphene ``nanoribbons'' that have a size-dependent band gap, e.g. [\onlinecite{son_prl_06, prezzi_prb_08}], giving rise to promising electronic and optical properties. For an overview on plasma synthesis of nanomaterials, see Refs. \onlinecite{meyyappan_jpd_11, ostrikov_aip_13}.
In these systems, finite size effects play an important role which is in particular true for the stopping power, as was demonstrated in the top part of Fig.~\ref{fig:stopping_vs_impact}. Finally, the NEGF approach also applies to strongly correlated materials
the potential of which for plasma applications has not been explored yet.

So far we have not clarified what is the energy loss mechanism for the projectile -- except for the fact that the projectile produces, within the above model, purely electronic excitations (lattice vibrations can be included straightforwardly, but this will not be of interest here, assuming that only times shorter than about 100 fs are considered).
In the following we use our time-dependent and space resolved simulations to investigate the electronic correlations effects that are excited by the projectile in a graphene-type nanostructure more in detail. Of particular interest is the local excitation of double occupancies (``doublons''), 
$d_i(t) =\langle \hat{n}_{i\uparrow}(t)\hat{n}_{i\downarrow}(t)\rangle\,$. These are pairs of electrons (with opposite spin projection) that occupy the same lattice site ``i'' and can be considered bound states, despite their repulsive Coulomb interaction.
If the initial state of the lattice is ``half filled'', i.e. there are as many electrons as lattice sites, this system will behave as an insulator, because any additional electron will have to sit on an already occupied site. This, however, costs an energy $U$, cf. the hamiltonian (\ref{eq:ham1}). An analysis of the energy spectrum reveals that this electron occupies an excited energy band ($U$ above the lower band). If an external electric field would applied, this electron would be  mobile, once it has overcome this energy gap. If the system is now hit by an energetic ion, part of this energy could be transformed into excitation of electrons to the upper Hubbard band, i.e. to the formation of doublons and of a conducting state.

This hypothesis is tested on an example in Fig.~\ref{fig:doublons} where a projectile penetrates a honeycomb cluster in the center. As a result, in the cluster the electrons are forced towards the center, which is seen by a density increase on the nearest site (B) and a corresponding decrease at the next nearest site (A). Quickly after the projectile has passed through the densities return to their initial values. At the same time, the double occupation (a correlation effect) of the site A, $d_A$,  is increased as well, which is due to a transition of the whole electronic system to a higher energy eigenstate. Most importantly, this excitation is stored in the system even after the projectile has left, it only redistributes equally among the sites A and B (cf. the average that is shown by the green curve). The largest effect is observed for energies of 120 eV, in the case of protons (480 eV, for alpha particles) [\onlinecite{balzer_prl_18}]. This way part of the energy of the projectile from the plasma is stored in non-trivial electronic excitations of the solid which might find interesting applications in the future.

We note that these simulations are very CPU time consuming. This allows to do simulations only for selected impact parameters (no averaging over trajectories), as  in TDDFT simulations [\onlinecite{ojanpera_prb_14}]. Also, long time scales (that describe, e.g. the coupling of electronic and phononic degrees of freedom) are not easily accessible. Also, a full quantum treatment of projectiles with internal degrees of freedom (atoms, larger ions, molecules) will require further developments. In this respect, these NEGF simulations are complementary to the quantum kinetic models of Sec.~\ref{s:qkin} or TDDFT simulations, cf. Sec.~\ref{ss:tddft}.
%
\subsection{\textit{Ab initio} NEGF (AI-NEGF)}\label{ss:ainegf}
The NEGF simulations of ion stopping were, so far, restricted to lattice model systems, in order to systematically explore the role of electronic correlations and to investigate finite size effects in nanostructures. In order to extend this approach to a broad range of realistic materials one can couple NEGF simulations to a Kohn-Sham basis that is precomputed by a ground state DFT simulation, as was discussed in Sec.~\ref{s:abinit}. In fact, such an approach was developed a few years ago by Marini and co-workers within their Yambo code [\onlinecite{marini_2009_yambo}] and provides a suitable basis for future improvements of the NEGF simulations towards realistic material properties. The public equilibrium part of the program is well suited to compute the dielectric function, optical properties of materials and was successfully tested in Kiel for graphene [\onlinecite{heese_bsc_17}]. In the following we plan to extend the nonequilibrium part by including stopping simulations as explained in Sec.~\ref{ss:negf-stopping}.

\section{Synergies of the surface simulation approaches}\label{s:synergies}
Let us now discuss what kind of synergies exist between the various surface simulation methods that were listed in the central box of Fig.~\ref{fig:theory} and were discussed in Secs.~\ref{s:mesoscopic}, \ref{s:qkin}, \ref{s:abinit} and \ref{s:negf}. These methods are very different in terms of accessible length and time scales and the nature of physical approximations, therefore, an efficient and reliable combination into a single single computer code (multi-scale approach) does not seem realistic, at the moment. Nevertheless, it is very interesting to explore what synergies these concepts may possess and how they may be combined for a better understanding of selected relevant surface simulation processes. Of particular interest are, of course, such processes that can be treated with more than one method which offers the potential of cross checks, benchmarks and improvements of approximations. In table~\ref{tab:processes-methods} we list such processes and discuss the potential of the surface simulation approaches.
\begin{table*}[]
    \centering
    \begin{tabular}{|l|l|l|l|}
    \hline
\textbf{Process}     & \textbf{Methods}  & \textbf{Pros} & \textbf{Cons}\\\hline
Ion stopping & MD &large system & el. adiabatic, unknown accuracy of empirical forces \\
& DFT & accounts for electronic structure & el. adiabatic, small system, limited accuracy of approx.XC\\
&&&kinetic energy limits due to pseudopotentials\\
         & TDDFT & electronically non-adiabatic& unknown accuracy of approx. XC\\
         &&& only electronic stopping from selected trajectories\\
         & NEGF  & electronic correlations& finite system, Hubbard-type model \\
         && Ref.~\onlinecite{balzer_prb16}, Sec.~\ref{ss:negf-stopping}& classical ion (Ehrenfest dynamics) \\
         & AI-NEGF & accounts for electronic structure, & small systems, approximate electronic correlations\\
         && possible with Yambo, Sec.~\ref{ss:ainegf} [\onlinecite{marini_2009_yambo}] & not yet tested for plasma-surface applications\\ \hline
Ion neutralization   & QBE & correlations in projectile, Ref.~\onlinecite{pamperin_2015_many}& model surface\\
                     & TDDFT & accounts for electronic structure & unknown/limited accuracy of adiabatic approx to XC\\\hline
Electron sticking/   & QBE &Ref.~\onlinecite{bronold_2015_absorption}& model surface\\
absorption 
& 
&& \\\hline
SEE & QBE & correlations in projectile, Sec.~\ref{s:qkin}  & model surface\\\hline
Atom sticking        & MD &Refs.~\onlinecite{filinov_psst18_1,filinov_psst18_2}, Sec.~\ref{ss:freezout}& classical, no electronic effects\\
                     & QBE  &Ref.~\onlinecite{brenig_zpb79} & model systems\\\hline
Cluster/layer growth & MD & Refs.~\onlinecite{abraham_jap16,abraham_cpp18}, Sec.~\ref{ss:spa}& classical, no electronic effects\\
                     & KMC & Refs.~\onlinecite{bonitz_cpp12,abraham_jap_15} & phenomenological\\\hline
Surface reactions    & DFT &accounts for electronic structure & el. adiabatic,  limited accuracy of approx.XC\\
                     &TDDFT &electron. non-adiabatic  ~[\onlinecite{brenig_2008_reaction}]& small system, few selected trajectories,\\
                     &&&
                     unknown/limited accuracy of adiabatic approx. to XC\\\hline
Sputtering           & MD & large system, Ref.~\onlinecite{brault_frontiers_18}& el. adiabatic, unknown accuracy of empirical forces\\
                     \hline
    \end{tabular}
    \caption{Selection of important plasma-surface processes, main surface science methods (as listed in Fig.~\ref{fig:theory}) as well as their quality and limitations. For applications discussed in this paper, the corresponding section number is given. ``XC'' denotes exchange correlation functional, ``el. adiabatic'': electronically adiabatic.}
\label{tab:processes-methods}
\end{table*}

The table shows that there exist a large range of applications of these methods to quite diverse topics. For method development and improvement the best candidates are topics where several methods have been applied in the past and where, therefore, cross checks are possible.

\section{Towards integrated plasma-surface modeling}\label{s:integrated}
After reviewing the various simulation methods for surface processes we now discuss possible ways how to integrate them into a simulation of the plasma-solid interface. In the first two parts of this section we consider a quantum kinetic theory approach to charge transfer processes across the interface. First, in Sec.~\ref{ss:intgrated_dl} we consider the description of the quasistationary electric double layer that was introduced in Sec.~\ref{ss:psi-processes} and Fig.~\ref{fig:interface-physics} and derive matching conditions between plasma and surface. After this, in Sec.~\ref{ss:intgrated_ion_impact} we outline a time-dependent approach to electron transfer between plasma ions and a solid, including electronic correlation effects. Finally, at the end of this section we briefly return to the coupling of plasma and surface simulations, in Sec.~\ref{ss:plasma-surface}, and to the issue of how to keep track of surface morphology changes, cf. Sec.~\ref{ss:morphology-surface}.

\subsection{Integrated modeling of the electric double layer}\label{ss:intgrated_dl}

In the previous sections
we discussed the interaction of a plasma with a solid surface in terms 
of elementary surface collision processes. Similar to collisions in the bulk
of the discharge they are characterized by collision probabilities (viz: 
sticking, reflection, and secondary electron emission coefficients) which have 
to be determined either by independent measurements or by separate quantum-mechanical 
calculations. In contrast to bulk processes, however, where the surrounding 
plasma does not affect the collision partners, in a surface collision the plasma affects the target. The parameters used to characterize a surface collision 
depend thus on the plasma. Taking the plasma-induced modifications of the 
surface into account requires a selfconsistent modeling of the mutual
influence between the plasma and the solid. Such an integral modeling approach
is of course very demanding involving, in general, processes and species acting 
on different scales. In this section we will thus exemplify the approach only 
for an idealistic situation, where the only physical consequence of the 
plasma-surface interaction is an electric response leading to the build-up of 
an electric double layer at the plasma-solid interface, as discussed in Sec.~\ref{ss:psi-processes}, cf. Fig.~\ref{fig:interface-physics}. First steps towards a 
selfconsistent description of this double layer, for a floating dielectric surface, have 
been made by Bronold and Fehske~[\onlinecite{BF17}]. In this subsection we summarize 
and comment on their work.

The formation of an electron-depleted region on the plasma side of the interface and 
an electron-rich region inside or on top of the solid, depending on its electronic
structure, is the most fundamental manifestation of the interaction of a solid surface
with an ionized gas. It arises because electrons in the plasma outrun heavy 
species leading to a more efficient electron deposition due to electron absorption 
than electron extraction by neutralization/de-excitation of ions/radicals. That the 
electric response of the plasma-solid interface leads to an electric double layer,
having a negative part inside the solid and a positive part inside the plasma, is 
known since the beginnings of gaseous electronics~[\onlinecite{LM24}]. Ever since, however, 
the focus of interest has been on the plasma-based electron-depleted part of the 
double layer--the plasma sheath--how it merges with the quasi-neutral bulk 
plasma~[\onlinecite{Robertson13,Brinkmann09,Franklin03,Riemann91,SB90}] and how it is 
affected by the emissive properties of the surface~[\onlinecite{CU16,LW15,SHK13,SKR09,TLC04,HZ66}].
The negative part of the double layer found essentially no attention. Yet it is 
an integral part of the electric response of the solid to the plasma.

\begin{figure}[t]
  \includegraphics[width=\linewidth]{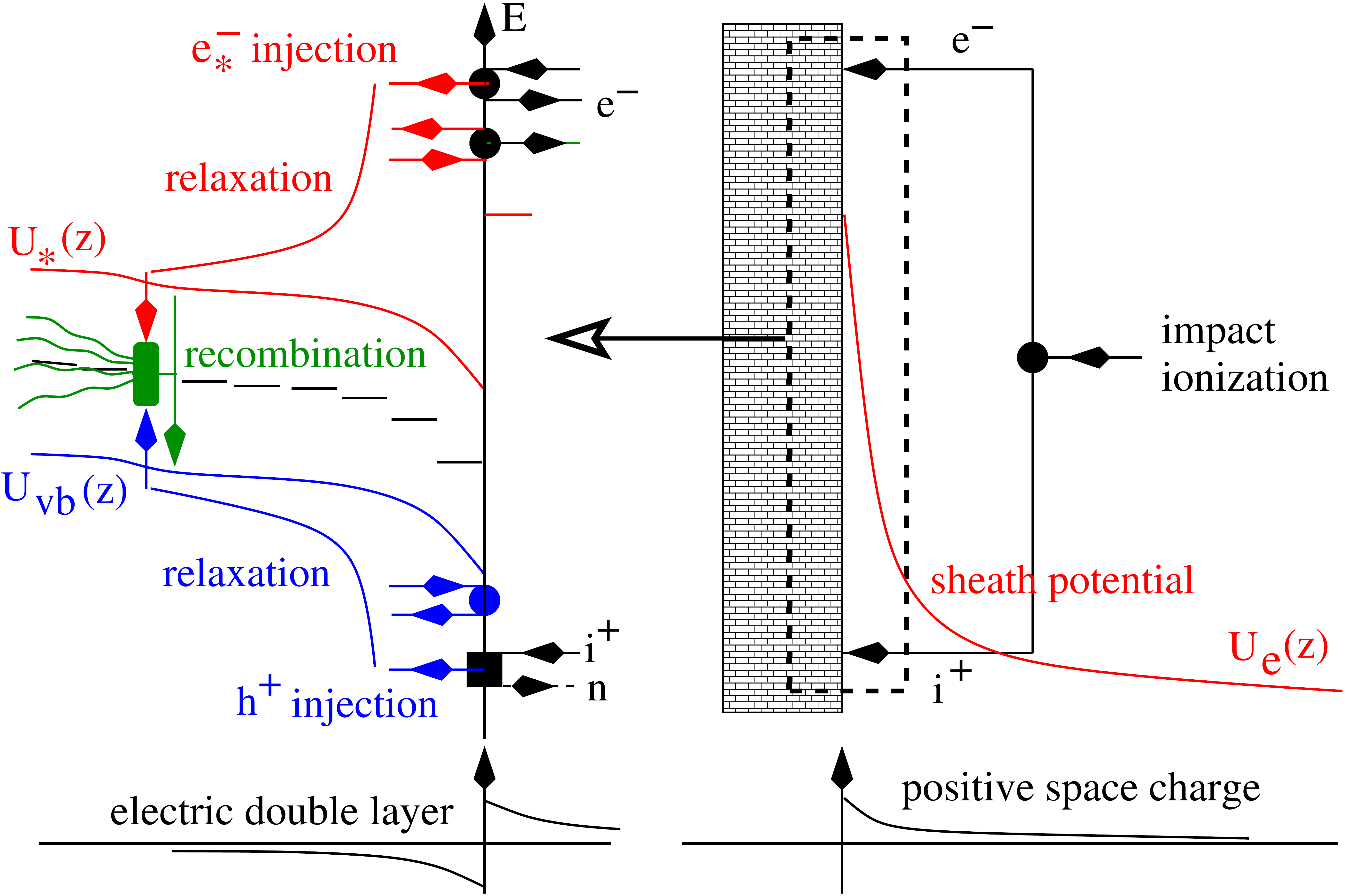}
  \caption{(Color online) 
        Illustration of the electric response of a floating dielectric plasma-solid
        interface. On the right is shown the traditional modeling treating the solid
        as a black box, characterized by surface parameters such as 
        the electron sticking coefficient and the ion wall recombination probability. 
        The left depicts the processes actually taking place inside the solid when 
        plasma is destroyed. Electrons ($e^-$) and ions ($i^+$), generated by 
        impact ionization in the plasma, hit the dielectric solid, thereby injecting 
        conduction band electrons ($e_*^-$) and valence band holes ($h^+$), which after 
        relaxation may either recombine non-radiatively or radiatively. As a result, 
        electron depletion occurs in front of the surface leading to a positive space 
        charge which in turn is balanced by a negative space charge inside the solid
        (electric double layer). The selfconsistent kinetic modeling~[\onlinecite{BF17}] treats 
        impact ionization in the plasma and recombination in the solid on an equal 
        footing by tracing the ambipolar charge transport across the interface, allowing
        for quantum-mechanical reflection/transmission and charge transport/relaxation
        on both sides of the interface.}
  \label{IntModelProcesses}
\end{figure}

Usually, the theoretical descriptions of the electric response of the plasma-solid 
interface assume that the processes inside the solid occur on time scales too fast and
length scales too small to affect the plasma~[\onlinecite{Franklin76}]. Based on such a view, it is thus sufficient to replace 
the plasma-facing solid by an object with a geometrical boundary and probabilities 
for electron sticking/reflection, ion neutralization, and secondary electron
emission~[\onlinecite{LL05}]. Within such an approach~[\onlinecite{BGL10,Kushner04,GMB02}] it is 
of course impossible to investigate the plasma-induced modifications of the electronic 
structure of the solid, which in turn however may strongly affect the probabilities 
for charge transfer. We consider this to be a particularly severe drawback for the 
modeling of microdischarges
on semiconducting substrates~[\onlinecite{DOL10,KSO12}]. Due to the continuing miniaturization 
of these structures~[\onlinecite{EPC13}] the electron transit times through the plasma sheath 
may become comparable to the electron relaxation time inside the solid. Between subsequent 
electron encounters the electrons inside the solid can thus no longer equilibrate. It is 
hence no longer viable to describe charge transfer across the plasma-solid interface
by a number of fixed surface parameters in such situations. Instead charge-transfer 
has to be considered as the linking part of a selfconsistent modeling of the charge 
dynamics on both sides of the interface. As illustrated in Fig.~\ref{IntModelProcesses}, 
the selfconsistent integral modeling of the electric response of the plasma-solid 
interface tracks electrons ($e^-$) and ions ($i^+$) generated by impact ionization
inside the plasma to the inside of the solid, where they recombine--after energy- 
and momentum-relaxation--either radiatively by band-to-band transitions or non-radiatively
via trap states as conduction band electrons ($e_*^-$) and valence band holes($h^+$).
The plasma source is thus linked to the plasma sink inside the solid. We expect 
an understanding of this link to be central for the progress of future efforts 
combining gaseous and semiconductor electronics~[\onlinecite{TWH11,OE05}].

\begin{figure}[b]
  \includegraphics[width=\linewidth]{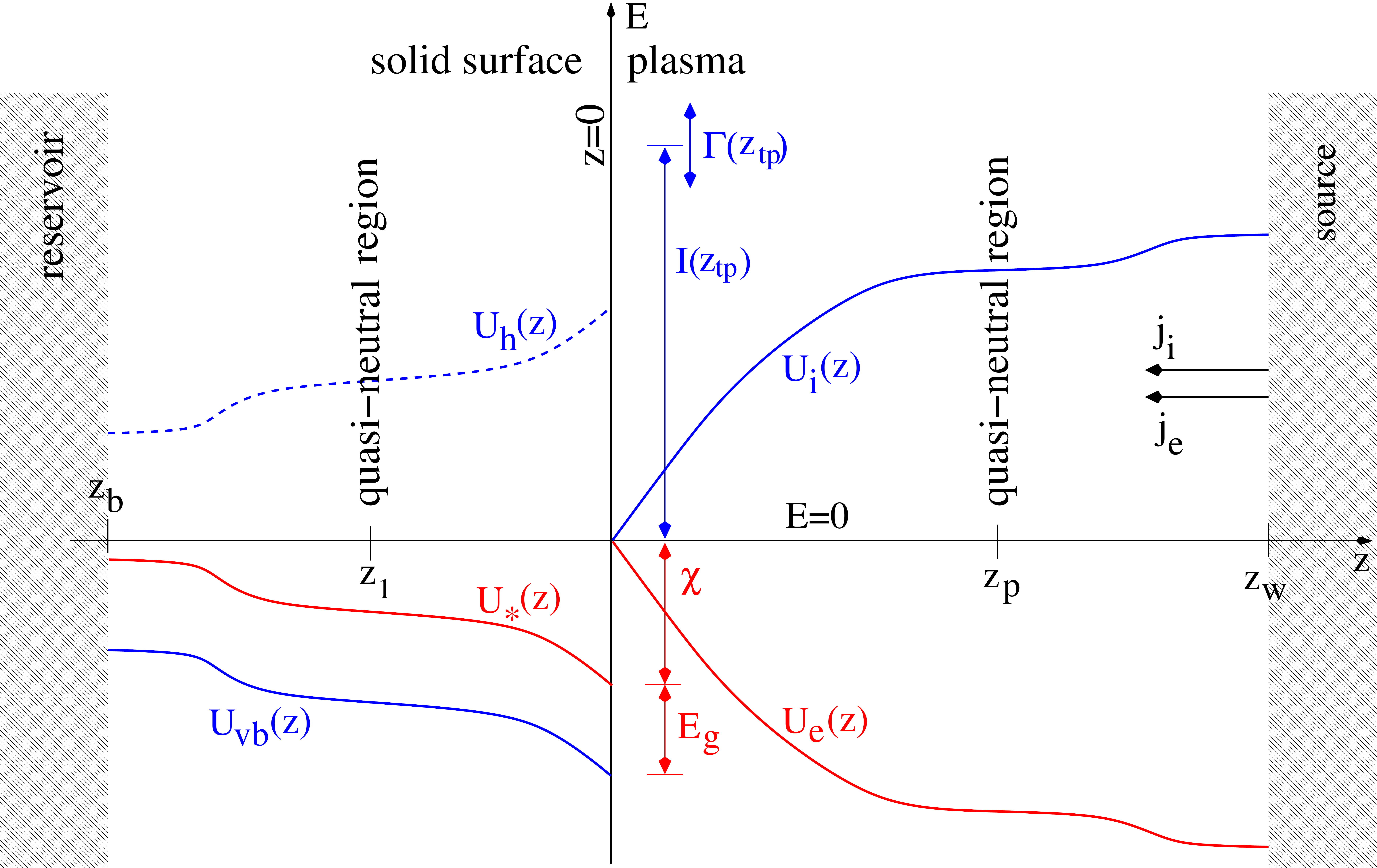}
  \caption{
        Notation used for the description of an electric double layer at a floating 
        dielectric plasma-solid interface with negative space charge inside the solid and 
        positive space  charge in front of it. Shown are the edges of the 
        conduction ($U_*$) and valence ($U_{vb}$) bands, the edge for the motion of
        valence band holes ($U_{h}$), the potential energies for electrons
        ($U_e$) and ions ($U_i$) on the plasma side, and the energetic range,
        specified by the ion's ionization energy $I$ and its broadening $\Gamma$,
        in which hole injection occurs due to the neutralization of ions at the
        interface. Source, reservoir, and quasi-neutral regions are indicated as they will 
        arise in the course of the calculation (adopted from Ref.~\onlinecite{BF17}).}
  \label{IntModelNotation}
\end{figure}

To get an idea about how to organize the self-consistent integral modeling of the electric 
response of the plasma-solid interface, let us consider a dielectric solid facing a 
plasma~[\onlinecite{BF17}]. The modeling then is based on two sets of spatially separated Boltzmann 
equations, one for the electrons and ions inside the plasma and one for the conduction band 
electrons and valence band holes inside the solid (using an electron-hole picture inside the 
solid simplifies the calculations). Defining a species index $s=e,i,*,h$ to denote electrons, 
ions, conduction band electrons, and valence band holes, the Boltzmann equations for the 
quasi-stationary distribution functions $F_s^\gtrless(z,E,\vec{K})$ for left and right 
moving particles can be generally written as (measuring length in Bohr radii and energy in 
Rydbergs)~[\onlinecite{BF17}]
\begin{widetext}
\begin{equation}
\left[\pm v_s(z,E,\vec{K})\frac{\partial}{\partial z}+\gamma^\gtrless_s[F_{s^\prime}^>,F_{s^\prime}^<]\right]
F_s^{\gtrless}(z,E,\vec{K})=\Phi^\gtrless_s[F_s^>,F_s^<,F_{s^\prime}^>,F_{s^\prime}^<]
\label{BTE}
\end{equation}
\end{widetext}
with
\begin{equation}
v_s(z,E,\vec{K}) = 2\bigg(\frac{m_e}{m_s}[E-U_s(z)] - (\frac{m_e}{m_s}\vec{K})^2 \bigg)^{1/2}~
\end{equation}
the velocity of the particles normal to the (planar) interface at $z=0$. Here, $z$, $E$, and
$\vec{K}$ denote the distance from the interface, the total energy, and the lateral momentum
of the particles. The functions $U_s(z)$ are shown in Fig.~\ref{IntModelNotation}
(together with other quantities relevant for the description of the electric double layer).
They define the regions in $z\!\!-\!\!E$ space where the respective species move freely~[\onlinecite{BF17}]. The alignment of the electronic energies of the solid with 
the ones of the plasma is controlled by the electron affinity $\chi$ and the energy gap $E_g$.
The functions $\gamma^\gtrless_s$ and $\Phi^\gtrless_s$ denoting, respectively, the rates for 
out-scattering and the in-scattering collision integrals depend on the scattering process.

The Boltzmann equations themselves are not sufficient. They have to be augmented by matching 
and boundary conditions for the distribution functions, the former applying to $z=0$ and 
the latter to the asymptotic regions of the interface. For a dielectric plasma-solid interface 
without interface/surface states the matching conditions
for the electron distribution 
functions become ($E>0$)
\begin{align}
F_{e,*}^{>,<}(0,E,\vec{K}) &= R(E,\vec{K}) F_{e,*}^{<,>}(0,E,\vec{K})\nonumber\\
                   &+ [1-R(E,\vec{K})]F_{*,e}^{>,<}(0,E,\vec{K})~,
\label{Fe_gtrless_match}
\end{align}
while the hole and ion distribution functions are connected by ($E>E_g+\chi$)
\begin{align}
F_h^<(0,E,\vec{K}) &= F_h^>(0,E,\vec{K}) + \alpha S_h^<(E,\vec{K})~~, \nonumber\\
F_i^>(0,E,\vec{K}) &= (1-\alpha)F_i^<(0,E,\vec{K})~.
\label{Fi_gtr_match}
\end{align}
The function $R(E,\vec{K})$ is the quantum-mechanical reflection coefficient for
electrons due to the surface potential which can be, for instance, modelled by a 
three-dimensional potential step with a mismatch in the electron mass arising from 
the difference between the effective electron mass $m_e^*$ inside the solid and the 
electron mass $m_e$ in the plasma, $\alpha$ is the neutralization probability for an ion 
at the interface, and $S_h^<(E,\vec{K})$ is a function specifying the accompanying hole 
injection into the valence band of the dielectric. An illustration of the matching 
conditions, coupling the distribution functions of the conduction band electrons ($e_*^-$) 
and valence band holes ($h^+$) in the solid with the distribution functions of the 
electrons ($e^-$) and ions ($i^+$) in the plasma, is included in Fig.~\ref{IntModelProcesses}.
The particular form of $S_h^<(E,\vec{K})$, into which $F_i^<$ enters~[\onlinecite{BF17}], depends
on the neutralization process. In case the neutralization induces also secondary electron 
emission, the matching condition for the electron distribution functions has to be augmented 
by a function $S_e^>(E,\vec{K})$. With the Poisson equation for the electric potential energy 
$U_c(z)$ (again given in atomic units),
\begin{align}
\frac{d}{dz}\varepsilon(z)\frac{d}{dz} U_c(z) = 
                                                8\pi\big[\rho_w(z)\theta(-z)-\rho_p(z)\theta(z)\big]~,
\label{PE}
\end{align}
where $\rho_w(z)=n_*(z)-n_h(z)-n_D+n_A>0$ and $\rho_p(z)=n_i(z)-n_e(z)>0$ are, respectively, the
charge densities inside the surface (assumed to contain donors and acceptors with concentrations $n_D$
and $n_A$) and the plasma, to be obtained from the distribution functions by integration,
\begin{eqnarray}
n_s(z)=\int\frac{dE d\,^2K}{(2\pi)^3} \frac{F^>_s(z,E,\vec{K})+F^<_s(z,E,\vec{K})}{v_s(z,E,\vec{K})}~,
\label{Density}
\end{eqnarray}
and the matching conditions for $U_c(z)$ at $z=0$, Eqs.~\ref{BTE}--\ref{Fi_gtr_match}
form a closed set of equations for the distribution functions and the electric potential energy
provided they are augmented by boundary conditions far away from the interface, at $z=z_b\!<\!0$ and
$z=z_w\!>\!0$ (see Fig.~\ref{IntModelNotation}).

\begin{figure}[t]
        \includegraphics[width=\linewidth]{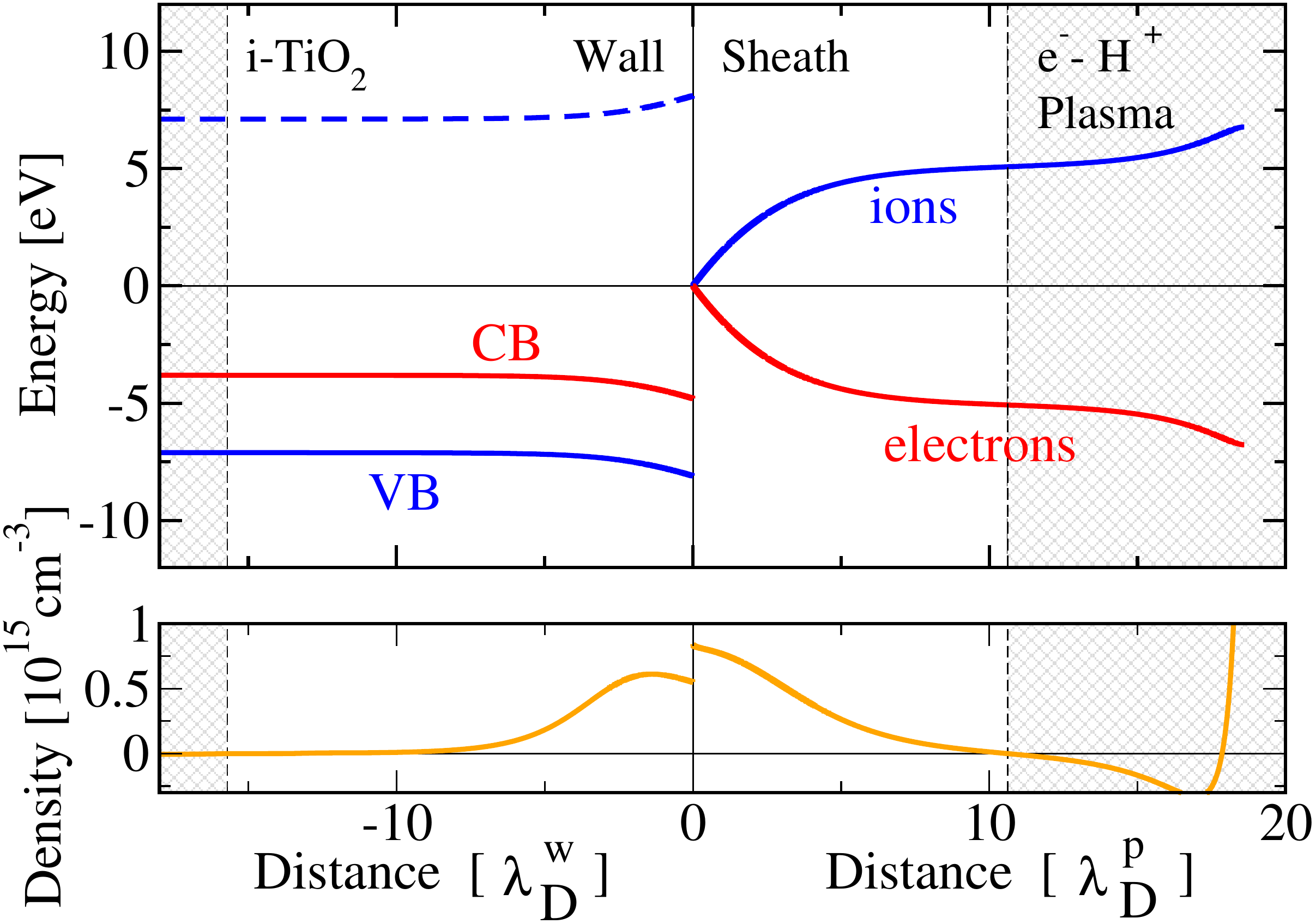}
        \caption{Upper panel: Edges for the conduction and valence bands (solid red and blue 
                 lines), the free hole motion (dashed blue lines), and the potential energies
                 for electrons and ions on the plasma side (also indicated by solid red and
                 blue lines) for an intrinsic ${\rm TiO}_2$ surface in contact with an hydrogen 
                 plasma. Shaded regions indicate respectively the reservoir and source which have 
                 been set up to provide the correct physical boundary conditions for the 
                 double layer. Lower panel: Density profiles $\rho_w(z)=[n_*(z)-n_h(z)]\theta(-z)$
                 and $\rho_p(z)=[n_i(z)-n_e(z)]\theta(z)$. By definition they are both 
                 positive outside the shaded regions, that is, in the regions which are 
                 physically relevant. The interface is collisionless on both sides and 
                 perfectly absorbing. Material and plasma parameters are $\chi=4.8\,{\rm eV}$, 
                 $E_g=3.3\,{\rm eV}$, $m_e^*=m_e$, $m_h=0.8\,m_e$, $\varepsilon=6$, $I=13.6\,{\rm eV}$, 
                 $\Gamma=2\,{\rm eV}$, $k_BT_*=k_BT_h=0.2\,{\rm eV}$, $k_BT_e=10\,k_BT_i=2\,{\rm eV}$.
                 Due to the somewhat unrealistic temperatures of the charges on both 
		 sides of the interface (required to stabilize the numerics) the Debye 
		 screening length on the solid side $\lambda_D^w=2.2\cdot 10^{-6}\,{\rm cm}$ is 
		 comparable to the Debye screening length on the plasma side 
		 $\lambda_D^p=1.6\cdot 10^{-6}\,{\rm cm}$. Due to the absence of collisions the 
                 numerical values, however, should not be taken literally. A collisional theory 
                 would produce different density and potential profiles. }
        \label{IntModelData}
\end{figure}

Essential for the selfconsistent description of the electric response is the implementation
of the boundary conditions far away from the interface. It depends on how the interface,
and hence the electric double layer, is electrically connected to the outside. For an 
electrically floating interface the double layer is embedded between two quasi-neutral,
field-free regions. The distribution functions in these regions are of course unknown.
They are themselves the result of the electric response and can thus not be used as 
boundary conditions for the Boltzmann equations. A possibility to overcome this complication
is to enforce two inflection points in the profile of $U_C(z)$ mimicking the quasi-neutral, 
field-free surroundings of the double layer,
see Fig.~\ref{IntModelNotation}. 
The price to be paid comes in the form of eight parameters: Four boundary densities 
($n_{b*}, n_{bh}, n_{se}$, and $n_{si}$), for each of the charged species considered, and four 
spatial coordinates ($z_b, z_1, z_p$, and $z_w$), two of them denote the positions of the 
inflection points ($z_1, z_p$), one the position of the particle reservoir in the solid ($z_b$),
and one the position of the source ($z_w$) in the plasma. The adjustment of the distribution
functions to the interface can then be emulated without simulating the bulk 
of the solid and the plasma.

The framework outlined indicates what is needed for an integral modeling of the
electric response. Besides specifying collision processes on both sides of the interface, 
matching and boundary conditions for the distribution functions have to be worked out--based 
on, respectively, the emissive properties of the plasma-solid interface and its asymptotics 
on both sides. To gain first insights into the selfconsistent electric response of a 
plasma-solid interface Bronold and Fehske~[\onlinecite{BF17}] applied the equations listed above 
to a collisionless, perfectly absorbing interface. By ignoring the collision integrals, the
lateral momentum $\vec{K}$ could be eliminated from all equations, including the matching
conditions. In addition, the solutions of the Boltzmann equations turned out to be no longer
explicit functions of $z$ but only functions of $E$ and $U_C(z)$. The first integral of the
Poisson equation could thus be obtained analytically, greatly simplifying the further numerical
treatment. The selfconsistent embedding finally led to four nonlinear algebraic equations for
the potential energies $U_C(z_b), U_C(z_1), U_C(z_p)$, and $U_C(z_w)$ which could be solved
numerically with moderate effort. Results for the perfectly absorbing, collisionless 
interface--for parameters applicable to intrinsic ${\rm TiO}_2$ in contact 
with an hydrogen plasma--are shown in Fig.~\ref{IntModelData}.

In a collisionless theory recombination and generation of electron-hole pairs inside
the solid are excluded. Nothing balances thus--strictly speaking--the permanent influx 
of electrons and ions onto the interface. For a quasi-stationary regime to develop it is 
necessary to introduce by hand a recombination condition. The condition employed in~Ref.~[\onlinecite{BF17}] 
utilized the fact that intraband energy and momentum relaxation is much faster than interband 
recombination. The electron and hole densities inside the solid can thus be split into 
thermalized/trapped [$n^t_{*,h}(z)$] and free parts [$n_{*,h}^j(z)$], with only the former 
acting as a source in the Poisson equation, and the latter assumed to cancel once integrated 
over a spatial strip extending from the inflection point $z_1<0$ inside the solid to the 
interface at $z=0$,
\begin{eqnarray}
\int_{z_1}^0 \!\! dz \, \rho_w^j(z) = \int_{z_1}^0 \!\!dz \, [\,n_*^j(z)-n_h^j(z)\,] = 0~.
\label{ReCombCond}
\end{eqnarray}
With this postulate the collisionless theory provided in total eight equations for the 
eight parameters mentioned above. A selfconsistent description of the electric response,
indicated by a selfconsistent embedding of the electric double layer between two 
quasi-neutral, field-free regions, could thus be realized without taking collisions 
into account. 

The absence of collisions can be at most justified on the plasma side, where electrons
are strongly depleted and hence nearly collisionless, and ions are only subject to
ion-neutral collisions which however have to be considered only in particular
situations~[\onlinecite{Sternovsky05,Riemann03,SG91}] or when a very precise description of the
plasma sheath is required~[\onlinecite{TC17}]. Hence, on the plasma side ignoring or including
collisions is essentially a question of how accurate one wants to describe the plasma
sheath. On the solid side of the interface, however, neglecting collisions makes the
theory conceptually incomplete. Without recombination, either radiatively or 
non-radiatively, both involving collisions between conduction band electrons and valence 
band holes, with the latter mediated by trap states (see Fig.~\ref{IntModelProcesses}), 
no quasi-stationarity can be achieved because nothing balances the injection of conduction 
band electrons and valence band holes from the plasma. The carrier concentration inside 
the solid would just grow indefinitely. Future work has thus to include collisions, at 
least on the solid side of the interface. The ad-hoc condition~\eqref{ReCombCond}
could then be replaced by a consistent truncation of the perturbation theory 
treatment of charge relaxation and recombination. Work in this direction is in progress.

In this subsection we discussed a general framework extending the kinetic modeling 
of the charge dynamics of the plasma to the inside of the solid. Within such an
approach the electric modification of the plasma-solid interface--the build-up of 
the electric double layer--can be described selfconsistently, treating the plasma-based 
positive and the solid-based negative part of the double layer on an equal footing. 
In particular for hybrid electronics, using arrays of microdischarges integrated on 
semiconducting substrates, it may be necessary to adopt this type of integral modeling. 
In discharges used for materials processing, chemical and structural responses of the 
interface, involving the formation of adsorption layers and/or the sputtering of the 
outer layers of the interface, are of course intimately coupled to the electric 
response. To include them into the framework we just described is in principle possible. 
Staying at the level of Boltzmann equations is rather advantageous at this point. For 
each species of interest one has to add a Boltzmann equation with appropriate collision 
terms and matching/boundary conditions. Ways to treat rough and disordered interfaces 
are also conceivable. The numerical effort however will be rather high. An integral 
modeling, taking the electric, chemical, and structural response of the plasma-solid 
interface selfconsistently into account will be possible using high performance 
computing.  

The limitations of the present approach are mostly due to the simplified treatment of the solid. The related processes in the surface material require \textit{ab initio} modeling. In the next section such an approach that is based on nonequilibrium Green functions will be outlined. At the same time, there the treatment of the plasma particles will be substantially simplified compared to the present treatment. So both methods have complementary features.



\subsection{Integrated modeling of ion-surface interaction}\label{ss:intgrated_ion_impact}
In the previous subsection the traditional plasma physics approach that assumes that the processes inside the solid occur on too short (small) time (length) scales to affect the plasma~[\onlinecite{Franklin76}]. However, if the plasma impact is spatially localized and due to rare events (such as the impact of a projectiles) substantial deviations from such an approach have to be expected. It is, therefore, of interest to consider a space and time resolved description which, in principle allows to verify the above hypothesis or, on the opposite, to capture processes that are missed by it. While the former approach can be understood as based on space and time averaging over the scales of the solid -- which is essentially a mean field description, the latter concept takes into account correlation effects and fluctuations around mean values, e.g.~[\onlinecite{lacroix_prb14}].

Here, we develop a quantum kinetic description of the coupled electron-ion dynamics across the plasma interface generalizing the methods described in Sec.~\ref{s:negf}. We start from the second-quantized many-body Hamiltonian for the electrons in the interface and separate the system into a plasma ($p$) and solid surface part ($s$) [we denote $\Omega=\{p,s\}$ and do not write the spin index explicitly],
\begin{eqnarray}
\label{eq.ham}
H_{\textup{interface}} &=&\sum_{\alpha\beta\in\Omega}\sum_{ij}H^{\alpha\beta}_{ij}(t)c^{\alpha\dagger}_ic^\beta_j+
\nonumber\\
&&\frac{1}{2}\sum_{\alpha\beta\gamma\delta\in\Omega}\sum_{ijkl}W^{\alpha\beta\gamma\delta}_{ijkl}c^{\alpha\dagger}_ic^{\beta\dagger}_jc^{\gamma}_kc^{\delta}_l\,.
\end{eqnarray}
Here, the operator $c^{\alpha\dagger}_i$ ($c^{\alpha}_i$) creates (annihilates) an electron in the state $i$ of part $\alpha$. The one-particle Hamiltonian~$H(t)$ contains the electron's kinetic and the time-dependent potential energy, and $W$ accounts for all possible electron-electron Coulomb interactions within and between the two parts.

Considering individual energetic plasma ions, which penetrate into the solid and undergo scattering and stopping in the surface layers, we describe the system~(\ref{eq.ham}) by a one-particle nonequilibrium Green function (NEGF) $G^{\alpha\beta}_{ij}(t,t')$, as introduced in Sec.~\ref{sss:negf-defs}, but here with an additional $2\times2$ matrix structure ($\alpha, \beta=\{p,s\}$), 
\begin{eqnarray}
\label{eq.negf}
 G^{\alpha\beta}_{ij}(t,t') &=&- i \langle T_C c^\alpha_{i}(t)c_{j}^{\beta\dagger}(t')\rangle\,, 
 \\
  \rho_{ij}^{\alpha\beta}(t) &=& - i G^{\beta\alpha}_{ji}(t,t^+)\,,
\label{eq:g-dm}  
\end{eqnarray}
e.g., Refs.~[\onlinecite{stefanucci_cambridge_2013, balzer-book}], and the time-diagonal elements provide the density matrix (\ref{eq:g-dm}), as discussed in Sec.~\ref{sss:negf-defs}.
The diagonal elements, $\rho_{ij}^{pp}$ [$\rho_{ij}^{ss}$], refer to the plasma part, describing the dynamics of free electrons and electrons bound in the ion [to the solid part,  describing electrons in bound states of the solid surface]. Moreover, the density matrix component $\rho_{ij}^{ps}$ is related to charge transfer processes between plasma and solid and will be of special interest in the following. 

The equations of motion for the NEGF are the generalization of Eq.~(\ref{eq.kbe})
to the interface 
%
\begin{eqnarray}
 i\,\partial_tG^{\alpha\beta}_{ij}(t,t') &-& \sum_{\delta\in\Omega,k} H^{\alpha\delta}_{ik}(t)G^{\delta\beta}_{kj}(t,t')=\delta^{\alpha\beta}_{ij}\delta_C(t,t')+
 \nonumber\\ 
 &&\sum_{\delta\in\Omega,k} \int_C\!\!\!d\bar{t}\,\Sigma^{\alpha\delta}_{ik}[W,G](t,\bar{t})G^{\delta\beta}_{kj}(\bar{t},t')\,,
\label{eq.kbe_interface}
\end{eqnarray}
where the self-energy $\Sigma^{\alpha\beta}(t,t')$ describes the interaction between the electrons and with phonons.
Even though a complete solution of the KBE~(\ref{eq.kbe_interface}) for real materials and with a full quantum treatment of the plasma electrons is out of reach, these equations provide the rigorous starting point for the development of consistent approximations. 
%
In the following we show how it is possible to include the electronic states of the ion via an embedding self-energy approach~[\onlinecite{stefanucci_cambridge_2013}], where resonant (neutralization and ionization) processes can be studied. While this embedding approach is based on a formal decoupling of the surface and plasma parts of the KBE, it retains one-electron charge transfer in the Hamiltonian $H^{sp}$, cf. Eq.~(\ref{eq.sigma.ct}), see below. 
A closed description of the solid can be maintained if correlations in the plasma part and the feedback of the solid on the plasma can be neglected, i.e., for $\Sigma^{sp}=\Sigma^{pp}=0$. This is usually well fulfilled, except for atmospheric pressure plasmas where small correlation corrections should be taken into account.  Then, the KBE~(\ref{eq.kbe}) for the plasma part simplify to
\begin{eqnarray}
  \sum_{k} \left\{i\partial_t\delta_{ik}-H^{pp}_{ik}(t)\right\}g^{pp}_{kj}(t,t')=\delta_{ij}\delta_C(t,t')\,,
\end{eqnarray}
 where, with the solution $g^{pp}(t,t')$ we denote the NEGF of the electrons inside the plasma ion, and the time dependence of $H^{pp}(t)$ accounts for possible parametric changes of the energy levels (e.g., as function of the distance of the ion from the surface). 
 
 The main result is a closed equation for $G^{ss}(t,t')$:
\begin{eqnarray}
\label{eq.kbe.embedding}
 \sum_{k} \left\{i\partial_t\delta_{ik}-H^{ss}_{ik}(t)\right\}G^{ss}_{kj}(t,t')=\delta_{ij}\delta_C(t,t')
 +
 \\\nonumber
 \sum_k\int_C\!\!\!d\bar{t}\,\left\{\Sigma^{ct}_{ik}(t,\bar{t})+\Sigma^{ss}_{ik}[G^{ss}](t,\bar{t})\right\}G^{ss}_{kj}(\bar{t},t')\,,
\end{eqnarray}
with the charge transfer (or embedding) self-energy that involves the charge transfer hamiltonian
\begin{eqnarray}
\label{eq.sigma.ct}
  \Sigma^{ct}_{ij}(t,t') &=& \sum_{kl}H^{sp}_{ik}(t)g_{kl}^{pp}(t,t') H^{ps}_{lj}(t')\,,
\\
H^{sp}_{ij}(t) &=& \int\!\!\! d^3r\,\phi^s_i(\vec{r})(T+V)\phi_j^p(\vec{r};t)\,.
\label{eq:hsp}
\end{eqnarray}
Equation~(\ref{eq.kbe.embedding}) shows how the many-body description of an isolated (but correlated) solid is altered by the presence of the electronic states of a plasma ion, with the latter giving rise to an additional self-energy $\Sigma^{\textup{ct}}(t,t')$. While,  for $\Sigma^{\textup{ct}}=0$, the KBE~(\ref{eq.kbe.embedding}) conserves the particle number [for a conserving approximation of the self-energy $\Sigma^{ss}$, such as Hartree-Fock, second order Born or GW], the inclusion of the embedding self-energy explicitly allows for time-dependent changes of the particle number in the solid and thus accounts for ion charging and neutralization effects. For the practical solution of Eq.~(\ref{eq.kbe.embedding}), the charge transfer Hamiltonian $H^{sp}(t)$ has to be computed by selecting the relevant electronic transitions between solid and plasma and computing the matrix elements of the kinetic and potential energy operators $T$ and $V$, with the electronic single-particle wave functions $\phi^{s}$ ($\phi^p$) in the solid (ion).

\begin{figure}[h]
  \begin{center} 
  \hspace{-0.cm}\includegraphics[width=0.25\textwidth]{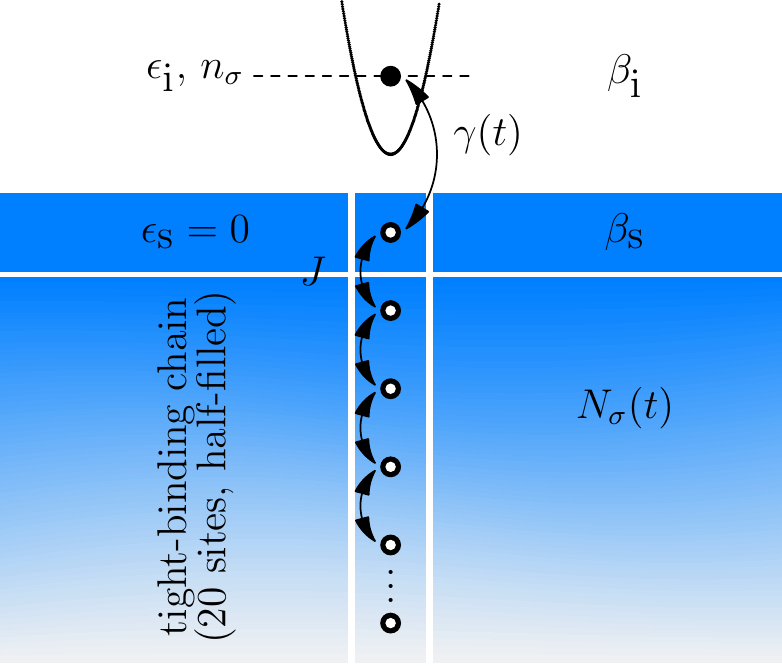}\hspace{0.1cm}
  \hspace{-0.cm}\includegraphics[width=0.22\textwidth]{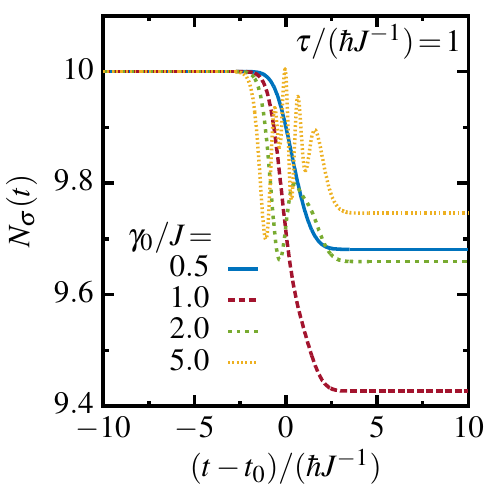}\\[0.1cm]
  \hspace{-0.cm}\includegraphics[width=0.485\textwidth]{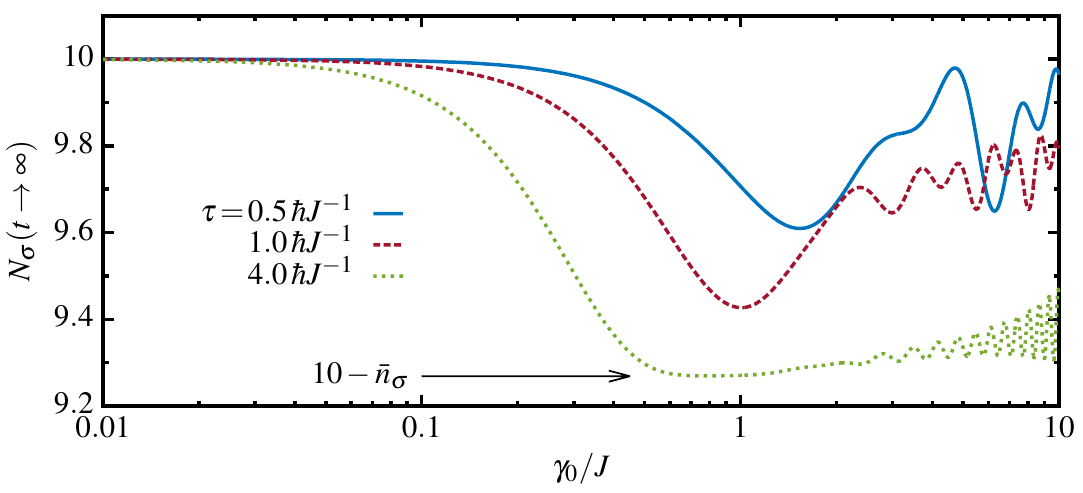}
  \end{center}
  \vspace{-.40cm}
  \caption{Example calculation illustrating the embedding scheme 
  An initially half-filled tight-binding chain ($20$ sites, nearest-neighbor hopping $J$, inverse temperature $\beta_\textup{s}=100J^{-1}$) is coupled via a time-dependent parameter $\gamma(t)=\gamma_0\textup{e}^{-(t-t_0)^2/(2\tau^2)}$ to an external energy level $\epsilon_\textup{i}=+J$ giving rise to the transfer of charge. The initial occupation of the energy level is given by $n_\sigma=0.269$ (corresponding to an inverse temperature of $\beta_\textup{i}=1J^{-1}$) and $\bar{n}_\sigma=1-n_\sigma$. Furthermore, $N_\sigma(t)=\sum_{i}n_{i\sigma}(t)$ denotes the total density on the chain.} 
  \label{fig:doublons1}
  \end{figure}

To test whether this approach can be applied to the plasma-solid interface we use the same concept as for the computation of the stopping power in Sec.~\ref{ss:negf-stopping}. We use, as a model, a nanostructured solid represented by a chain of 20 lattice sites at half filling (i.e. containing 20 electrons). The charge transfer selfenergy (\ref{eq.sigma.ct}) is parametrized by a time-dependent Gaussian coupling term, $\gamma(t)=\gamma_0\textup{e}^{-(t-t_0)^2/(2\tau^2)}$ that models the approach and elastic scattering of a projcetile. In Figure~\ref{fig:doublons1} we analyze how the charge transfere changes upon variation of the amplitude $\gamma_0$ and interaction duration $\tau$. Interestingly, the charge transfer dependes non-monotonically on the amplitude (see top right part of the figure) and, a large couplings, undergoes strong oscillations. On the other hand, increasing the duration of the interaction (bottom right figure) enhances the charge transfer.

With this we have formulated the general NEGF framework for integrated plasma-surface modeling. This approach is largely complementary to the one developed in Sec.~\ref{ss:intgrated_dl}: There the electronic double layer and charge transfer processes along the interface were studied. Here we focused on resonant electron transfer processes attempting an accurate description of the solid. However, the plasma particles were treated on a simplified level. Ultimately, a combination of both approaches will be needed.

\subsection{Coupling plasma and surface simulations}\label{ss:plasma-surface}
Returning to the overview given in Fig.~\ref{fig:theory}, the coupling between surface and plasma simulations proceeds via the fluxes of particles, momentum and energy. Fluxes of neutrals, electrons and ions from the plasma, $\textbf{J}^p_a$, have to be provided by plasma simulations and serve as an input for surface simulations. Vice versa, surface simulations, ultimately have to provide the the energy or momentum resolved fluxes,
 $\textbf{J}^s_a$, of atoms, electrons and ions that leave the surface.

In the previous sections we have discussed examples of particle fluxes that are produced by surface science simulations. The first example are the fluxes of neutral gas atoms that are scattered from a metal surface and which are obtained from a combined MD--rate equations model as discussed in Sec.~\ref{ss:freezout}.
The result of these simulations is the sticking probability in dependence of impact energy and angle of incidence and lattice temperature, $R_{st}(E_{in},\theta_{in};T_s)$,  [\onlinecite{filinov_psst18_2}]. Furthermore, these simulations also yield the energy distribution of the reflected atoms. These results can be directly used in PIC-MCC simulations that trace the dynamics of (fast) neutrals, in addition to electrons and ions. Based on the sticking probabilities, a Monte Carlo procedure can be used to determine if the particle is reflected back into the plasma or if it is adsorbed on the surface. In the former case, the energy distribution of the reflected atoms provides valuable information on the collision process. 

The second example is secondary electron emission (SEE) which is of key relevance, as we discussed in Sec.~\ref{s:intro}, cf. Fig.~\ref{fig:see}. In this paper (Sec.~\ref{s:qkin}) we have presented quantum kinetic results for the energy resolved SEE coefficient. The present approach is expected to be more accurate and potentially more general than previous models and thus may serve as a valuable input for PIC-MCC simulations. 

\subsection{Taking into account plasma-induced  surface morphology changes}\label{ss:morphology-surface} 
In Sec.~\ref{ss:spa} simulations of metal cluster growth on a polymer substrate were described. Using selective process acceleration the MD simulations could be extended to several minutes.
The resulting plasma-induced adsorbate layer strongly differs from the original ``clean'' surface and can be used to re-compute the electronic structure of bulk and surface states via DFT as indicated in Fig.~\ref{fig:theory}.
Furthermore, as was discussed in Sec.~\ref{s:influences}, such a modified (``dirty'') surface reacts very differently on the plasma contact leading, in particular, to strongly enhanced secondary electron emission.
This could be investigated in detail, e.g. by means of TDDFT (Sec.~\ref{ss:tddft}) using the pre-computed surface morphology as the starting point. 
Alternatively, this information can be used as input for quantum kinetic simulations as described in Sec.~\ref{s:qkin} which requires to modify the employed model of the solid.
Moreover, the MD simulation of neutral atom scattering, cf. Sec.~\ref{ss:freezout} can be repeated using the plasma-modified surface as an input. This will directly impact the plasma properties via the modified fluxes $\textbf{J}^s_a$, as discussed in Sec.~\ref{ss:plasma-surface}.

\section{Conclusion and outlook}\label{s:conclusion1}
In this paper we considered the interaction of low-temperature plasmas with a solid surface, discussed the broad variety of physical and chemical processes [Fig.~\ref{fig:interface-physics}] and argued that the mutual interaction between both sides requires to develop closely coupled plasma-surface simulations. We presented an overview on a research poject that is under way at Kiel University, in collaboration with scientists from Greifswald, that aims, aomng others, at developing such simulations.
Such an approach has the potential for major advances of this field because most of the currently used models are phenomenological using surface coefficients that are poorly known, both experimentally and theoretically. Moreover, these parameters--even if they exist--may carry an (unknown) dependence on the surface conditions or the plasma parameters.
We discussed the simulation approaches that are required to treat the plasma and the solid side of the interface and then listed the methods that are suitable to simulate particle, energy and momentum fluxes across the interface, cf. Fig.~\ref{fig:theory}.
We highlighted four groups of methods: semiclassical approaches such as MD and KMC, quantum Boltzmann equation-based models and \textit{ab initio} approaches that are either based on density functional theory or on nonequilibrium Green functions.

In Sec.~\ref{s:mesoscopic} we discussed two examples of semiclassical molecular dynamics that use accurate force fields to simulate surface processes involving neutral plasma particles. The main challenge here is to extend these simulations that typcially use a time step on the order of one femtosecond to experimentally relevant times of minutes [\onlinecite{bonitz_psst18}]. This can be achieved via selective process acceleration -- which was demonstrated for the metal cluster growth on a polymer surface. Here synchronized acceleration of metal atom diffusion and deposition allows to extend the simulations by more than nine orders of magnitude up to several minutes [\onlinecite{abraham_jap16}]. The second approach that was discussed was dynamical freezout of collective modes (DFCM) [\onlinecite{filinov_psst18_1, bonitz_psst18}]. Here MD simulations were used to reduce the dynamics of atom desorption on a metal surface to a small set of collective modes that obey coupled rate equations. These rate equations completely describe the sticking behavior at time scales larger than a few tens of picoseconds.
There are various ways how to extend the present idea. If the surface is 
inhomogeneous,  a straightforward generalization would be to include the space  dependence into the densities and the rates. Then, the rate equations turn into 
hydrodynamic equations. Furthermore, the effect of a plasma environment, such as 
characteristic particle fluxes or an adsorbate-covered surface, are 
straightforwardly included into our scheme, as discussed in Ref.~\onlinecite{filinov_psst18_1}.

However, semiclassical MD has a limited sphere of applicabiity.
In particular for the description of electrons and ions crossing the interface, semiclassical MD fails, and quantum approaches are necessary. This concerns the neutralization of low-energy ions, e.g. [\onlinecite{pamperin_2015_many}] and their stopping in the solid, as well as the electron dynamics across the interface, e.g. [\onlinecite{bronold_2015_absorption}]. Here nonequlibrium quantum methods such as the quantum Boltzmann equation, density functional theory and nonequlibrium Green functions simulations, e.g. [\onlinecite{zhao_2015_comparison, balzer_prb16, balzer-book}] are the methods of choice which we discussed in Secs.~\ref{s:qkin}, \ref{s:abinit} and \ref{s:negf}. In the present article we attempted at giving a comprehensive overview on these methods regarding their existing applications to and future potential for accurately simulating plasma surface interaction. 

At the same time, these methods have been developed, so far, almost independently of each other, but we hope that the present work will stimulate future comparisons and combinations.
Each of these quantum methods, in particular time-dependent DFT and NEGF, is computationally extremely expensive, and still each of them has significant limitations, for an overview and examples, cf. table~\ref{tab:processes-methods}. Interestingly, TDDFT and NEGF are highly complementary, so it will be important to develop suitable combinations that allow one to overcome bottlenecks. Here we mention a recently proposed hybrid scheme [\onlinecite{hopian_prl_16}], as an example.

Another important goal will have to be to use the results of TDDFT and NEGF as input to simpler approaches such as the quantum Boltzmann equation, e.g. Ref. \onlinecite{bonitz_qkt}. Moreover, these simulations can also be extended to longer times using the coupling to a reduced system of rate equations (DFDM) as was explained in the context of MD simulations above. In fact, in Sec.~\ref{s:qkin} for the computation of the secondary electron emission coefficient also a system of rate equations was derived that allows to capture the relevant degrees of freedom (electronic states of the helium projectile). 

Finally, to properly capture the influence of the plasma on the solid, the above surface simulations have to be linked to fluid or kinetic simulations of the plasma, as indicated by the arrows in Fig.~\ref{fig:theory}.
Ultimately, an integrated modeling of the plasma and the solid surface will be required [\onlinecite{interface}] to overcome the trial and error character of many experiments and to achieve a predictive modeling of the relevant processes. In Sec.~\ref{s:integrated} we presented two possible approaches that are based on the quantum Boltzmann equation and nonequilibrium Green functions, within an embedding approach, respectively.
Even though these have been rather simple examples and model systems, they indicate the way how such an integrated modeling can be constructed in the future.

We expect that our results will not only be of relevance for ``traditional'' materials embedded in a plasma such as metals or semiconductors but also for new materials. Of particular interest, in the near future, could be \textit{nanomaterials}, are such as boron nitride structures or carbon nanotubes or graphene sheets and nanoribbons. For these nanomaterials both finitie size effects and electronic correlations will be of particular importance [\onlinecite{son_prl_06, prezzi_prb_08}] and the array of methods oulined in this article should be, in their combination, suitable to describe plasma-surface interaction with such exciting novel materials.


\section*{Acknowledgements}
The authors acknowledge many fruitful discussions with our colleagues in Kiel -- in particular, M. Bauer, J. Benedikt, J. Golda, B. Hartke, H. Kersten, O. Magnussen, K. Rossnagel, J. Stettner, and T. Trottenberg -- with whom the present concept has been developed.
EP is grateful to L. Deuchler for insightful discussions concerning 
TDDFT-MD simulations. MB and KB are grateful to A. Marini and D. Sangalli for their support in using their \textit{ab initio} NEGF-code \textit{Yambo}.
MB is grateful to K. Becker, A. Bogaerts, P. Bruggeman, Z. Donko, J.G. Eden, U. Fantz, I. Kaganovich, M. Kushner, E. Neyts, G. Oehrlein, K. Ostrikov, and Y. Raitses for many stimulating remarks during presentation of early versions of this work. 




\begin{thebibliography}{241}
\expandafter\ifx\csname natexlab\endcsname\relax\def\natexlab#1{#1}\fi
\expandafter\ifx\csname bibnamefont\endcsname\relax
  \def\bibnamefont#1{#1}\fi
\expandafter\ifx\csname bibfnamefont\endcsname\relax
  \def\bibfnamefont#1{#1}\fi
\expandafter\ifx\csname citenamefont\endcsname\relax
  \def\citenamefont#1{#1}\fi
\expandafter\ifx\csname url\endcsname\relax
  \def\url#1{\texttt{#1}}\fi
\expandafter\ifx\csname urlprefix\endcsname\relax\def\urlprefix{URL }\fi
\providecommand{\bibinfo}[2]{#2}
\providecommand{\eprint}[2][]{\url{#2}}

\bibitem[{\citenamefont{Skiff and Wurtele}(2017)}]{doe-report-17}
\bibinfo{author}{\bibfnamefont{F.}~\bibnamefont{Skiff}} \bibnamefont{and}
  \bibinfo{author}{\bibfnamefont{J.}~\bibnamefont{Wurtele}},
  \bibinfo{type}{Tech. Rep.}, \bibinfo{institution}{U.S. Department of Energy,
  Office of Sciences} (\bibinfo{year}{2017}), \bibinfo{note}{report of the
  panel on Frontiers of Plasma Science}.

\bibitem[{\citenamefont{Meyyappan}(2011)}]{meyyappan_jpd_11}
\bibinfo{author}{\bibfnamefont{M.}~\bibnamefont{Meyyappan}},
  \bibinfo{journal}{Journal of Physics D: Applied Physics}
  \textbf{\bibinfo{volume}{44}}, \bibinfo{pages}{174002}
  (\bibinfo{year}{2011}),
  \urlprefix\url{http://stacks.iop.org/0022-3727/44/i=17/a=174002}.

\bibitem[{\citenamefont{Ostrikov et~al.}(2013)\citenamefont{Ostrikov, Neyts,
  and Meyyappan}}]{ostrikov_aip_13}
\bibinfo{author}{\bibfnamefont{K.}~\bibnamefont{Ostrikov}},
  \bibinfo{author}{\bibfnamefont{E.~C.} \bibnamefont{Neyts}}, \bibnamefont{and}
  \bibinfo{author}{\bibfnamefont{M.}~\bibnamefont{Meyyappan}},
  \bibinfo{journal}{Advances in Physics} \textbf{\bibinfo{volume}{62}},
  \bibinfo{pages}{113} (\bibinfo{year}{2013}),
  \eprint{https://doi.org/10.1080/00018732.2013.808047},
  \urlprefix\url{https://doi.org/10.1080/00018732.2013.808047}.

\bibitem[{\citenamefont{Son et~al.}(2006)\citenamefont{Son, Cohen, and
  Louie}}]{son_prl_06}
\bibinfo{author}{\bibfnamefont{Y.-W.} \bibnamefont{Son}},
  \bibinfo{author}{\bibfnamefont{M.~L.} \bibnamefont{Cohen}}, \bibnamefont{and}
  \bibinfo{author}{\bibfnamefont{S.~G.} \bibnamefont{Louie}},
  \bibinfo{journal}{Phys. Rev. Lett.} \textbf{\bibinfo{volume}{97}},
  \bibinfo{pages}{216803} (\bibinfo{year}{2006}),
  \urlprefix\url{https://link.aps.org/doi/10.1103/PhysRevLett.97.216803}.

\bibitem[{\citenamefont{Prezzi et~al.}(2008)\citenamefont{Prezzi, Varsano,
  Ruini, Marini, and Molinari}}]{prezzi_prb_08}
\bibinfo{author}{\bibfnamefont{D.}~\bibnamefont{Prezzi}},
  \bibinfo{author}{\bibfnamefont{D.}~\bibnamefont{Varsano}},
  \bibinfo{author}{\bibfnamefont{A.}~\bibnamefont{Ruini}},
  \bibinfo{author}{\bibfnamefont{A.}~\bibnamefont{Marini}}, \bibnamefont{and}
  \bibinfo{author}{\bibfnamefont{E.}~\bibnamefont{Molinari}},
  \bibinfo{journal}{Phys. Rev. B} \textbf{\bibinfo{volume}{77}},
  \bibinfo{pages}{041404} (\bibinfo{year}{2008}),
  \urlprefix\url{https://link.aps.org/doi/10.1103/PhysRevB.77.041404}.

\bibitem[{\citenamefont{Adamovich et~al.}(2017)\citenamefont{Adamovich,
  Baalrud, Bogaerts, Bruggeman, Cappelli, Colombo, Czarnetzki, Ebert, Eden,
  Favia et~al.}}]{adamovich_2017_plasma}
\bibinfo{author}{\bibfnamefont{I.}~\bibnamefont{Adamovich}},
  \bibinfo{author}{\bibfnamefont{S.~D.} \bibnamefont{Baalrud}},
  \bibinfo{author}{\bibfnamefont{A.}~\bibnamefont{Bogaerts}},
  \bibinfo{author}{\bibfnamefont{P.~J.} \bibnamefont{Bruggeman}},
  \bibinfo{author}{\bibfnamefont{M.}~\bibnamefont{Cappelli}},
  \bibinfo{author}{\bibfnamefont{V.}~\bibnamefont{Colombo}},
  \bibinfo{author}{\bibfnamefont{U.}~\bibnamefont{Czarnetzki}},
  \bibinfo{author}{\bibfnamefont{U.}~\bibnamefont{Ebert}},
  \bibinfo{author}{\bibfnamefont{J.~G.} \bibnamefont{Eden}},
  \bibinfo{author}{\bibfnamefont{P.}~\bibnamefont{Favia}},
  \bibnamefont{et~al.}, \bibinfo{journal}{J. Phys. D: Appl. Phys.}
  \textbf{\bibinfo{volume}{50}}, \bibinfo{pages}{323001}
  (\bibinfo{year}{2017}),
  \urlprefix\url{http://stacks.iop.org/0022-3727/50/i=32/a=323001}.

\bibitem[{\citenamefont{Hagelaar and Pitchford}(2005)}]{hagelaar_2005}
\bibinfo{author}{\bibfnamefont{G.~J.~M.} \bibnamefont{Hagelaar}}
  \bibnamefont{and} \bibinfo{author}{\bibfnamefont{L.~C.}
  \bibnamefont{Pitchford}}, \bibinfo{journal}{Plasma Sources Science and
  Technology} \textbf{\bibinfo{volume}{14}}, \bibinfo{pages}{722}
  (\bibinfo{year}{2005}),
  \urlprefix\url{http://stacks.iop.org/0963-0252/14/i=4/a=011}.

\bibitem[{\citenamefont{Donko and Dyatko}(2016)}]{donko_2016}
\bibinfo{author}{\bibfnamefont{Z.}~\bibnamefont{Donko}} \bibnamefont{and}
  \bibinfo{author}{\bibfnamefont{N.}~\bibnamefont{Dyatko}},
  \bibinfo{journal}{The European Physical Journal D}
  \textbf{\bibinfo{volume}{70}}, \bibinfo{pages}{135} (\bibinfo{year}{2016}),
  ISSN \bibinfo{issn}{1434-6079},
  \urlprefix\url{https://doi.org/10.1140/epjd/e2016-60726-4}.

\bibitem[{\citenamefont{Teunissen and Ebert}(2016)}]{ebert_2016}
\bibinfo{author}{\bibfnamefont{J.}~\bibnamefont{Teunissen}} \bibnamefont{and}
  \bibinfo{author}{\bibfnamefont{U.}~\bibnamefont{Ebert}},
  \bibinfo{journal}{Plasma Sources Science and Technology}
  \textbf{\bibinfo{volume}{25}}, \bibinfo{pages}{044005}
  (\bibinfo{year}{2016}),
  \urlprefix\url{http://stacks.iop.org/0963-0252/25/i=4/a=044005}.

\bibitem[{\citenamefont{Becker et~al.}(2017)\citenamefont{Becker, Kählert,
  Sun, Bonitz, and Loffhagen}}]{becker_psst17}
\bibinfo{author}{\bibfnamefont{M.~M.} \bibnamefont{Becker}},
  \bibinfo{author}{\bibfnamefont{H.}~\bibnamefont{Kählert}},
  \bibinfo{author}{\bibfnamefont{A.}~\bibnamefont{Sun}},
  \bibinfo{author}{\bibfnamefont{M.}~\bibnamefont{Bonitz}}, \bibnamefont{and}
  \bibinfo{author}{\bibfnamefont{D.}~\bibnamefont{Loffhagen}},
  \bibinfo{journal}{Plasma Sources Science and Technology}
  \textbf{\bibinfo{volume}{26}}, \bibinfo{pages}{044001}
  (\bibinfo{year}{2017}),
  \urlprefix\url{http://stacks.iop.org/0963-0252/26/i=4/a=044001}.

\bibitem[{\citenamefont{Derzsi et~al.}(2015)\citenamefont{Derzsi, Korolov,
  Schüngel, Donk{\'{o}}, and Schulze}}]{derzi_2015_effects}
\bibinfo{author}{\bibfnamefont{A.}~\bibnamefont{Derzsi}},
  \bibinfo{author}{\bibfnamefont{I.}~\bibnamefont{Korolov}},
  \bibinfo{author}{\bibfnamefont{E.}~\bibnamefont{Schüngel}},
  \bibinfo{author}{\bibfnamefont{Z.}~\bibnamefont{Donk{\'{o}}}},
  \bibnamefont{and} \bibinfo{author}{\bibfnamefont{J.}~\bibnamefont{Schulze}},
  \bibinfo{journal}{Plasma Sources Sci. Technol.}
  \textbf{\bibinfo{volume}{24}}, \bibinfo{pages}{034002}
  (\bibinfo{year}{2015}),
  \urlprefix\url{https://doi.org/10.1088%2F0963-0252%2F24%2F3%2F034002}.

\bibitem[{\citenamefont{Phelps and {Z. Lj. Petrovic}}(1999)}]{phelps_1999_cold}
\bibinfo{author}{\bibfnamefont{A.~V.} \bibnamefont{Phelps}} \bibnamefont{and}
  \bibinfo{author}{\bibnamefont{{Z. Lj. Petrovic}}}, \bibinfo{journal}{Plasma
  Sources Sci. Technol.} \textbf{\bibinfo{volume}{8}}, \bibinfo{pages}{R21}
  (\bibinfo{year}{1999}).

\bibitem[{\citenamefont{Brault}(2018)}]{brault_frontiers_18}
\bibinfo{author}{\bibfnamefont{P.}~\bibnamefont{Brault}},
  \bibinfo{journal}{Frontiers in Physics} \textbf{\bibinfo{volume}{6}},
  \bibinfo{pages}{59} (\bibinfo{year}{2018}), ISSN \bibinfo{issn}{2296-424X},
  \urlprefix\url{https://www.frontiersin.org/article/10.3389/fphy.2018.00059}.

\bibitem[{\citenamefont{Zhao et~al.}(2015{\natexlab{a}})\citenamefont{Zhao,
  Kang, Xue, Zhang, and Zhang}}]{zhao_2015_comparison}
\bibinfo{author}{\bibfnamefont{S.}~\bibnamefont{Zhao}},
  \bibinfo{author}{\bibfnamefont{W.}~\bibnamefont{Kang}},
  \bibinfo{author}{\bibfnamefont{J.}~\bibnamefont{Xue}},
  \bibinfo{author}{\bibfnamefont{X.}~\bibnamefont{Zhang}}, \bibnamefont{and}
  \bibinfo{author}{\bibfnamefont{P.}~\bibnamefont{Zhang}}, \bibinfo{journal}{J.
  Phys. Condens. Matter} \textbf{\bibinfo{volume}{27}}, \bibinfo{pages}{025401}
  (\bibinfo{year}{2015}{\natexlab{a}}).

\bibitem[{\citenamefont{Balzer et~al.}(2016)\citenamefont{Balzer, Schl\"unzen,
  and Bonitz}}]{balzer_prb16}
\bibinfo{author}{\bibfnamefont{K.}~\bibnamefont{Balzer}},
  \bibinfo{author}{\bibfnamefont{N.}~\bibnamefont{Schl\"unzen}},
  \bibnamefont{and} \bibinfo{author}{\bibfnamefont{M.}~\bibnamefont{Bonitz}},
  \bibinfo{journal}{Phys. Rev. B} \textbf{\bibinfo{volume}{94}},
  \bibinfo{pages}{245118} (\bibinfo{year}{2016}),
  \urlprefix\url{https://link.aps.org/doi/10.1103/PhysRevB.94.245118}.

\bibitem[{int()}]{interface}
\bibinfo{note}{This concept is based on a research project devoted to
  plasma-surface physics that is being developed at Kiel University and was
  first presented by M. Bonitz at the Conference \textit{Quo vadis--complex
  plasmas}, Hamburg, August 2015, and at the \textit{GEC} in Bochum, October
  2015.}

\bibitem[{\citenamefont{Graves and Brault}(2009)}]{graves_2009_molecular}
\bibinfo{author}{\bibfnamefont{D.~B.} \bibnamefont{Graves}} \bibnamefont{and}
  \bibinfo{author}{\bibfnamefont{P.}~\bibnamefont{Brault}},
  \bibinfo{journal}{J. Phys. D: Appl. Phys.} \textbf{\bibinfo{volume}{42}},
  \bibinfo{pages}{194011} (\bibinfo{year}{2009}).

\bibitem[{\citenamefont{Neyts and Brault}(2017)}]{neyts2017molecular}
\bibinfo{author}{\bibfnamefont{E.~C.} \bibnamefont{Neyts}} \bibnamefont{and}
  \bibinfo{author}{\bibfnamefont{P.}~\bibnamefont{Brault}},
  \bibinfo{journal}{Plasma Processes and Polymers}
  \textbf{\bibinfo{volume}{14}}, \bibinfo{pages}{1600145}
  (\bibinfo{year}{2017}),
  \eprint{https://onlinelibrary.wiley.com/doi/pdf/10.1002/ppap.201600145},
  \urlprefix\url{https://onlinelibrary.wiley.com/doi/abs/10.1002/ppap.201600145}.

\bibitem[{\citenamefont{Sheehan
  et~al.}(2013{\natexlab{a}})\citenamefont{Sheehan, Hershkowitz, Kaganovich,
  Wang, Raitses, Barbat, Weatherfor, and Sydorenko}}]{sheehan_2013_kinetic}
\bibinfo{author}{\bibfnamefont{J.~P.} \bibnamefont{Sheehan}},
  \bibinfo{author}{\bibfnamefont{N.}~\bibnamefont{Hershkowitz}},
  \bibinfo{author}{\bibfnamefont{I.~D.} \bibnamefont{Kaganovich}},
  \bibinfo{author}{\bibfnamefont{H.}~\bibnamefont{Wang}},
  \bibinfo{author}{\bibfnamefont{Y.}~\bibnamefont{Raitses}},
  \bibinfo{author}{\bibfnamefont{E.~V.} \bibnamefont{Barbat}},
  \bibinfo{author}{\bibfnamefont{B.~R.} \bibnamefont{Weatherfor}},
  \bibnamefont{and}
  \bibinfo{author}{\bibfnamefont{D.}~\bibnamefont{Sydorenko}},
  \bibinfo{journal}{Phys. Rev. Lett.} \textbf{\bibinfo{volume}{111}},
  \bibinfo{pages}{075002} (\bibinfo{year}{2013}{\natexlab{a}}).

\bibitem[{\citenamefont{Bronold and Fehske}(2015)}]{bronold_2015_absorption}
\bibinfo{author}{\bibfnamefont{F.~X.} \bibnamefont{Bronold}} \bibnamefont{and}
  \bibinfo{author}{\bibfnamefont{H.}~\bibnamefont{Fehske}},
  \bibinfo{journal}{Phys. Rev. Lett.} \textbf{\bibinfo{volume}{115}},
  \bibinfo{pages}{225001} (\bibinfo{year}{2015}).

\bibitem[{\citenamefont{Sun et~al.}(2018)\citenamefont{Sun, Becker, and
  Loffhagen}}]{sun_psst_2018}
\bibinfo{author}{\bibfnamefont{A.}~\bibnamefont{Sun}},
  \bibinfo{author}{\bibfnamefont{M.~M.} \bibnamefont{Becker}},
  \bibnamefont{and}
  \bibinfo{author}{\bibfnamefont{D.}~\bibnamefont{Loffhagen}},
  \bibinfo{journal}{Plasma Sources Science and Technology}
  \textbf{\bibinfo{volume}{27}}, \bibinfo{pages}{054002}
  (\bibinfo{year}{2018}),
  \urlprefix\url{http://stacks.iop.org/0963-0252/27/i=5/a=054002}.

\bibitem[{\citenamefont{Li and Go}(2013)}]{li_2013}
\bibinfo{author}{\bibfnamefont{Y.}~\bibnamefont{Li}} \bibnamefont{and}
  \bibinfo{author}{\bibfnamefont{D.~B.} \bibnamefont{Go}},
  \bibinfo{journal}{Applied Physics Letters} \textbf{\bibinfo{volume}{103}},
  \bibinfo{pages}{234104} (\bibinfo{year}{2013}),
  \eprint{https://doi.org/10.1063/1.4841495},
  \urlprefix\url{https://doi.org/10.1063/1.4841495}.

\bibitem[{\citenamefont{von Helmholtz}(1853)}]{helmholtz}
\bibinfo{author}{\bibfnamefont{H.}~\bibnamefont{von Helmholtz}},
  \bibinfo{journal}{Ann. Phys. (Berlin)} \textbf{\bibinfo{volume}{165}},
  \bibinfo{pages}{211} (\bibinfo{year}{1853}).

\bibitem[{\citenamefont{Heinisch et~al.}(2012)\citenamefont{Heinisch, Bronold,
  and Fehske}}]{heinisch_2012_electron}
\bibinfo{author}{\bibfnamefont{R.~L.} \bibnamefont{Heinisch}},
  \bibinfo{author}{\bibfnamefont{F.~X.} \bibnamefont{Bronold}},
  \bibnamefont{and} \bibinfo{author}{\bibfnamefont{H.}~\bibnamefont{Fehske}},
  \bibinfo{journal}{Phys. Rev. B} \textbf{\bibinfo{volume}{85}},
  \bibinfo{pages}{075323} (\bibinfo{year}{2012}).

\bibitem[{\citenamefont{Onida et~al.}(2002)\citenamefont{Onida, Reining, and
  Rubio}}]{onida_rmp_02}
\bibinfo{author}{\bibfnamefont{G.}~\bibnamefont{Onida}},
  \bibinfo{author}{\bibfnamefont{L.}~\bibnamefont{Reining}}, \bibnamefont{and}
  \bibinfo{author}{\bibfnamefont{A.}~\bibnamefont{Rubio}},
  \bibinfo{journal}{Rev. Mod. Phys.} \textbf{\bibinfo{volume}{74}},
  \bibinfo{pages}{601} (\bibinfo{year}{2002}),
  \urlprefix\url{https://link.aps.org/doi/10.1103/RevModPhys.74.601}.

\bibitem[{\citenamefont{Kotliar et~al.}(2006)\citenamefont{Kotliar, Savrasov,
  Haule, Oudovenko, Parcollet, and Marianetti}}]{kotliar_rmp_06}
\bibinfo{author}{\bibfnamefont{G.}~\bibnamefont{Kotliar}},
  \bibinfo{author}{\bibfnamefont{S.~Y.} \bibnamefont{Savrasov}},
  \bibinfo{author}{\bibfnamefont{K.}~\bibnamefont{Haule}},
  \bibinfo{author}{\bibfnamefont{V.~S.} \bibnamefont{Oudovenko}},
  \bibinfo{author}{\bibfnamefont{O.}~\bibnamefont{Parcollet}},
  \bibnamefont{and} \bibinfo{author}{\bibfnamefont{C.~A.}
  \bibnamefont{Marianetti}}, \bibinfo{journal}{Rev. Mod. Phys.}
  \textbf{\bibinfo{volume}{78}}, \bibinfo{pages}{865} (\bibinfo{year}{2006}),
  \urlprefix\url{https://link.aps.org/doi/10.1103/RevModPhys.78.865}.

\bibitem[{\citenamefont{Foulkes et~al.}(2001)\citenamefont{Foulkes, Mitas,
  Needs, and Rajagopal}}]{foulkes_rmp_01}
\bibinfo{author}{\bibfnamefont{W.~M.~C.} \bibnamefont{Foulkes}},
  \bibinfo{author}{\bibfnamefont{L.}~\bibnamefont{Mitas}},
  \bibinfo{author}{\bibfnamefont{R.~J.} \bibnamefont{Needs}}, \bibnamefont{and}
  \bibinfo{author}{\bibfnamefont{G.}~\bibnamefont{Rajagopal}},
  \bibinfo{journal}{Rev. Mod. Phys.} \textbf{\bibinfo{volume}{73}},
  \bibinfo{pages}{33} (\bibinfo{year}{2001}),
  \urlprefix\url{https://link.aps.org/doi/10.1103/RevModPhys.73.33}.

\bibitem[{\citenamefont{Dornheim et~al.}(2018)\citenamefont{Dornheim, Groth,
  and Bonitz}}]{DORNHEIM_physrep18}
\bibinfo{author}{\bibfnamefont{T.}~\bibnamefont{Dornheim}},
  \bibinfo{author}{\bibfnamefont{S.}~\bibnamefont{Groth}}, \bibnamefont{and}
  \bibinfo{author}{\bibfnamefont{M.}~\bibnamefont{Bonitz}},
  \bibinfo{journal}{Physics Reports} \textbf{\bibinfo{volume}{744}},
  \bibinfo{pages}{1 } (\bibinfo{year}{2018}), ISSN \bibinfo{issn}{0370-1573},
  \urlprefix\url{http://www.sciencedirect.com/science/article/pii/S0370157318300516}.

\bibitem[{\citenamefont{Abraham}(2018)}]{abraham_phd_18}
\bibinfo{author}{\bibfnamefont{J.-W.} \bibnamefont{Abraham}}, Ph.D. thesis,
  \bibinfo{school}{Kiel University}, \bibinfo{address}{Kiel, FRG}
  (\bibinfo{year}{2018}), \bibinfo{note}{unpublished}.

\bibitem[{\citenamefont{Daniil et~al.}(2016)\citenamefont{Daniil, Carlos, and
  Vasco}}]{marinov_2016}
\bibinfo{author}{\bibfnamefont{M.}~\bibnamefont{Daniil}},
  \bibinfo{author}{\bibfnamefont{T.}~\bibnamefont{Carlos}}, \bibnamefont{and}
  \bibinfo{author}{\bibfnamefont{G.}~\bibnamefont{Vasco}},
  \bibinfo{journal}{Plasma Processes and Polymers}
  \textbf{\bibinfo{volume}{14}}, \bibinfo{pages}{1600175}
  (\bibinfo{year}{2016}),
  \eprint{https://onlinelibrary.wiley.com/doi/pdf/10.1002/ppap.201600175},
  \urlprefix\url{https://onlinelibrary.wiley.com/doi/abs/10.1002/ppap.201600175}.

\bibitem[{\citenamefont{Guerra and Loureiro}(2004)}]{guerra}
\bibinfo{author}{\bibfnamefont{V.}~\bibnamefont{Guerra}} \bibnamefont{and}
  \bibinfo{author}{\bibfnamefont{J.}~\bibnamefont{Loureiro}},
  \bibinfo{journal}{Plasma Sources Science and Technology}
  \textbf{\bibinfo{volume}{13}}, \bibinfo{pages}{85} (\bibinfo{year}{2004}),
  \urlprefix\url{http://stacks.iop.org/0963-0252/13/i=1/a=011}.

\bibitem[{\citenamefont{Abraham et~al.}(2015)\citenamefont{Abraham, Kongsuwan,
  Strunskus, Faupel, and Bonitz}}]{abraham_jap_15}
\bibinfo{author}{\bibfnamefont{J.~W.} \bibnamefont{Abraham}},
  \bibinfo{author}{\bibfnamefont{N.}~\bibnamefont{Kongsuwan}},
  \bibinfo{author}{\bibfnamefont{T.}~\bibnamefont{Strunskus}},
  \bibinfo{author}{\bibfnamefont{F.}~\bibnamefont{Faupel}}, \bibnamefont{and}
  \bibinfo{author}{\bibfnamefont{M.}~\bibnamefont{Bonitz}},
  \bibinfo{journal}{Journal of Applied Physics} \textbf{\bibinfo{volume}{117}},
  \bibinfo{pages}{014305} (\bibinfo{year}{2015}),
  \urlprefix\url{http://scitation.aip.org/content/aip/journal/jap/117/1/10.1063/1.4905255}.

\bibitem[{\citenamefont{Fujioka}(2015)}]{fujioka_phd_15}
\bibinfo{author}{\bibfnamefont{K.}~\bibnamefont{Fujioka}}, Ph.D. thesis,
  \bibinfo{school}{Kiel University}, \bibinfo{address}{Kiel, FRG}
  (\bibinfo{year}{2015}), \bibinfo{note}{unpublished}.

\bibitem[{\citenamefont{Polonskyi et~al.}(2018)\citenamefont{Polonskyi, Ahadi,
  Peter, Fujioka, Abraham, Vasiliauskaite, Hinz, Strunskus, Wolf, Bonitz
  et~al.}}]{polonsky_epjd18}
\bibinfo{author}{\bibfnamefont{O.}~\bibnamefont{Polonskyi}},
  \bibinfo{author}{\bibfnamefont{A.~M.} \bibnamefont{Ahadi}},
  \bibinfo{author}{\bibfnamefont{T.}~\bibnamefont{Peter}},
  \bibinfo{author}{\bibfnamefont{K.}~\bibnamefont{Fujioka}},
  \bibinfo{author}{\bibfnamefont{J.~W.} \bibnamefont{Abraham}},
  \bibinfo{author}{\bibfnamefont{E.}~\bibnamefont{Vasiliauskaite}},
  \bibinfo{author}{\bibfnamefont{A.}~\bibnamefont{Hinz}},
  \bibinfo{author}{\bibfnamefont{T.}~\bibnamefont{Strunskus}},
  \bibinfo{author}{\bibfnamefont{S.}~\bibnamefont{Wolf}},
  \bibinfo{author}{\bibfnamefont{M.}~\bibnamefont{Bonitz}},
  \bibnamefont{et~al.}, \bibinfo{journal}{The European Physical Journal D}
  \textbf{\bibinfo{volume}{72}}, \bibinfo{pages}{93} (\bibinfo{year}{2018}),
  ISSN \bibinfo{issn}{1434-6079},
  \urlprefix\url{https://doi.org/10.1140/epjd/e2017-80419-8}.

\bibitem[{\citenamefont{Rosenthal}(2013)}]{rosenthal_phd_13}
\bibinfo{author}{\bibfnamefont{L.}~\bibnamefont{Rosenthal}}, Ph.D. thesis,
  \bibinfo{school}{Kiel University}, \bibinfo{address}{Kiel, FRG}
  (\bibinfo{year}{2013}), \bibinfo{note}{unpublished}.

\bibitem[{\citenamefont{Runge and Gross}(1984{\natexlab{a}})}]{runge_gross}
\bibinfo{author}{\bibfnamefont{E.}~\bibnamefont{Runge}} \bibnamefont{and}
  \bibinfo{author}{\bibfnamefont{E.~K.~U.} \bibnamefont{Gross}},
  \bibinfo{journal}{Phys. Rev. Lett.} \textbf{\bibinfo{volume}{52}},
  \bibinfo{pages}{997} (\bibinfo{year}{1984}{\natexlab{a}}),
  \urlprefix\url{https://link.aps.org/doi/10.1103/PhysRevLett.52.997}.

\bibitem[{\citenamefont{Balzer and Bonitz}(2013)}]{balzer-book}
\bibinfo{author}{\bibfnamefont{K.}~\bibnamefont{Balzer}} \bibnamefont{and}
  \bibinfo{author}{\bibfnamefont{M.}~\bibnamefont{Bonitz}},
  \emph{\bibinfo{title}{Nonequilibrium {G}reen's {F}unctions Approach to
  Inhomogeneous Systems}} (\bibinfo{publisher}{Springer},
  \bibinfo{address}{Berlin Heidelberg}, \bibinfo{year}{2013}).

\bibitem[{\citenamefont{Schlünzen and Bonitz}(2016)}]{schluenzen_cpp16}
\bibinfo{author}{\bibfnamefont{N.}~\bibnamefont{Schlünzen}} \bibnamefont{and}
  \bibinfo{author}{\bibfnamefont{M.}~\bibnamefont{Bonitz}},
  \bibinfo{journal}{Contributions to Plasma Physics}
  \textbf{\bibinfo{volume}{56}}, \bibinfo{pages}{5} (\bibinfo{year}{2016}),
  \eprint{https://onlinelibrary.wiley.com/doi/pdf/10.1002/ctpp.201610003},
  \urlprefix\url{https://onlinelibrary.wiley.com/doi/abs/10.1002/ctpp.201610003}.

\bibitem[{\citenamefont{Marini et~al.}(2009)\citenamefont{Marini, Hogan,
  Grüning, and Varsano}}]{marini_2009_yambo}
\bibinfo{author}{\bibfnamefont{A.}~\bibnamefont{Marini}},
  \bibinfo{author}{\bibfnamefont{C.}~\bibnamefont{Hogan}},
  \bibinfo{author}{\bibfnamefont{M.}~\bibnamefont{Grüning}}, \bibnamefont{and}
  \bibinfo{author}{\bibfnamefont{D.}~\bibnamefont{Varsano}},
  \bibinfo{journal}{Comp. Phys. Commun.} \textbf{\bibinfo{volume}{180}},
  \bibinfo{pages}{1392} (\bibinfo{year}{2009}).

\bibitem[{\citenamefont{Jürg}(2012)}]{hutter_wires_12}
\bibinfo{author}{\bibfnamefont{H.}~\bibnamefont{Jürg}},
  \bibinfo{journal}{Wiley Interdisciplinary Reviews: Computational Molecular
  Science} \textbf{\bibinfo{volume}{2}}, \bibinfo{pages}{604}
  (\bibinfo{year}{2012}),
  \eprint{https://onlinelibrary.wiley.com/doi/pdf/10.1002/wcms.90},
  \urlprefix\url{https://onlinelibrary.wiley.com/doi/abs/10.1002/wcms.90}.

\bibitem[{\citenamefont{Gross}(2009)}]{gross-book}
\bibinfo{author}{\bibfnamefont{A.}~\bibnamefont{Gross}},
  \emph{\bibinfo{title}{Theoretical Surface Science}}
  (\bibinfo{publisher}{Cambridge University Press},
  \bibinfo{address}{Springer}, \bibinfo{year}{2009}), \bibinfo{edition}{2nd}
  ed.

\bibitem[{\citenamefont{Bonitz et~al.}(2014)\citenamefont{Bonitz, Lopez,
  Becker, and Thomsen}}]{com-plasma_springer_14}
\bibinfo{editor}{\bibfnamefont{M.}~\bibnamefont{Bonitz}},
  \bibinfo{editor}{\bibfnamefont{J.}~\bibnamefont{Lopez}},
  \bibinfo{editor}{\bibfnamefont{K.}~\bibnamefont{Becker}}, \bibnamefont{and}
  \bibinfo{editor}{\bibfnamefont{H.}~\bibnamefont{Thomsen}}, eds.,
  \emph{\bibinfo{title}{Complex Plasmas: Scientific Challnges and Technological
  Opportunities}} (\bibinfo{publisher}{Springer}, \bibinfo{year}{2014}).

\bibitem[{\citenamefont{Ott and Bonitz}(2011)}]{ott_prl_11}
\bibinfo{author}{\bibfnamefont{T.}~\bibnamefont{Ott}} \bibnamefont{and}
  \bibinfo{author}{\bibfnamefont{M.}~\bibnamefont{Bonitz}},
  \bibinfo{journal}{Phys. Rev. Lett.} \textbf{\bibinfo{volume}{107}},
  \bibinfo{pages}{135003} (\bibinfo{year}{2011}),
  \urlprefix\url{http://link.aps.org/doi/10.1103/PhysRevLett.107.135003}.

\bibitem[{\citenamefont{Abraham et~al.}(2016)\citenamefont{Abraham, Strunskus,
  Faupel, and Bonitz}}]{abraham_jap16}
\bibinfo{author}{\bibfnamefont{J.~W.} \bibnamefont{Abraham}},
  \bibinfo{author}{\bibfnamefont{T.}~\bibnamefont{Strunskus}},
  \bibinfo{author}{\bibfnamefont{F.}~\bibnamefont{Faupel}}, \bibnamefont{and}
  \bibinfo{author}{\bibfnamefont{M.}~\bibnamefont{Bonitz}},
  \bibinfo{journal}{Journal of Applied Physics} \textbf{\bibinfo{volume}{119}},
  \bibinfo{pages}{185301} (\bibinfo{year}{2016}),
  \eprint{https://doi.org/10.1063/1.4948375},
  \urlprefix\url{https://doi.org/10.1063/1.4948375}.

\bibitem[{\citenamefont{Nakano et~al.}(2008)\citenamefont{Nakano, Kalia, ichi
  Nomura, Sharma, Vashishta, Shimojo, van Duin, Goddard, Biswas, Srivastava
  et~al.}}]{nakano_08}
\bibinfo{author}{\bibfnamefont{A.}~\bibnamefont{Nakano}},
  \bibinfo{author}{\bibfnamefont{R.~K.} \bibnamefont{Kalia}},
  \bibinfo{author}{\bibfnamefont{K.}~\bibnamefont{ichi Nomura}},
  \bibinfo{author}{\bibfnamefont{A.}~\bibnamefont{Sharma}},
  \bibinfo{author}{\bibfnamefont{P.}~\bibnamefont{Vashishta}},
  \bibinfo{author}{\bibfnamefont{F.}~\bibnamefont{Shimojo}},
  \bibinfo{author}{\bibfnamefont{A.~C.~T.} \bibnamefont{van Duin}},
  \bibinfo{author}{\bibfnamefont{W.~A.} \bibnamefont{Goddard}},
  \bibinfo{author}{\bibfnamefont{R.}~\bibnamefont{Biswas}},
  \bibinfo{author}{\bibfnamefont{D.}~\bibnamefont{Srivastava}},
  \bibnamefont{et~al.}, \bibinfo{journal}{The International Journal of High
  Performance Computing Applications} \textbf{\bibinfo{volume}{22}},
  \bibinfo{pages}{113} (\bibinfo{year}{2008}),
  \eprint{https://doi.org/10.1177/1094342007085015},
  \urlprefix\url{https://doi.org/10.1177/1094342007085015}.

\bibitem[{\citenamefont{Piana et~al.}(2013)\citenamefont{Piana,
  Lindorff-Larsen, and Shaw}}]{piana_13}
\bibinfo{author}{\bibfnamefont{S.}~\bibnamefont{Piana}},
  \bibinfo{author}{\bibfnamefont{K.}~\bibnamefont{Lindorff-Larsen}},
  \bibnamefont{and} \bibinfo{author}{\bibfnamefont{D.~E.} \bibnamefont{Shaw}},
  \bibinfo{journal}{Proceedings of the National Academy of Sciences}
  \textbf{\bibinfo{volume}{110}}, \bibinfo{pages}{5915} (\bibinfo{year}{2013}),
  ISSN \bibinfo{issn}{0027-8424},
  \eprint{http://www.pnas.org/content/110/15/5915.full.pdf},
  \urlprefix\url{http://www.pnas.org/content/110/15/5915}.

\bibitem[{\citenamefont{Voter}(1997)}]{voter_97}
\bibinfo{author}{\bibfnamefont{A.~F.} \bibnamefont{Voter}},
  \bibinfo{journal}{Phys. Rev. Lett.} \textbf{\bibinfo{volume}{78}},
  \bibinfo{pages}{3908} (\bibinfo{year}{1997}),
  \urlprefix\url{https://link.aps.org/doi/10.1103/PhysRevLett.78.3908}.

\bibitem[{\citenamefont{Laio and Parrinello}(2002)}]{LaioParrinello2002}
\bibinfo{author}{\bibfnamefont{A.}~\bibnamefont{Laio}} \bibnamefont{and}
  \bibinfo{author}{\bibfnamefont{M.}~\bibnamefont{Parrinello}},
  \bibinfo{journal}{PNAS} \textbf{\bibinfo{volume}{99}}, \bibinfo{pages}{12562}
  (\bibinfo{year}{2002}).

\bibitem[{\citenamefont{So/rensen and Voter}(2000)}]{Sorensen2000}
\bibinfo{author}{\bibfnamefont{M.~R.} \bibnamefont{So/rensen}}
  \bibnamefont{and} \bibinfo{author}{\bibfnamefont{A.~F.} \bibnamefont{Voter}},
  \bibinfo{journal}{The Journal of Chemical Physics}
  \textbf{\bibinfo{volume}{112}}, \bibinfo{pages}{9599} (\bibinfo{year}{2000}),
  \eprint{https://doi.org/10.1063/1.481576},
  \urlprefix\url{https://doi.org/10.1063/1.481576}.

\bibitem[{\citenamefont{Bal and Neyts}(2015)}]{Bal2015}
\bibinfo{author}{\bibfnamefont{K.~M.} \bibnamefont{Bal}} \bibnamefont{and}
  \bibinfo{author}{\bibfnamefont{E.~C.} \bibnamefont{Neyts}},
  \bibinfo{journal}{Journal of Chemical Theory and Computation}
  \textbf{\bibinfo{volume}{11}}, \bibinfo{pages}{4545} (\bibinfo{year}{2015}),
  \eprint{https://doi.org/10.1021/acs.jctc.5b00597},
  \urlprefix\url{https://doi.org/10.1021/acs.jctc.5b00597}.

\bibitem[{\citenamefont{Bonitz et~al.}(2018{\natexlab{a}})\citenamefont{Bonitz,
  Filinov, Abraham, and Loffhagen}}]{bonitz_psst18}
\bibinfo{author}{\bibfnamefont{M.}~\bibnamefont{Bonitz}},
  \bibinfo{author}{\bibfnamefont{A.~V.} \bibnamefont{Filinov}},
  \bibinfo{author}{\bibfnamefont{J.~W.} \bibnamefont{Abraham}},
  \bibnamefont{and}
  \bibinfo{author}{\bibfnamefont{D.}~\bibnamefont{Loffhagen}},
  \bibinfo{journal}{Plasma Sources Science and Technology}
  (\bibinfo{year}{2018}{\natexlab{a}}),
  \urlprefix\url{http://iopscience.iop.org/10.1088/1361-6595/aaca75}.

\bibitem[{\citenamefont{Abraham and Bonitz}(2018)}]{abraham_cpp18}
\bibinfo{author}{\bibfnamefont{J.~W.} \bibnamefont{Abraham}} \bibnamefont{and}
  \bibinfo{author}{\bibfnamefont{M.}~\bibnamefont{Bonitz}},
  \bibinfo{journal}{Contributions to Plasma Physics}
  \textbf{\bibinfo{volume}{58}}, \bibinfo{pages}{164} (\bibinfo{year}{2018}),
  \eprint{https://onlinelibrary.wiley.com/doi/pdf/10.1002/ctpp.201700151},
  \urlprefix\url{https://onlinelibrary.wiley.com/doi/abs/10.1002/ctpp.201700151}.

\bibitem[{\citenamefont{Franke and Pehlke}(2010)}]{franke_2010_diffusion}
\bibinfo{author}{\bibfnamefont{A.}~\bibnamefont{Franke}} \bibnamefont{and}
  \bibinfo{author}{\bibfnamefont{E.}~\bibnamefont{Pehlke}},
  \bibinfo{journal}{Phys. Rev. B} \textbf{\bibinfo{volume}{82}},
  \bibinfo{pages}{205423} (\bibinfo{year}{2010}).

\bibitem[{\citenamefont{Filinov
  et~al.}(2018{\natexlab{a}})\citenamefont{Filinov, Bonitz, and
  Loffhagen}}]{filinov_psst18_1}
\bibinfo{author}{\bibfnamefont{A.}~\bibnamefont{Filinov}},
  \bibinfo{author}{\bibfnamefont{M.}~\bibnamefont{Bonitz}}, \bibnamefont{and}
  \bibinfo{author}{\bibfnamefont{D.}~\bibnamefont{Loffhagen}},
  \bibinfo{journal}{Plasma Sources Science and Technology}
  \textbf{\bibinfo{volume}{27}}, \bibinfo{pages}{064003}
  (\bibinfo{year}{2018}{\natexlab{a}}),
  \urlprefix\url{http://stacks.iop.org/0963-0252/27/i=6/a=064003}.

\bibitem[{\citenamefont{Schwartzkopf et~al.}(2015)\citenamefont{Schwartzkopf,
  Santoro, Brett, Rothkirch, Polonskyi, Hinz, Metwalli, Yao, Strunskus, Faupel
  et~al.}}]{schwartzkopf_2015_real}
\bibinfo{author}{\bibfnamefont{M.}~\bibnamefont{Schwartzkopf}},
  \bibinfo{author}{\bibfnamefont{G.}~\bibnamefont{Santoro}},
  \bibinfo{author}{\bibfnamefont{C.~J.} \bibnamefont{Brett}},
  \bibinfo{author}{\bibfnamefont{A.}~\bibnamefont{Rothkirch}},
  \bibinfo{author}{\bibfnamefont{O.}~\bibnamefont{Polonskyi}},
  \bibinfo{author}{\bibfnamefont{A.}~\bibnamefont{Hinz}},
  \bibinfo{author}{\bibfnamefont{E.}~\bibnamefont{Metwalli}},
  \bibinfo{author}{\bibfnamefont{Y.}~\bibnamefont{Yao}},
  \bibinfo{author}{\bibfnamefont{T.}~\bibnamefont{Strunskus}},
  \bibinfo{author}{\bibfnamefont{F.}~\bibnamefont{Faupel}},
  \bibnamefont{et~al.}, \bibinfo{journal}{ACS Appl. Mater. Interfaces}
  \textbf{\bibinfo{volume}{7}}, \bibinfo{pages}{13547} (\bibinfo{year}{2015}).

\bibitem[{\citenamefont{Abraham et~al.}(2018)\citenamefont{Abraham, Hinz,
  Strunskus, Faupel, and Bonitz}}]{abraham_epjd18}
\bibinfo{author}{\bibfnamefont{J.~W.} \bibnamefont{Abraham}},
  \bibinfo{author}{\bibfnamefont{A.}~\bibnamefont{Hinz}},
  \bibinfo{author}{\bibfnamefont{T.}~\bibnamefont{Strunskus}},
  \bibinfo{author}{\bibfnamefont{F.}~\bibnamefont{Faupel}}, \bibnamefont{and}
  \bibinfo{author}{\bibfnamefont{M.}~\bibnamefont{Bonitz}},
  \bibinfo{journal}{The European Physical Journal D}
  \textbf{\bibinfo{volume}{72}}, \bibinfo{pages}{92} (\bibinfo{year}{2018}),
  ISSN \bibinfo{issn}{1434-6079},
  \urlprefix\url{https://doi.org/10.1140/epjd/e2017-80426-9}.

\bibitem[{\citenamefont{Bonitz}(2016)}]{bonitz_qkt}
\bibinfo{author}{\bibfnamefont{M.}~\bibnamefont{Bonitz}},
  \emph{\bibinfo{title}{Quantum Kinetic Theory}}, Teubner-Texte zur Physik
  (\bibinfo{publisher}{Springer}, \bibinfo{year}{2016}), \bibinfo{edition}{2nd}
  ed.

\bibitem[{\citenamefont{Filinov
  et~al.}(2018{\natexlab{b}})\citenamefont{Filinov, Bonitz, and
  Loffhagen}}]{filinov_psst18_2}
\bibinfo{author}{\bibfnamefont{A.}~\bibnamefont{Filinov}},
  \bibinfo{author}{\bibfnamefont{M.}~\bibnamefont{Bonitz}}, \bibnamefont{and}
  \bibinfo{author}{\bibfnamefont{D.}~\bibnamefont{Loffhagen}},
  \bibinfo{journal}{Plasma Sources Science and Technology}
  \textbf{\bibinfo{volume}{27}}, \bibinfo{pages}{064002}
  (\bibinfo{year}{2018}{\natexlab{b}}),
  \urlprefix\url{http://stacks.iop.org/0963-0252/27/i=6/a=064002}.

\bibitem[{\citenamefont{Lieberman and Lichtenberg}(2005)}]{LL05}
\bibinfo{author}{\bibfnamefont{M.~A.} \bibnamefont{Lieberman}}
  \bibnamefont{and} \bibinfo{author}{\bibfnamefont{A.~J.}
  \bibnamefont{Lichtenberg}}, \emph{\bibinfo{title}{{Principles of plasma
  discharges and materials processing}}}
  (\bibinfo{publisher}{Wiley-Interscience}, \bibinfo{address}{New York},
  \bibinfo{year}{2005}).

\bibitem[{\citenamefont{Rabalais}(1994)}]{Rabalais94}
\bibinfo{editor}{\bibfnamefont{J.~W.} \bibnamefont{Rabalais}}, ed.,
  \emph{\bibinfo{title}{{Low energy ion-surface interaction}}}
  (\bibinfo{publisher}{Wiley and Sons}, \bibinfo{address}{New York},
  \bibinfo{year}{1994}).

\bibitem[{\citenamefont{Winter}(2002{\natexlab{a}})}]{Winter02}
\bibinfo{author}{\bibfnamefont{H.}~\bibnamefont{Winter}},
  \bibinfo{journal}{Phys. Rep.} \textbf{\bibinfo{volume}{367}},
  \bibinfo{pages}{387} (\bibinfo{year}{2002}{\natexlab{a}}).

\bibitem[{\citenamefont{{HP. Winter and J. Burgd\"orfer}}(2007)}]{Winter07}
\bibinfo{editor}{\bibnamefont{{HP. Winter and J. Burgd\"orfer}}}, ed.,
  \emph{\bibinfo{title}{{Slow heavy-particle induced electron emission from
  solid surface}}} (\bibinfo{publisher}{Springer Verlag},
  \bibinfo{address}{Berlin}, \bibinfo{year}{2007}).

\bibitem[{\citenamefont{{A. V. Phelps and Z. Lj. Petrovi\'c}}(1999)}]{PP99}
\bibinfo{author}{\bibnamefont{{A. V. Phelps and Z. Lj. Petrovi\'c}}},
  \bibinfo{journal}{Plasma Sources Sci. Technol.} \textbf{\bibinfo{volume}{8}},
  \bibinfo{pages}{R21} (\bibinfo{year}{1999}).

\bibitem[{\citenamefont{Daksha et~al.}(2016)\citenamefont{Daksha, Berger,
  Schuengel, Korolov, Derzsi, Koepke, Donko, and Schulze}}]{DBS16}
\bibinfo{author}{\bibfnamefont{M.}~\bibnamefont{Daksha}},
  \bibinfo{author}{\bibfnamefont{B.}~\bibnamefont{Berger}},
  \bibinfo{author}{\bibfnamefont{E.}~\bibnamefont{Schuengel}},
  \bibinfo{author}{\bibfnamefont{I.}~\bibnamefont{Korolov}},
  \bibinfo{author}{\bibfnamefont{A.}~\bibnamefont{Derzsi}},
  \bibinfo{author}{\bibfnamefont{M.}~\bibnamefont{Koepke}},
  \bibinfo{author}{\bibfnamefont{Z.}~\bibnamefont{Donko}}, \bibnamefont{and}
  \bibinfo{author}{\bibfnamefont{J.}~\bibnamefont{Schulze}},
  \bibinfo{journal}{J. Phys. D: Appl. Phys.} \textbf{\bibinfo{volume}{49}},
  \bibinfo{pages}{234001} (\bibinfo{year}{2016}).

\bibitem[{\citenamefont{Marcak et~al.}(2015)\citenamefont{Marcak, Corbella,
  de~los Arcos, and von Keudell}}]{MCA15}
\bibinfo{author}{\bibfnamefont{A.}~\bibnamefont{Marcak}},
  \bibinfo{author}{\bibfnamefont{C.}~\bibnamefont{Corbella}},
  \bibinfo{author}{\bibfnamefont{T.}~\bibnamefont{de~los Arcos}},
  \bibnamefont{and} \bibinfo{author}{\bibfnamefont{A.}~\bibnamefont{von
  Keudell}}, \bibinfo{journal}{Rev. Sci. Instrum.}
  \textbf{\bibinfo{volume}{86}}, \bibinfo{pages}{106102}
  (\bibinfo{year}{2015}).

\bibitem[{\citenamefont{Pamperin et~al.}(2018)\citenamefont{Pamperin, Bronold,
  and Fehske}}]{PBF18}
\bibinfo{author}{\bibfnamefont{M.}~\bibnamefont{Pamperin}},
  \bibinfo{author}{\bibfnamefont{F.~X.} \bibnamefont{Bronold}},
  \bibnamefont{and} \bibinfo{author}{\bibfnamefont{H.}~\bibnamefont{Fehske}},
  \bibinfo{journal}{Plasma Sources Sci. Technol.}  (\bibinfo{year}{2018}).

\bibitem[{\citenamefont{More et~al.}(1998)\citenamefont{More, Merino, Monreal,
  Pou, and Flores}}]{MMM98}
\bibinfo{author}{\bibfnamefont{W.}~\bibnamefont{More}},
  \bibinfo{author}{\bibfnamefont{J.}~\bibnamefont{Merino}},
  \bibinfo{author}{\bibfnamefont{R.}~\bibnamefont{Monreal}},
  \bibinfo{author}{\bibfnamefont{P.}~\bibnamefont{Pou}}, \bibnamefont{and}
  \bibinfo{author}{\bibfnamefont{F.}~\bibnamefont{Flores}},
  \bibinfo{journal}{Phys. Rev. B} \textbf{\bibinfo{volume}{58}},
  \bibinfo{pages}{7385} (\bibinfo{year}{1998}).

\bibitem[{\citenamefont{Newns et~al.}(1983)\citenamefont{Newns, Makoshi, Brako,
  and van Wunnik}}]{NMB83}
\bibinfo{author}{\bibfnamefont{D.~M.} \bibnamefont{Newns}},
  \bibinfo{author}{\bibfnamefont{K.}~\bibnamefont{Makoshi}},
  \bibinfo{author}{\bibfnamefont{R.}~\bibnamefont{Brako}}, \bibnamefont{and}
  \bibinfo{author}{\bibfnamefont{J.~N.~M.} \bibnamefont{van Wunnik}},
  \bibinfo{journal}{Phys. Scr.} \textbf{\bibinfo{volume}{T6}},
  \bibinfo{pages}{5} (\bibinfo{year}{1983}).

\bibitem[{\citenamefont{Yoshimori and Makoshi}(1986)}]{YM86}
\bibinfo{author}{\bibfnamefont{A.}~\bibnamefont{Yoshimori}} \bibnamefont{and}
  \bibinfo{author}{\bibfnamefont{K.}~\bibnamefont{Makoshi}},
  \bibinfo{journal}{Prog. Surf. Sci.} \textbf{\bibinfo{volume}{21}},
  \bibinfo{pages}{251} (\bibinfo{year}{1986}).

\bibitem[{\citenamefont{Los and Geerlings}(1990)}]{LG90}
\bibinfo{author}{\bibfnamefont{J.}~\bibnamefont{Los}} \bibnamefont{and}
  \bibinfo{author}{\bibfnamefont{J.~J.~C.} \bibnamefont{Geerlings}},
  \bibinfo{journal}{Phys. Rep.} \textbf{\bibinfo{volume}{190}},
  \bibinfo{pages}{133} (\bibinfo{year}{1990}).

\bibitem[{\citenamefont{Wang et~al.}(2001)\citenamefont{Wang, Garc\'{\i}a,
  Monreal, Flores, Goldberg, Brongersma, and Bauer}}]{WGM01}
\bibinfo{author}{\bibfnamefont{N.~P.} \bibnamefont{Wang}},
  \bibinfo{author}{\bibfnamefont{E.~A.} \bibnamefont{Garc\'{\i}a}},
  \bibinfo{author}{\bibfnamefont{R.}~\bibnamefont{Monreal}},
  \bibinfo{author}{\bibfnamefont{F.}~\bibnamefont{Flores}},
  \bibinfo{author}{\bibfnamefont{E.~C.} \bibnamefont{Goldberg}},
  \bibinfo{author}{\bibfnamefont{H.~H.} \bibnamefont{Brongersma}},
  \bibnamefont{and} \bibinfo{author}{\bibfnamefont{P.}~\bibnamefont{Bauer}},
  \bibinfo{journal}{Phys. Rev. A} \textbf{\bibinfo{volume}{64}},
  \bibinfo{pages}{012901} (\bibinfo{year}{2001}).

\bibitem[{\citenamefont{Vald\'{e}s et~al.}(2005)\citenamefont{Vald\'{e}s,
  Goldberg, Blanco, and Monreal}}]{VGB05}
\bibinfo{author}{\bibfnamefont{D.}~\bibnamefont{Vald\'{e}s}},
  \bibinfo{author}{\bibfnamefont{E.~C.} \bibnamefont{Goldberg}},
  \bibinfo{author}{\bibfnamefont{J.~M.} \bibnamefont{Blanco}},
  \bibnamefont{and} \bibinfo{author}{\bibfnamefont{R.~C.}
  \bibnamefont{Monreal}}, \bibinfo{journal}{Phys. Rev. B}
  \textbf{\bibinfo{volume}{71}}, \bibinfo{pages}{245417}
  (\bibinfo{year}{2005}).

\bibitem[{\citenamefont{Marbach
  et~al.}(2012{\natexlab{a}})\citenamefont{Marbach, Bronold, and
  Fehske}}]{MBF12a}
\bibinfo{author}{\bibfnamefont{J.}~\bibnamefont{Marbach}},
  \bibinfo{author}{\bibfnamefont{F.~X.} \bibnamefont{Bronold}},
  \bibnamefont{and} \bibinfo{author}{\bibfnamefont{H.}~\bibnamefont{Fehske}},
  \bibinfo{journal}{Eur. Phys. J. D} \textbf{\bibinfo{volume}{66}},
  \bibinfo{pages}{106} (\bibinfo{year}{2012}{\natexlab{a}}).

\bibitem[{\citenamefont{Marbach
  et~al.}(2012{\natexlab{b}})\citenamefont{Marbach, Bronold, and
  Fehske}}]{MBF12b}
\bibinfo{author}{\bibfnamefont{J.}~\bibnamefont{Marbach}},
  \bibinfo{author}{\bibfnamefont{F.~X.} \bibnamefont{Bronold}},
  \bibnamefont{and} \bibinfo{author}{\bibfnamefont{H.}~\bibnamefont{Fehske}},
  \bibinfo{journal}{Phys. Rev. B} \textbf{\bibinfo{volume}{86}},
  \bibinfo{pages}{115417} (\bibinfo{year}{2012}{\natexlab{b}}).

\bibitem[{\citenamefont{Iglesias-Garc\'{\i}a
  et~al.}(2013)\citenamefont{Iglesias-Garc\'{\i}a, Garc\'{\i}a, and
  Goldberg}}]{IGG13}
\bibinfo{author}{\bibfnamefont{A.}~\bibnamefont{Iglesias-Garc\'{\i}a}},
  \bibinfo{author}{\bibfnamefont{E.~A.} \bibnamefont{Garc\'{\i}a}},
  \bibnamefont{and} \bibinfo{author}{\bibfnamefont{E.~C.}
  \bibnamefont{Goldberg}}, \bibinfo{journal}{Phys. Rev. B}
  \textbf{\bibinfo{volume}{87}}, \bibinfo{pages}{075434}
  (\bibinfo{year}{2013}).

\bibitem[{\citenamefont{Iglesias-Garc\'ia
  et~al.}(2014)\citenamefont{Iglesias-Garc\'ia, Garc\'ia, and
  Goldberg}}]{IGG14}
\bibinfo{author}{\bibfnamefont{A.}~\bibnamefont{Iglesias-Garc\'ia}},
  \bibinfo{author}{\bibfnamefont{E.~A.} \bibnamefont{Garc\'ia}},
  \bibnamefont{and} \bibinfo{author}{\bibfnamefont{E.~C.}
  \bibnamefont{Goldberg}}, \bibinfo{journal}{Phys. Rev. B}
  \textbf{\bibinfo{volume}{90}}, \bibinfo{pages}{195416}
  (\bibinfo{year}{2014}).

\bibitem[{\citenamefont{Pamperin
  et~al.}(2015{\natexlab{a}})\citenamefont{Pamperin, Bronold, and
  Fehske}}]{PBF15a}
\bibinfo{author}{\bibfnamefont{M.}~\bibnamefont{Pamperin}},
  \bibinfo{author}{\bibfnamefont{F.~X.} \bibnamefont{Bronold}},
  \bibnamefont{and} \bibinfo{author}{\bibfnamefont{H.}~\bibnamefont{Fehske}},
  \bibinfo{journal}{Phys. Rev. B} \textbf{\bibinfo{volume}{91}},
  \bibinfo{pages}{035440} (\bibinfo{year}{2015}{\natexlab{a}}).

\bibitem[{\citenamefont{Pamperin
  et~al.}(2015{\natexlab{b}})\citenamefont{Pamperin, Bronold, and
  Fehske}}]{PBF15b}
\bibinfo{author}{\bibfnamefont{M.}~\bibnamefont{Pamperin}},
  \bibinfo{author}{\bibfnamefont{F.~X.} \bibnamefont{Bronold}},
  \bibnamefont{and} \bibinfo{author}{\bibfnamefont{H.}~\bibnamefont{Fehske}},
  \bibinfo{journal}{Phys. Scr.} \textbf{\bibinfo{volume}{T165}},
  \bibinfo{pages}{014008} (\bibinfo{year}{2015}{\natexlab{b}}).

\bibitem[{\citenamefont{Gadzuk}(1967{\natexlab{a}})}]{Gadzuk67a}
\bibinfo{author}{\bibfnamefont{J.~W.} \bibnamefont{Gadzuk}},
  \bibinfo{journal}{Surf. Sci.} \textbf{\bibinfo{volume}{6}},
  \bibinfo{pages}{133} (\bibinfo{year}{1967}{\natexlab{a}}).

\bibitem[{\citenamefont{Gadzuk}(1967{\natexlab{b}})}]{Gadzuk67b}
\bibinfo{author}{\bibfnamefont{J.~W.} \bibnamefont{Gadzuk}},
  \bibinfo{journal}{Surf. Sci.} \textbf{\bibinfo{volume}{6}},
  \bibinfo{pages}{159} (\bibinfo{year}{1967}{\natexlab{b}}).

\bibitem[{\citenamefont{Propst}(1963)}]{Propst63}
\bibinfo{author}{\bibfnamefont{F.~M.} \bibnamefont{Propst}},
  \bibinfo{journal}{Phys. Rev.} \textbf{\bibinfo{volume}{129}},
  \bibinfo{pages}{7} (\bibinfo{year}{1963}).

\bibitem[{\citenamefont{Penn and Apell}(1990)}]{PA90}
\bibinfo{author}{\bibfnamefont{D.~R.} \bibnamefont{Penn}} \bibnamefont{and}
  \bibinfo{author}{\bibfnamefont{P.}~\bibnamefont{Apell}},
  \bibinfo{journal}{Phys. Rev. B} \textbf{\bibinfo{volume}{41}},
  \bibinfo{pages}{3303} (\bibinfo{year}{1990}).

\bibitem[{\citenamefont{Langreth and Nordlander}(1991)}]{LN91}
\bibinfo{author}{\bibfnamefont{D.~C.} \bibnamefont{Langreth}} \bibnamefont{and}
  \bibinfo{author}{\bibfnamefont{P.}~\bibnamefont{Nordlander}},
  \bibinfo{journal}{Phys. Rev. B} \textbf{\bibinfo{volume}{43}},
  \bibinfo{pages}{2541} (\bibinfo{year}{1991}).

\bibitem[{\citenamefont{Shao et~al.}(1994{\natexlab{a}})\citenamefont{Shao,
  Langreth, and Nordlander}}]{SLN94a}
\bibinfo{author}{\bibfnamefont{H.}~\bibnamefont{Shao}},
  \bibinfo{author}{\bibfnamefont{D.~C.} \bibnamefont{Langreth}},
  \bibnamefont{and}
  \bibinfo{author}{\bibfnamefont{P.}~\bibnamefont{Nordlander}},
  \bibinfo{journal}{Phys. Rev. B} \textbf{\bibinfo{volume}{49}},
  \bibinfo{pages}{13929} (\bibinfo{year}{1994}{\natexlab{a}}).

\bibitem[{\citenamefont{Shao et~al.}(1994{\natexlab{b}})\citenamefont{Shao,
  Langreth, and Nordlander}}]{SLN94b}
\bibinfo{author}{\bibfnamefont{H.}~\bibnamefont{Shao}},
  \bibinfo{author}{\bibfnamefont{D.~C.} \bibnamefont{Langreth}},
  \bibnamefont{and}
  \bibinfo{author}{\bibfnamefont{P.}~\bibnamefont{Nordlander}}, in
  \emph{\bibinfo{booktitle}{{Low energy ion-surface interaction}}}, edited by
  \bibinfo{editor}{\bibfnamefont{J.~W.} \bibnamefont{Rabalais}}
  (\bibinfo{publisher}{Wiley and Sons}, \bibinfo{address}{New York},
  \bibinfo{year}{1994}{\natexlab{b}}), p. \bibinfo{pages}{117}.

\bibitem[{\citenamefont{Marx and Hutter}(2009)}]{Marx2009}
\bibinfo{author}{\bibfnamefont{D.}~\bibnamefont{Marx}} \bibnamefont{and}
  \bibinfo{author}{\bibfnamefont{J.}~\bibnamefont{Hutter}},
  \emph{\bibinfo{title}{Ab Initio Molecular Dynamics: Basic Theory and Advanced
  Methods}} (\bibinfo{publisher}{Cambridge University Press},
  \bibinfo{year}{2009}).

\bibitem[{\citenamefont{Hafner}(2008)}]{hafner_2008_initio}
\bibinfo{author}{\bibfnamefont{J.}~\bibnamefont{Hafner}}, \bibinfo{journal}{J.
  Comput. Chem.} \textbf{\bibinfo{volume}{29}}, \bibinfo{pages}{2044}
  (\bibinfo{year}{2008}).

\bibitem[{\citenamefont{Groß}(2016)}]{Gross_O2Pt_2016}
\bibinfo{author}{\bibfnamefont{A.}~\bibnamefont{Groß}},
  \bibinfo{journal}{Catalysis Today} \textbf{\bibinfo{volume}{260}},
  \bibinfo{pages}{60 } (\bibinfo{year}{2016}).

\bibitem[{\citenamefont{K\"uhne}(2014)}]{Kuehne2014}
\bibinfo{author}{\bibfnamefont{T.~D.} \bibnamefont{K\"uhne}},
  \bibinfo{journal}{Wiley Interdisciplinary Reviews: Computational Molecular
  Science} \textbf{\bibinfo{volume}{4}}, \bibinfo{pages}{391}
  (\bibinfo{year}{2014}).

\bibitem[{\citenamefont{Baer}(2006)}]{Baer2006}
\bibinfo{author}{\bibfnamefont{M.}~\bibnamefont{Baer}},
  \emph{\bibinfo{title}{Beyond Born-Oppenheimer: Electronic Nonadiabatic
  Coupling Terms and Conical Intersections}}
  (\bibinfo{publisher}{Wiley-Interscience}, \bibinfo{year}{2006}).

\bibitem[{\citenamefont{Nos\'e}(1984)}]{Nose1984}
\bibinfo{author}{\bibfnamefont{S.}~\bibnamefont{Nos\'e}},
  \bibinfo{journal}{Journal of Chemical Physics} \textbf{\bibinfo{volume}{81}},
  \bibinfo{pages}{511} (\bibinfo{year}{1984}).

\bibitem[{\citenamefont{Henkelman and J\'onsson}(2000)}]{JCP:H00_TAN}
\bibinfo{author}{\bibfnamefont{G.}~\bibnamefont{Henkelman}} \bibnamefont{and}
  \bibinfo{author}{\bibfnamefont{H.}~\bibnamefont{J\'onsson}},
  \bibinfo{journal}{J. Chem. Phys.} \textbf{\bibinfo{volume}{113}},
  \bibinfo{pages}{9978} (\bibinfo{year}{2000}).

\bibitem[{\citenamefont{Henkelman et~al.}(2000)\citenamefont{Henkelman,
  Uberuaga, and J\'onsson}}]{JCP:H00_CI}
\bibinfo{author}{\bibfnamefont{G.}~\bibnamefont{Henkelman}},
  \bibinfo{author}{\bibfnamefont{B.~P.} \bibnamefont{Uberuaga}},
  \bibnamefont{and}
  \bibinfo{author}{\bibfnamefont{H.}~\bibnamefont{J\'onsson}},
  \bibinfo{journal}{J. Chem. Phys.} \textbf{\bibinfo{volume}{113}},
  \bibinfo{pages}{9901} (\bibinfo{year}{2000}).

\bibitem[{\citenamefont{Vineyard}(1957)}]{JPCS:VIN57}
\bibinfo{author}{\bibfnamefont{G.~H.} \bibnamefont{Vineyard}},
  \bibinfo{journal}{J. Phys. Chem. Solids} \textbf{\bibinfo{volume}{3}},
  \bibinfo{pages}{121} (\bibinfo{year}{1957}).

\bibitem[{\citenamefont{Laio and Gervasio}(2008)}]{Laio2008}
\bibinfo{author}{\bibfnamefont{A.}~\bibnamefont{Laio}} \bibnamefont{and}
  \bibinfo{author}{\bibfnamefont{F.~L.} \bibnamefont{Gervasio}},
  \bibinfo{journal}{Rep. Prog. Phys.} \textbf{\bibinfo{volume}{71}},
  \bibinfo{pages}{126601} (\bibinfo{year}{2008}).

\bibitem[{\citenamefont{Martin}(2004)}]{Martin2004}
\bibinfo{author}{\bibfnamefont{R.~M.} \bibnamefont{Martin}},
  \emph{\bibinfo{title}{Electronic Structure: Basic Theory and Practical
  Methods}} (\bibinfo{publisher}{Cambridge University Press},
  \bibinfo{year}{2004}).

\bibitem[{\citenamefont{Burke}(2012)}]{Burke2012}
\bibinfo{author}{\bibfnamefont{K.}~\bibnamefont{Burke}},
  \bibinfo{journal}{Journal of Chemical Physics}
  \textbf{\bibinfo{volume}{136}}, \bibinfo{pages}{150901}
  (\bibinfo{year}{2012}).

\bibitem[{\citenamefont{Becke}(2014)}]{Becke_Perspective_2014}
\bibinfo{author}{\bibfnamefont{A.~D.} \bibnamefont{Becke}},
  \bibinfo{journal}{Journal of Chemical Physics}
  \textbf{\bibinfo{volume}{140}}, \bibinfo{pages}{18A301}
  (\bibinfo{year}{2014}).

\bibitem[{\citenamefont{Yu et~al.}(2016)\citenamefont{Yu, Li, and
  Truhlar}}]{Yu_Perspective_2016}
\bibinfo{author}{\bibfnamefont{H.~S.} \bibnamefont{Yu}},
  \bibinfo{author}{\bibfnamefont{S.~L.} \bibnamefont{Li}}, \bibnamefont{and}
  \bibinfo{author}{\bibfnamefont{D.~G.} \bibnamefont{Truhlar}},
  \bibinfo{journal}{Journal of Chemical Physics}
  \textbf{\bibinfo{volume}{145}}, \bibinfo{pages}{130901}
  (\bibinfo{year}{2016}).

\bibitem[{\citenamefont{Hohenberg and Kohn}(1964)}]{Hohenberg1964}
\bibinfo{author}{\bibfnamefont{P.}~\bibnamefont{Hohenberg}} \bibnamefont{and}
  \bibinfo{author}{\bibfnamefont{W.}~\bibnamefont{Kohn}},
  \bibinfo{journal}{Phys. Rev.} \textbf{\bibinfo{volume}{136}},
  \bibinfo{pages}{B864} (\bibinfo{year}{1964}).

\bibitem[{\citenamefont{Kohn and Sham}(1965)}]{Kohn1965}
\bibinfo{author}{\bibfnamefont{W.}~\bibnamefont{Kohn}} \bibnamefont{and}
  \bibinfo{author}{\bibfnamefont{L.~J.} \bibnamefont{Sham}},
  \bibinfo{journal}{Phys. Rev.} \textbf{\bibinfo{volume}{140}},
  \bibinfo{pages}{A1133} (\bibinfo{year}{1965}).

\bibitem[{\citenamefont{Klimes and Michaelides}(2012)}]{Klimes2012}
\bibinfo{author}{\bibfnamefont{J.}~\bibnamefont{Klimes}} \bibnamefont{and}
  \bibinfo{author}{\bibfnamefont{A.}~\bibnamefont{Michaelides}},
  \bibinfo{journal}{Journal of Chemical Physics}
  \textbf{\bibinfo{volume}{137}}, \bibinfo{pages}{120901}
  (\bibinfo{year}{2012}).

\bibitem[{\citenamefont{Perdew et~al.}(1992)\citenamefont{Perdew, Chevary,
  Vosko, Jackson, Pederson, Singh, and Fiolhais}}]{Perdew1992_PW91}
\bibinfo{author}{\bibfnamefont{J.}~\bibnamefont{Perdew}},
  \bibinfo{author}{\bibfnamefont{J.}~\bibnamefont{Chevary}},
  \bibinfo{author}{\bibfnamefont{S.}~\bibnamefont{Vosko}},
  \bibinfo{author}{\bibfnamefont{K.}~\bibnamefont{Jackson}},
  \bibinfo{author}{\bibfnamefont{M.}~\bibnamefont{Pederson}},
  \bibinfo{author}{\bibfnamefont{D.}~\bibnamefont{Singh}}, \bibnamefont{and}
  \bibinfo{author}{\bibfnamefont{C.}~\bibnamefont{Fiolhais}},
  \bibinfo{journal}{Phys. Rev. B} \textbf{\bibinfo{volume}{46}},
  \bibinfo{pages}{6671} (\bibinfo{year}{1992}).

\bibitem[{\citenamefont{Perdew et~al.}(1996)\citenamefont{Perdew, Burke, and
  Ernzerhof}}]{Perdew1996_PBE}
\bibinfo{author}{\bibfnamefont{J.~P.} \bibnamefont{Perdew}},
  \bibinfo{author}{\bibfnamefont{K.}~\bibnamefont{Burke}}, \bibnamefont{and}
  \bibinfo{author}{\bibfnamefont{M.}~\bibnamefont{Ernzerhof}},
  \bibinfo{journal}{Phys. Rev. Lett.} \textbf{\bibinfo{volume}{77}},
  \bibinfo{pages}{3865} (\bibinfo{year}{1996}).

\bibitem[{\citenamefont{Tao et~al.}(2003)\citenamefont{Tao, Perdew, Staroverov,
  and Scuseria}}]{Tao2003}
\bibinfo{author}{\bibfnamefont{J.}~\bibnamefont{Tao}},
  \bibinfo{author}{\bibfnamefont{J.~P.} \bibnamefont{Perdew}},
  \bibinfo{author}{\bibfnamefont{V.~N.} \bibnamefont{Staroverov}},
  \bibnamefont{and} \bibinfo{author}{\bibfnamefont{G.~E.}
  \bibnamefont{Scuseria}}, \bibinfo{journal}{Phys. Rev. Lett.}
  \textbf{\bibinfo{volume}{91}}, \bibinfo{pages}{146401}
  (\bibinfo{year}{2003}).

\bibitem[{\citenamefont{Kresse and
  Furthm\"{u}ller}(1996{\natexlab{a}})}]{CMS:VASP96}
\bibinfo{author}{\bibfnamefont{G.}~\bibnamefont{Kresse}} \bibnamefont{and}
  \bibinfo{author}{\bibfnamefont{J.}~\bibnamefont{Furthm\"{u}ller}},
  \bibinfo{journal}{Comp.\ Mat.\ Sci.} \textbf{\bibinfo{volume}{6}},
  \bibinfo{pages}{15} (\bibinfo{year}{1996}{\natexlab{a}}).

\bibitem[{\citenamefont{Kresse and Hafner}(1993)}]{PRB:VASP93}
\bibinfo{author}{\bibfnamefont{G.}~\bibnamefont{Kresse}} \bibnamefont{and}
  \bibinfo{author}{\bibfnamefont{J.}~\bibnamefont{Hafner}},
  \bibinfo{journal}{Phys. Rev. B} \textbf{\bibinfo{volume}{47}},
  \bibinfo{pages}{558} (\bibinfo{year}{1993}).

\bibitem[{\citenamefont{Kresse and Hafner}(1994)}]{PRB:VASP94}
\bibinfo{author}{\bibfnamefont{G.}~\bibnamefont{Kresse}} \bibnamefont{and}
  \bibinfo{author}{\bibfnamefont{J.}~\bibnamefont{Hafner}},
  \bibinfo{journal}{Phys. Rev. B} \textbf{\bibinfo{volume}{49}},
  \bibinfo{pages}{14251} (\bibinfo{year}{1994}).

\bibitem[{\citenamefont{Kresse and
  Furthm\"{u}ller}(1996{\natexlab{b}})}]{PRB:VASP96}
\bibinfo{author}{\bibfnamefont{G.}~\bibnamefont{Kresse}} \bibnamefont{and}
  \bibinfo{author}{\bibfnamefont{J.}~\bibnamefont{Furthm\"{u}ller}},
  \bibinfo{journal}{Phys. Rev. B} \textbf{\bibinfo{volume}{54}},
  \bibinfo{pages}{11169} (\bibinfo{year}{1996}{\natexlab{b}}).

\bibitem[{\citenamefont{Giannozzi et~al.}(2009)\citenamefont{Giannozzi, Baroni,
  Bonini, Calandra, Car, Cavazzoni, Ceresoli, Chiarotti, Cococcioni, Dabo
  et~al.}}]{Giannozzi_QE_2009}
\bibinfo{author}{\bibfnamefont{P.}~\bibnamefont{Giannozzi}},
  \bibinfo{author}{\bibfnamefont{S.}~\bibnamefont{Baroni}},
  \bibinfo{author}{\bibfnamefont{N.}~\bibnamefont{Bonini}},
  \bibinfo{author}{\bibfnamefont{M.}~\bibnamefont{Calandra}},
  \bibinfo{author}{\bibfnamefont{R.}~\bibnamefont{Car}},
  \bibinfo{author}{\bibfnamefont{C.}~\bibnamefont{Cavazzoni}},
  \bibinfo{author}{\bibfnamefont{D.}~\bibnamefont{Ceresoli}},
  \bibinfo{author}{\bibfnamefont{G.~L.} \bibnamefont{Chiarotti}},
  \bibinfo{author}{\bibfnamefont{M.}~\bibnamefont{Cococcioni}},
  \bibinfo{author}{\bibfnamefont{I.}~\bibnamefont{Dabo}}, \bibnamefont{et~al.},
  \bibinfo{journal}{Journal of Physics: Condensed Matter}
  \textbf{\bibinfo{volume}{21}}, \bibinfo{pages}{395502}
  (\bibinfo{year}{2009}).

\bibitem[{\citenamefont{Giannozzi et~al.}(2017)\citenamefont{Giannozzi,
  Andreussi, Brumme, Bunau, Nardelli, Calandra, Car, Cavazzoni, Ceresoli,
  Cococcioni et~al.}}]{Giannozzi_QE_2017}
\bibinfo{author}{\bibfnamefont{P.}~\bibnamefont{Giannozzi}},
  \bibinfo{author}{\bibfnamefont{O.}~\bibnamefont{Andreussi}},
  \bibinfo{author}{\bibfnamefont{T.}~\bibnamefont{Brumme}},
  \bibinfo{author}{\bibfnamefont{O.}~\bibnamefont{Bunau}},
  \bibinfo{author}{\bibfnamefont{M.~B.} \bibnamefont{Nardelli}},
  \bibinfo{author}{\bibfnamefont{M.}~\bibnamefont{Calandra}},
  \bibinfo{author}{\bibfnamefont{R.}~\bibnamefont{Car}},
  \bibinfo{author}{\bibfnamefont{C.}~\bibnamefont{Cavazzoni}},
  \bibinfo{author}{\bibfnamefont{D.}~\bibnamefont{Ceresoli}},
  \bibinfo{author}{\bibfnamefont{M.}~\bibnamefont{Cococcioni}},
  \bibnamefont{et~al.}, \bibinfo{journal}{Journal of Physics: Condensed Matter}
  \textbf{\bibinfo{volume}{29}}, \bibinfo{pages}{465901}
  (\bibinfo{year}{2017}).

\bibitem[{\citenamefont{Mattsson et~al.}(2005)\citenamefont{Mattsson, Schultz,
  Desjarlais, Mattsson, and Leung}}]{Mattsson2005}
\bibinfo{author}{\bibfnamefont{A.~E.} \bibnamefont{Mattsson}},
  \bibinfo{author}{\bibfnamefont{P.~A.} \bibnamefont{Schultz}},
  \bibinfo{author}{\bibfnamefont{M.~P.} \bibnamefont{Desjarlais}},
  \bibinfo{author}{\bibfnamefont{T.~J.} \bibnamefont{Mattsson}},
  \bibnamefont{and} \bibinfo{author}{\bibfnamefont{K.}~\bibnamefont{Leung}},
  \bibinfo{journal}{Modelling Simul. Mater. Sci. Eng.}
  \textbf{\bibinfo{volume}{13}}, \bibinfo{pages}{R1} (\bibinfo{year}{2005}).

\bibitem[{\citenamefont{Hamann}(1989)}]{Hamann1989}
\bibinfo{author}{\bibfnamefont{D.~R.} \bibnamefont{Hamann}},
  \bibinfo{journal}{Phys. Rev. B} \textbf{\bibinfo{volume}{40}},
  \bibinfo{pages}{2980} (\bibinfo{year}{1989}).

\bibitem[{\citenamefont{Trouiller and Martins}(1991)}]{Trouiller1991}
\bibinfo{author}{\bibfnamefont{N.}~\bibnamefont{Trouiller}} \bibnamefont{and}
  \bibinfo{author}{\bibfnamefont{J.~L.} \bibnamefont{Martins}},
  \bibinfo{journal}{Phys. Rev. B} \textbf{\bibinfo{volume}{43}},
  \bibinfo{pages}{1993} (\bibinfo{year}{1991}).

\bibitem[{\citenamefont{Fuchs and Scheffler}(1999)}]{Fuchs1999}
\bibinfo{author}{\bibfnamefont{M.}~\bibnamefont{Fuchs}} \bibnamefont{and}
  \bibinfo{author}{\bibfnamefont{M.}~\bibnamefont{Scheffler}},
  \bibinfo{journal}{Comput. Phys. Commun.} \textbf{\bibinfo{volume}{119}},
  \bibinfo{pages}{67} (\bibinfo{year}{1999}).

\bibitem[{\citenamefont{Vanderbilt}(1990)}]{Vanderbilt1990}
\bibinfo{author}{\bibfnamefont{D.}~\bibnamefont{Vanderbilt}},
  \bibinfo{journal}{Phys. Rev. B} \textbf{\bibinfo{volume}{41}},
  \bibinfo{pages}{7892} (\bibinfo{year}{1990}).

\bibitem[{\citenamefont{Bl{\"o}chl}(1994)}]{PRB:PAW94}
\bibinfo{author}{\bibfnamefont{P.~E.} \bibnamefont{Bl{\"o}chl}},
  \bibinfo{journal}{Phys. Rev. B} \textbf{\bibinfo{volume}{50}},
  \bibinfo{pages}{17953} (\bibinfo{year}{1994}).

\bibitem[{\citenamefont{Kresse and Joubert}(1999)}]{PRB:PAW99}
\bibinfo{author}{\bibfnamefont{G.}~\bibnamefont{Kresse}} \bibnamefont{and}
  \bibinfo{author}{\bibfnamefont{D.}~\bibnamefont{Joubert}},
  \bibinfo{journal}{Phys. Rev. B} \textbf{\bibinfo{volume}{59}},
  \bibinfo{pages}{1758} (\bibinfo{year}{1999}).

\bibitem[{\citenamefont{Modinos}(1987)}]{Modinos1987}
\bibinfo{author}{\bibfnamefont{A.}~\bibnamefont{Modinos}},
  \bibinfo{journal}{Progress in Surface Science} \textbf{\bibinfo{volume}{26}},
  \bibinfo{pages}{19 } (\bibinfo{year}{1987}).

\bibitem[{\citenamefont{Brako and Newns}(1989)}]{Brako1989}
\bibinfo{author}{\bibfnamefont{R.}~\bibnamefont{Brako}} \bibnamefont{and}
  \bibinfo{author}{\bibfnamefont{D.~M.} \bibnamefont{Newns}},
  \bibinfo{journal}{Rep. Prog. Phys.} \textbf{\bibinfo{volume}{52}},
  \bibinfo{pages}{655} (\bibinfo{year}{1989}).

\bibitem[{\citenamefont{Kimmel and
  Cooper}(1993)}]{Kimmel_1993_alkali_ion_Cu_resonant_charge_transfer}
\bibinfo{author}{\bibfnamefont{G.~A.} \bibnamefont{Kimmel}} \bibnamefont{and}
  \bibinfo{author}{\bibfnamefont{B.~H.} \bibnamefont{Cooper}},
  \bibinfo{journal}{Phys. Rev. B} \textbf{\bibinfo{volume}{48}},
  \bibinfo{pages}{12164} (\bibinfo{year}{1993}).

\bibitem[{\citenamefont{Winter}(2002{\natexlab{b}})}]{Winter_PhysRep_2002}
\bibinfo{author}{\bibfnamefont{H.}~\bibnamefont{Winter}},
  \bibinfo{journal}{Phys. Rep.} \textbf{\bibinfo{volume}{367}},
  \bibinfo{pages}{387} (\bibinfo{year}{2002}{\natexlab{b}}).

\bibitem[{\citenamefont{Race et~al.}(2010)\citenamefont{Race, Mason, Finnis,
  Foulkes, Horsfield, and Sutton}}]{Race_RepProgPhys_2010}
\bibinfo{author}{\bibfnamefont{C.~P.} \bibnamefont{Race}},
  \bibinfo{author}{\bibfnamefont{D.~R.} \bibnamefont{Mason}},
  \bibinfo{author}{\bibfnamefont{M.~W.} \bibnamefont{Finnis}},
  \bibinfo{author}{\bibfnamefont{W.~M.~C.} \bibnamefont{Foulkes}},
  \bibinfo{author}{\bibfnamefont{A.~P.} \bibnamefont{Horsfield}},
  \bibnamefont{and} \bibinfo{author}{\bibfnamefont{A.~P.}
  \bibnamefont{Sutton}}, \bibinfo{journal}{Rep. Prog. Phys.}
  \textbf{\bibinfo{volume}{73}}, \bibinfo{pages}{116501}
  (\bibinfo{year}{2010}).

\bibitem[{\citenamefont{Wucher and Duvenbeck}(2011)}]{wucher_2011_kinetic}
\bibinfo{author}{\bibfnamefont{A.}~\bibnamefont{Wucher}} \bibnamefont{and}
  \bibinfo{author}{\bibfnamefont{A.}~\bibnamefont{Duvenbeck}},
  \bibinfo{journal}{Nucl. Instrum. Methods Phys. Res. B}
  \textbf{\bibinfo{volume}{269}}, \bibinfo{pages}{1655} (\bibinfo{year}{2011}).

\bibitem[{\citenamefont{Lindenblatt
  et~al.}(2006{\natexlab{a}})\citenamefont{Lindenblatt, Pehlke, Duvenbeck,
  Rethfeld, and Wucher}}]{Lindenblatt2006b}
\bibinfo{author}{\bibfnamefont{M.}~\bibnamefont{Lindenblatt}},
  \bibinfo{author}{\bibfnamefont{E.}~\bibnamefont{Pehlke}},
  \bibinfo{author}{\bibfnamefont{A.}~\bibnamefont{Duvenbeck}},
  \bibinfo{author}{\bibfnamefont{B.}~\bibnamefont{Rethfeld}}, \bibnamefont{and}
  \bibinfo{author}{\bibfnamefont{A.}~\bibnamefont{Wucher}},
  \bibinfo{journal}{Nuclear Instruments and Methods in Physics Research Section
  B: Beam Interactions with Materials and Atoms}
  \textbf{\bibinfo{volume}{246}}, \bibinfo{pages}{333 }
  (\bibinfo{year}{2006}{\natexlab{a}}).

\bibitem[{\citenamefont{Nienhaus}(2002)}]{Nienhaus_2002_chemicurrents}
\bibinfo{author}{\bibfnamefont{H.}~\bibnamefont{Nienhaus}},
  \bibinfo{journal}{Surf. Sci. Rep.} \textbf{\bibinfo{volume}{45}},
  \bibinfo{pages}{1} (\bibinfo{year}{2002}).

\bibitem[{\citenamefont{Diesing and Hasselbrink}(2016)}]{Diesing2016}
\bibinfo{author}{\bibfnamefont{D.}~\bibnamefont{Diesing}} \bibnamefont{and}
  \bibinfo{author}{\bibfnamefont{E.}~\bibnamefont{Hasselbrink}},
  \bibinfo{journal}{Chem. Soc. Rev.} \textbf{\bibinfo{volume}{45}},
  \bibinfo{pages}{3747} (\bibinfo{year}{2016}).

\bibitem[{\citenamefont{B{\"u}nermann et~al.}(2015)\citenamefont{B{\"u}nermann,
  Jiang, Dorenkamp, Kandratsenka, Janke, Auerbach, and
  Wodtke}}]{Buenermann1346_2015}
\bibinfo{author}{\bibfnamefont{O.}~\bibnamefont{B{\"u}nermann}},
  \bibinfo{author}{\bibfnamefont{H.}~\bibnamefont{Jiang}},
  \bibinfo{author}{\bibfnamefont{Y.}~\bibnamefont{Dorenkamp}},
  \bibinfo{author}{\bibfnamefont{A.}~\bibnamefont{Kandratsenka}},
  \bibinfo{author}{\bibfnamefont{S.~M.} \bibnamefont{Janke}},
  \bibinfo{author}{\bibfnamefont{D.~J.} \bibnamefont{Auerbach}},
  \bibnamefont{and} \bibinfo{author}{\bibfnamefont{A.~M.}
  \bibnamefont{Wodtke}}, \bibinfo{journal}{Science}
  \textbf{\bibinfo{volume}{350}}, \bibinfo{pages}{1346} (\bibinfo{year}{2015}).

\bibitem[{\citenamefont{Wodtke}(2016)}]{Wodtke2016}
\bibinfo{author}{\bibfnamefont{A.~M.} \bibnamefont{Wodtke}},
  \bibinfo{journal}{Chem. Soc. Rev.} \textbf{\bibinfo{volume}{45}},
  \bibinfo{pages}{3641} (\bibinfo{year}{2016}).

\bibitem[{\citenamefont{Rittmeyer et~al.}(2015)\citenamefont{Rittmeyer, Meyer,
  Juaristi, and Reuter}}]{Rittmeyer_2015_LDFA}
\bibinfo{author}{\bibfnamefont{S.~P.} \bibnamefont{Rittmeyer}},
  \bibinfo{author}{\bibfnamefont{J.}~\bibnamefont{Meyer}},
  \bibinfo{author}{\bibfnamefont{J.~I.} \bibnamefont{Juaristi}},
  \bibnamefont{and} \bibinfo{author}{\bibfnamefont{K.}~\bibnamefont{Reuter}},
  \bibinfo{journal}{Phys. Rev. Lett.} \textbf{\bibinfo{volume}{115}},
  \bibinfo{pages}{046102} (\bibinfo{year}{2015}).

\bibitem[{\citenamefont{Rittmeyer et~al.}(2018)\citenamefont{Rittmeyer, Bukas,
  and Reuter}}]{Rittmeyer2018}
\bibinfo{author}{\bibfnamefont{S.~P.} \bibnamefont{Rittmeyer}},
  \bibinfo{author}{\bibfnamefont{V.~J.} \bibnamefont{Bukas}}, \bibnamefont{and}
  \bibinfo{author}{\bibfnamefont{K.}~\bibnamefont{Reuter}},
  \bibinfo{journal}{Advances in Physics: X} \textbf{\bibinfo{volume}{3}},
  \bibinfo{pages}{1381574} (\bibinfo{year}{2018}).

\bibitem[{\citenamefont{Alducin et~al.}(2017)\citenamefont{Alducin, Muiño, and
  Juaristi}}]{Alducin2017}
\bibinfo{author}{\bibfnamefont{M.}~\bibnamefont{Alducin}},
  \bibinfo{author}{\bibfnamefont{R.~D.} \bibnamefont{Muiño}},
  \bibnamefont{and} \bibinfo{author}{\bibfnamefont{J.~I.}
  \bibnamefont{Juaristi}}, \bibinfo{journal}{Progress in Surface Science}
  \textbf{\bibinfo{volume}{92}}, \bibinfo{pages}{317 } (\bibinfo{year}{2017}).

\bibitem[{\citenamefont{Janke et~al.}(2015)\citenamefont{Janke, Auerbach,
  Wodtke, and Kandratsenka}}]{Janke2015}
\bibinfo{author}{\bibfnamefont{S.~M.} \bibnamefont{Janke}},
  \bibinfo{author}{\bibfnamefont{D.~J.} \bibnamefont{Auerbach}},
  \bibinfo{author}{\bibfnamefont{A.~M.} \bibnamefont{Wodtke}},
  \bibnamefont{and}
  \bibinfo{author}{\bibfnamefont{A.}~\bibnamefont{Kandratsenka}},
  \bibinfo{journal}{The Journal of Chemical Physics}
  \textbf{\bibinfo{volume}{143}}, \bibinfo{pages}{124708}
  (\bibinfo{year}{2015}).

\bibitem[{\citenamefont{Kroes et~al.}(2014)\citenamefont{Kroes, Pavanello,
  Blanco-Rey, Alducin, and Auerbach}}]{Kroes2014}
\bibinfo{author}{\bibfnamefont{G.-J.} \bibnamefont{Kroes}},
  \bibinfo{author}{\bibfnamefont{M.}~\bibnamefont{Pavanello}},
  \bibinfo{author}{\bibfnamefont{M.}~\bibnamefont{Blanco-Rey}},
  \bibinfo{author}{\bibfnamefont{M.}~\bibnamefont{Alducin}}, \bibnamefont{and}
  \bibinfo{author}{\bibfnamefont{D.~J.} \bibnamefont{Auerbach}},
  \bibinfo{journal}{The Journal of Chemical Physics}
  \textbf{\bibinfo{volume}{141}}, \bibinfo{pages}{054705}
  (\bibinfo{year}{2014}).

\bibitem[{\citenamefont{Monturet and Saalfrank}(2010)}]{Monturet2010}
\bibinfo{author}{\bibfnamefont{S.}~\bibnamefont{Monturet}} \bibnamefont{and}
  \bibinfo{author}{\bibfnamefont{P.}~\bibnamefont{Saalfrank}},
  \bibinfo{journal}{Phys. Rev. B} \textbf{\bibinfo{volume}{82}},
  \bibinfo{pages}{075404} (\bibinfo{year}{2010}).

\bibitem[{\citenamefont{Mizielinski et~al.}(2005)\citenamefont{Mizielinski,
  Bird, Persson, and Holloway}}]{Mizielinski2005}
\bibinfo{author}{\bibfnamefont{M.~S.} \bibnamefont{Mizielinski}},
  \bibinfo{author}{\bibfnamefont{D.~M.} \bibnamefont{Bird}},
  \bibinfo{author}{\bibfnamefont{M.}~\bibnamefont{Persson}}, \bibnamefont{and}
  \bibinfo{author}{\bibfnamefont{S.}~\bibnamefont{Holloway}},
  \bibinfo{journal}{J. Chem. Phys.} \textbf{\bibinfo{volume}{122}},
  \bibinfo{pages}{084710} (\bibinfo{year}{2005}).

\bibitem[{\citenamefont{Mizielinski et~al.}(2007)\citenamefont{Mizielinski,
  Bird, Persson, and Holloway}}]{Mizielinski2007}
\bibinfo{author}{\bibfnamefont{M.~S.} \bibnamefont{Mizielinski}},
  \bibinfo{author}{\bibfnamefont{D.~M.} \bibnamefont{Bird}},
  \bibinfo{author}{\bibfnamefont{M.}~\bibnamefont{Persson}}, \bibnamefont{and}
  \bibinfo{author}{\bibfnamefont{S.}~\bibnamefont{Holloway}},
  \bibinfo{journal}{J. Chem. Phys.} \textbf{\bibinfo{volume}{126}},
  \bibinfo{pages}{034705} (\bibinfo{year}{2007}).

\bibitem[{\citenamefont{Mizielinski et~al.}(2008)\citenamefont{Mizielinski,
  Bird, Persson, and Holloway}}]{Mizielinski2008}
\bibinfo{author}{\bibfnamefont{M.~S.} \bibnamefont{Mizielinski}},
  \bibinfo{author}{\bibfnamefont{D.~M.} \bibnamefont{Bird}},
  \bibinfo{author}{\bibfnamefont{M.}~\bibnamefont{Persson}}, \bibnamefont{and}
  \bibinfo{author}{\bibfnamefont{S.}~\bibnamefont{Holloway}},
  \bibinfo{journal}{Surf. Sci.} \textbf{\bibinfo{volume}{602}},
  \bibinfo{pages}{2617 } (\bibinfo{year}{2008}).

\bibitem[{\citenamefont{Mizielinski and Bird}(2010)}]{Mizielinski2010}
\bibinfo{author}{\bibfnamefont{M.~S.} \bibnamefont{Mizielinski}}
  \bibnamefont{and} \bibinfo{author}{\bibfnamefont{D.~M.} \bibnamefont{Bird}},
  \bibinfo{journal}{J. Chem. Phys.} \textbf{\bibinfo{volume}{132}},
  \bibinfo{pages}{184704} (\bibinfo{year}{2010}).

\bibitem[{\citenamefont{Bird et~al.}(2008)\citenamefont{Bird, Mizielinski,
  Lindenblatt, and Pehlke}}]{Bird2008}
\bibinfo{author}{\bibfnamefont{D.~M.} \bibnamefont{Bird}},
  \bibinfo{author}{\bibfnamefont{M.~S.} \bibnamefont{Mizielinski}},
  \bibinfo{author}{\bibfnamefont{M.}~\bibnamefont{Lindenblatt}},
  \bibnamefont{and} \bibinfo{author}{\bibfnamefont{E.}~\bibnamefont{Pehlke}},
  \bibinfo{journal}{Surface Science} \textbf{\bibinfo{volume}{602}},
  \bibinfo{pages}{1212} (\bibinfo{year}{2008}).

\bibitem[{\citenamefont{Lindenblatt
  et~al.}(2006{\natexlab{b}})\citenamefont{Lindenblatt, van Heys, and
  Pehlke}}]{Lindenblatt2006}
\bibinfo{author}{\bibfnamefont{M.}~\bibnamefont{Lindenblatt}},
  \bibinfo{author}{\bibfnamefont{J.}~\bibnamefont{van Heys}}, \bibnamefont{and}
  \bibinfo{author}{\bibfnamefont{E.}~\bibnamefont{Pehlke}},
  \bibinfo{journal}{Surface Science} \textbf{\bibinfo{volume}{600}},
  \bibinfo{pages}{3624 } (\bibinfo{year}{2006}{\natexlab{b}}).

\bibitem[{\citenamefont{Lindenblatt and
  Pehlke}(2006{\natexlab{a}})}]{Lindenblatt2006a}
\bibinfo{author}{\bibfnamefont{M.}~\bibnamefont{Lindenblatt}} \bibnamefont{and}
  \bibinfo{author}{\bibfnamefont{E.}~\bibnamefont{Pehlke}},
  \bibinfo{journal}{Surface Science} \textbf{\bibinfo{volume}{600}},
  \bibinfo{pages}{5068 } (\bibinfo{year}{2006}{\natexlab{a}}).

\bibitem[{\citenamefont{Lindenblatt and
  Pehlke}(2006{\natexlab{b}})}]{Lindenblatt2006PRL}
\bibinfo{author}{\bibfnamefont{M.}~\bibnamefont{Lindenblatt}} \bibnamefont{and}
  \bibinfo{author}{\bibfnamefont{E.}~\bibnamefont{Pehlke}},
  \bibinfo{journal}{Phys. Rev. Lett.} \textbf{\bibinfo{volume}{97}},
  \bibinfo{pages}{216101} (\bibinfo{year}{2006}{\natexlab{b}}).

\bibitem[{\citenamefont{Grotemeyer and Pehlke}(2014)}]{Grotemeyer2014}
\bibinfo{author}{\bibfnamefont{M.}~\bibnamefont{Grotemeyer}} \bibnamefont{and}
  \bibinfo{author}{\bibfnamefont{E.}~\bibnamefont{Pehlke}},
  \bibinfo{journal}{Phys. Rev. Lett.} \textbf{\bibinfo{volume}{112}},
  \bibinfo{pages}{043201} (\bibinfo{year}{2014}).

\bibitem[{\citenamefont{Timmer and
  Kratzer}(2009)}]{Timmer_2009_H_Al111_perturbation}
\bibinfo{author}{\bibfnamefont{M.}~\bibnamefont{Timmer}} \bibnamefont{and}
  \bibinfo{author}{\bibfnamefont{P.}~\bibnamefont{Kratzer}},
  \bibinfo{journal}{Phys. Rev. B} \textbf{\bibinfo{volume}{79}},
  \bibinfo{pages}{165407} (\bibinfo{year}{2009}).

\bibitem[{\citenamefont{Zhao et~al.}(2015{\natexlab{b}})\citenamefont{Zhao,
  Kang, Xue, Zhang, and Zhang}}]{Zhao_2015_H+Be}
\bibinfo{author}{\bibfnamefont{S.}~\bibnamefont{Zhao}},
  \bibinfo{author}{\bibfnamefont{W.}~\bibnamefont{Kang}},
  \bibinfo{author}{\bibfnamefont{J.}~\bibnamefont{Xue}},
  \bibinfo{author}{\bibfnamefont{X.}~\bibnamefont{Zhang}}, \bibnamefont{and}
  \bibinfo{author}{\bibfnamefont{P.}~\bibnamefont{Zhang}},
  \bibinfo{journal}{Physics Letters A} \textbf{\bibinfo{volume}{379}},
  \bibinfo{pages}{319} (\bibinfo{year}{2015}{\natexlab{b}}).

\bibitem[{\citenamefont{Moss et~al.}(2009)\citenamefont{Moss, Isborn, and
  Li}}]{Moss_2009_Li+_AlCluster}
\bibinfo{author}{\bibfnamefont{C.~L.} \bibnamefont{Moss}},
  \bibinfo{author}{\bibfnamefont{C.~M.} \bibnamefont{Isborn}},
  \bibnamefont{and} \bibinfo{author}{\bibfnamefont{X.}~\bibnamefont{Li}},
  \bibinfo{journal}{Phys. Rev. A} \textbf{\bibinfo{volume}{80}},
  \bibinfo{pages}{024503} (\bibinfo{year}{2009}).

\bibitem[{\citenamefont{Castro et~al.}(2012)\citenamefont{Castro, Isla,
  Mart\'inez, and Alonso}}]{Castro2012_H+Li4}
\bibinfo{author}{\bibfnamefont{A.}~\bibnamefont{Castro}},
  \bibinfo{author}{\bibfnamefont{M.}~\bibnamefont{Isla}},
  \bibinfo{author}{\bibfnamefont{J.~I.} \bibnamefont{Mart\'inez}},
  \bibnamefont{and} \bibinfo{author}{\bibfnamefont{J.~A.}
  \bibnamefont{Alonso}}, \bibinfo{journal}{Chemical Physics}
  \textbf{\bibinfo{volume}{399}}, \bibinfo{pages}{130} (\bibinfo{year}{2012}).

\bibitem[{\citenamefont{Krasheninnikov
  et~al.}(2007)\citenamefont{Krasheninnikov, Miyamoto, and
  Tom\'anek}}]{Krasheninnikov2007}
\bibinfo{author}{\bibfnamefont{A.~V.} \bibnamefont{Krasheninnikov}},
  \bibinfo{author}{\bibfnamefont{Y.}~\bibnamefont{Miyamoto}}, \bibnamefont{and}
  \bibinfo{author}{\bibfnamefont{D.}~\bibnamefont{Tom\'anek}},
  \bibinfo{journal}{Phys. Rev. Lett.} \textbf{\bibinfo{volume}{99}},
  \bibinfo{pages}{016104} (\bibinfo{year}{2007}).

\bibitem[{\citenamefont{Bubin et~al.}(2012)\citenamefont{Bubin, Wang,
  Pantelides, and Varga}}]{Bubin2012}
\bibinfo{author}{\bibfnamefont{S.}~\bibnamefont{Bubin}},
  \bibinfo{author}{\bibfnamefont{B.}~\bibnamefont{Wang}},
  \bibinfo{author}{\bibfnamefont{S.}~\bibnamefont{Pantelides}},
  \bibnamefont{and} \bibinfo{author}{\bibfnamefont{K.}~\bibnamefont{Varga}},
  \bibinfo{journal}{Phys. Rev. B} \textbf{\bibinfo{volume}{85}},
  \bibinfo{pages}{235435} (\bibinfo{year}{2012}).

\bibitem[{\citenamefont{Ojanper\"a et~al.}(2014)\citenamefont{Ojanper\"a,
  Krasheninnikov, and Puska}}]{ojanpera_prb_14}
\bibinfo{author}{\bibfnamefont{A.}~\bibnamefont{Ojanper\"a}},
  \bibinfo{author}{\bibfnamefont{A.~V.} \bibnamefont{Krasheninnikov}},
  \bibnamefont{and} \bibinfo{author}{\bibfnamefont{M.}~\bibnamefont{Puska}},
  \bibinfo{journal}{Phys. Rev. B} \textbf{\bibinfo{volume}{89}},
  \bibinfo{pages}{035120} (\bibinfo{year}{2014}),
  \urlprefix\url{https://link.aps.org/doi/10.1103/PhysRevB.89.035120}.

\bibitem[{\citenamefont{Wang et~al.}(2015)\citenamefont{Wang, Li, and
  Wang}}]{Wang2015}
\bibinfo{author}{\bibfnamefont{Z.}~\bibnamefont{Wang}},
  \bibinfo{author}{\bibfnamefont{S.-S.} \bibnamefont{Li}}, \bibnamefont{and}
  \bibinfo{author}{\bibfnamefont{L.-W.} \bibnamefont{Wang}},
  \bibinfo{journal}{Phys. Rev. Lett.} \textbf{\bibinfo{volume}{114}},
  \bibinfo{pages}{063004} (\bibinfo{year}{2015}).

\bibitem[{\citenamefont{Yost et~al.}(2017)\citenamefont{Yost, Yao, and
  Kanai}}]{Yost2017}
\bibinfo{author}{\bibfnamefont{D.~C.} \bibnamefont{Yost}},
  \bibinfo{author}{\bibfnamefont{Y.}~\bibnamefont{Yao}}, \bibnamefont{and}
  \bibinfo{author}{\bibfnamefont{Y.}~\bibnamefont{Kanai}},
  \bibinfo{journal}{Phys. Rev. B} \textbf{\bibinfo{volume}{96}},
  \bibinfo{pages}{115134} (\bibinfo{year}{2017}).

\bibitem[{\citenamefont{Schleife
  et~al.}(2015{\natexlab{a}})\citenamefont{Schleife, Kanai, and
  Correa}}]{Schleife_2015_H_He_Al}
\bibinfo{author}{\bibfnamefont{A.}~\bibnamefont{Schleife}},
  \bibinfo{author}{\bibfnamefont{Y.}~\bibnamefont{Kanai}}, \bibnamefont{and}
  \bibinfo{author}{\bibfnamefont{A.~A.} \bibnamefont{Correa}},
  \bibinfo{journal}{Phys. Rev. B} \textbf{\bibinfo{volume}{91}},
  \bibinfo{pages}{014306} (\bibinfo{year}{2015}{\natexlab{a}}).

\bibitem[{\citenamefont{Correa et~al.}(2012)\citenamefont{Correa, Kohanoff,
  Artacho, S\'anchez-Portal, and Caro}}]{Correa_2012_H_in_Al}
\bibinfo{author}{\bibfnamefont{A.~A.} \bibnamefont{Correa}},
  \bibinfo{author}{\bibfnamefont{J.}~\bibnamefont{Kohanoff}},
  \bibinfo{author}{\bibfnamefont{E.}~\bibnamefont{Artacho}},
  \bibinfo{author}{\bibfnamefont{D.}~\bibnamefont{S\'anchez-Portal}},
  \bibnamefont{and} \bibinfo{author}{\bibfnamefont{A.}~\bibnamefont{Caro}},
  \bibinfo{journal}{Phys. Rev. Lett.} \textbf{\bibinfo{volume}{108}},
  \bibinfo{pages}{213201} (\bibinfo{year}{2012}).

\bibitem[{\citenamefont{Zeb et~al.}(2012)\citenamefont{Zeb, Kohanoff,
  S\'anchez-Portal, Arnau, Juaristi, and Artacho}}]{Zeb2012}
\bibinfo{author}{\bibfnamefont{M.~A.} \bibnamefont{Zeb}},
  \bibinfo{author}{\bibfnamefont{J.}~\bibnamefont{Kohanoff}},
  \bibinfo{author}{\bibfnamefont{D.}~\bibnamefont{S\'anchez-Portal}},
  \bibinfo{author}{\bibfnamefont{A.}~\bibnamefont{Arnau}},
  \bibinfo{author}{\bibfnamefont{J.~I.} \bibnamefont{Juaristi}},
  \bibnamefont{and} \bibinfo{author}{\bibfnamefont{E.}~\bibnamefont{Artacho}},
  \bibinfo{journal}{Phys. Rev. Lett.} \textbf{\bibinfo{volume}{108}},
  \bibinfo{pages}{225504} (\bibinfo{year}{2012}).

\bibitem[{\citenamefont{Ullah et~al.}(2015)\citenamefont{Ullah, Corsetti,
  S\'anchez-Portal, and Artacho}}]{Ullah2015}
\bibinfo{author}{\bibfnamefont{R.}~\bibnamefont{Ullah}},
  \bibinfo{author}{\bibfnamefont{F.}~\bibnamefont{Corsetti}},
  \bibinfo{author}{\bibfnamefont{D.}~\bibnamefont{S\'anchez-Portal}},
  \bibnamefont{and} \bibinfo{author}{\bibfnamefont{E.}~\bibnamefont{Artacho}},
  \bibinfo{journal}{Phys. Rev. B} \textbf{\bibinfo{volume}{91}},
  \bibinfo{pages}{125203} (\bibinfo{year}{2015}).

\bibitem[{\citenamefont{Marques et~al.}(2006)\citenamefont{Marques, Ullrich,
  Nogueira, Rubio, Burke, and Gross}}]{Marques_TDDFT_I}
\bibinfo{editor}{\bibfnamefont{M.~A.~L.} \bibnamefont{Marques}},
  \bibinfo{editor}{\bibfnamefont{C.~A.} \bibnamefont{Ullrich}},
  \bibinfo{editor}{\bibfnamefont{F.}~\bibnamefont{Nogueira}},
  \bibinfo{editor}{\bibfnamefont{A.}~\bibnamefont{Rubio}},
  \bibinfo{editor}{\bibfnamefont{K.}~\bibnamefont{Burke}}, \bibnamefont{and}
  \bibinfo{editor}{\bibfnamefont{E.~K.~U.} \bibnamefont{Gross}}, eds.,
  \emph{\bibinfo{title}{Time-Dependent Density Functional Theory}}, vol.
  \bibinfo{volume}{706} of \emph{\bibinfo{series}{Lecture Notes in Physics}}
  (\bibinfo{publisher}{Springer}, \bibinfo{year}{2006}).

\bibitem[{\citenamefont{Marques et~al.}(2012)\citenamefont{Marques, Maitra,
  Nogueira, Gross, and Rubio}}]{Marques_TDDFT_II}
\bibinfo{editor}{\bibfnamefont{M.~A.~L.} \bibnamefont{Marques}},
  \bibinfo{editor}{\bibfnamefont{N.~T.} \bibnamefont{Maitra}},
  \bibinfo{editor}{\bibfnamefont{F.~M.~S.} \bibnamefont{Nogueira}},
  \bibinfo{editor}{\bibfnamefont{E.~K.~U.} \bibnamefont{Gross}},
  \bibnamefont{and} \bibinfo{editor}{\bibfnamefont{A.}~\bibnamefont{Rubio}},
  eds., \emph{\bibinfo{title}{Fundamentals of Time-Dependent Density Functional
  Theory}}, vol. \bibinfo{volume}{837} of \emph{\bibinfo{series}{Lecture Notes
  in Physics}} (\bibinfo{publisher}{Springer}, \bibinfo{year}{2012}).

\bibitem[{\citenamefont{Ullrich}(2012)}]{Ullrich_TDDFT}
\bibinfo{author}{\bibfnamefont{C.~A.} \bibnamefont{Ullrich}},
  \emph{\bibinfo{title}{Time-Dependent Density-Functional Theory}}
  (\bibinfo{publisher}{Oxford University Press}, \bibinfo{year}{2012}).

\bibitem[{\citenamefont{Ullrich and Yang}(2014)}]{Ullrich2014}
\bibinfo{author}{\bibfnamefont{C.~A.} \bibnamefont{Ullrich}} \bibnamefont{and}
  \bibinfo{author}{\bibfnamefont{Z.-h.} \bibnamefont{Yang}},
  \bibinfo{journal}{Braz. J. Phys.} \textbf{\bibinfo{volume}{44}},
  \bibinfo{pages}{154} (\bibinfo{year}{2014}).

\bibitem[{\citenamefont{Maitra}(2016)}]{Maitra_Perspective_2016}
\bibinfo{author}{\bibfnamefont{N.~T.} \bibnamefont{Maitra}},
  \bibinfo{journal}{Journal of Chemical Physics}
  \textbf{\bibinfo{volume}{144}}, \bibinfo{pages}{220901}
  (\bibinfo{year}{2016}).

\bibitem[{\citenamefont{Runge and Gross}(1984{\natexlab{b}})}]{Runge1984}
\bibinfo{author}{\bibfnamefont{E.}~\bibnamefont{Runge}} \bibnamefont{and}
  \bibinfo{author}{\bibfnamefont{E.~K.~U.} \bibnamefont{Gross}},
  \bibinfo{journal}{Phys. Rev. Lett.} \textbf{\bibinfo{volume}{52}},
  \bibinfo{pages}{997} (\bibinfo{year}{1984}{\natexlab{b}}).

\bibitem[{\citenamefont{Gross and Kohn}(1985)}]{Gross_Kohn_1985}
\bibinfo{author}{\bibfnamefont{E.~K.~U.} \bibnamefont{Gross}} \bibnamefont{and}
  \bibinfo{author}{\bibfnamefont{W.}~\bibnamefont{Kohn}},
  \bibinfo{journal}{Phys. Rev. Lett.} \textbf{\bibinfo{volume}{55}},
  \bibinfo{pages}{2850} (\bibinfo{year}{1985}).

\bibitem[{\citenamefont{Provorse and Isborn}(2016)}]{Provorse2016}
\bibinfo{author}{\bibfnamefont{M.~R.} \bibnamefont{Provorse}} \bibnamefont{and}
  \bibinfo{author}{\bibfnamefont{C.~M.} \bibnamefont{Isborn}},
  \bibinfo{journal}{International Journal of Quantum Chemistry}
  \textbf{\bibinfo{volume}{116}}, \bibinfo{pages}{739} (\bibinfo{year}{2016}).

\bibitem[{\citenamefont{Nagano et~al.}(2000)\citenamefont{Nagano, Yabana,
  Tazawa, and Abe}}]{Nagano2000}
\bibinfo{author}{\bibfnamefont{R.}~\bibnamefont{Nagano}},
  \bibinfo{author}{\bibfnamefont{K.}~\bibnamefont{Yabana}},
  \bibinfo{author}{\bibfnamefont{T.}~\bibnamefont{Tazawa}}, \bibnamefont{and}
  \bibinfo{author}{\bibfnamefont{Y.}~\bibnamefont{Abe}},
  \bibinfo{journal}{Phys. Rev. A} \textbf{\bibinfo{volume}{62}},
  \bibinfo{pages}{062721} (\bibinfo{year}{2000}).

\bibitem[{\citenamefont{Nazarov et~al.}(2007)\citenamefont{Nazarov, Pitarke,
  Takada, Vignale, and Chang}}]{Nazarov2007}
\bibinfo{author}{\bibfnamefont{V.~U.} \bibnamefont{Nazarov}},
  \bibinfo{author}{\bibfnamefont{J.~M.} \bibnamefont{Pitarke}},
  \bibinfo{author}{\bibfnamefont{Y.}~\bibnamefont{Takada}},
  \bibinfo{author}{\bibfnamefont{G.}~\bibnamefont{Vignale}}, \bibnamefont{and}
  \bibinfo{author}{\bibfnamefont{Y.-C.} \bibnamefont{Chang}},
  \bibinfo{journal}{Phys. Rev. B} \textbf{\bibinfo{volume}{76}},
  \bibinfo{pages}{205103} (\bibinfo{year}{2007}).

\bibitem[{\citenamefont{Tully}(1990)}]{Tully1990}
\bibinfo{author}{\bibfnamefont{J.~C.} \bibnamefont{Tully}},
  \bibinfo{journal}{J. Chem. Phys.} \textbf{\bibinfo{volume}{93}},
  \bibinfo{pages}{1061} (\bibinfo{year}{1990}).

\bibitem[{\citenamefont{Shenvi et~al.}(2009)\citenamefont{Shenvi, Roy, and
  Tully}}]{Shenvi_2009}
\bibinfo{author}{\bibfnamefont{N.}~\bibnamefont{Shenvi}},
  \bibinfo{author}{\bibfnamefont{S.}~\bibnamefont{Roy}}, \bibnamefont{and}
  \bibinfo{author}{\bibfnamefont{J.~C.} \bibnamefont{Tully}},
  \bibinfo{journal}{The Journal of Chemical Physics}
  \textbf{\bibinfo{volume}{130}}, \bibinfo{pages}{174107}
  (\bibinfo{year}{2009}).

\bibitem[{\citenamefont{Marques et~al.}(2003)\citenamefont{Marques, Castro,
  Bertsch, and Rubio}}]{Marques_octopus_2003}
\bibinfo{author}{\bibfnamefont{M.~A.~L.} \bibnamefont{Marques}},
  \bibinfo{author}{\bibfnamefont{A.}~\bibnamefont{Castro}},
  \bibinfo{author}{\bibfnamefont{G.~F.} \bibnamefont{Bertsch}},
  \bibnamefont{and} \bibinfo{author}{\bibfnamefont{A.}~\bibnamefont{Rubio}},
  \bibinfo{journal}{Comput. Phys. Commun.} \textbf{\bibinfo{volume}{151}},
  \bibinfo{pages}{60} (\bibinfo{year}{2003}).

\bibitem[{\citenamefont{Foglia et~al.}(2017)\citenamefont{Foglia, Morzan,
  Estrin, Scherlies, and Lebrero}}]{Foglia2017}
\bibinfo{author}{\bibfnamefont{N.~O.} \bibnamefont{Foglia}},
  \bibinfo{author}{\bibfnamefont{U.~N.} \bibnamefont{Morzan}},
  \bibinfo{author}{\bibfnamefont{D.~A.} \bibnamefont{Estrin}},
  \bibinfo{author}{\bibfnamefont{D.~A.} \bibnamefont{Scherlies}},
  \bibnamefont{and} \bibinfo{author}{\bibfnamefont{M.~C.~G.}
  \bibnamefont{Lebrero}}, \bibinfo{journal}{J. Chem. Theory Comput.}
  \textbf{\bibinfo{volume}{13}}, \bibinfo{pages}{77} (\bibinfo{year}{2017}).

\bibitem[{\citenamefont{Avenda\~no Franco}(2013)}]{AvendanoFranko_Thesis_2013}
\bibinfo{author}{\bibfnamefont{G.}~\bibnamefont{Avenda\~no Franco}}, Ph.D.
  thesis, \bibinfo{school}{Universit\'e Catholique de Louvain}
  (\bibinfo{year}{2013}).

\bibitem[{\citenamefont{German et~al.}(1994)\citenamefont{German, Weare, and
  Yarmoff}}]{German1994}
\bibinfo{author}{\bibfnamefont{K.~A.~H.} \bibnamefont{German}},
  \bibinfo{author}{\bibfnamefont{C.~B.} \bibnamefont{Weare}}, \bibnamefont{and}
  \bibinfo{author}{\bibfnamefont{J.~A.} \bibnamefont{Yarmoff}},
  \bibinfo{journal}{Phys. Rev. B} \textbf{\bibinfo{volume}{50}},
  \bibinfo{pages}{14452} (\bibinfo{year}{1994}).

\bibitem[{\citenamefont{Castro et~al.}(2006)\citenamefont{Castro, Appel,
  Oliveira, Rozzi, Andrade, Lorenzen, Marques, Gross, and
  Rubio}}]{Castro_octopus_2006}
\bibinfo{author}{\bibfnamefont{A.}~\bibnamefont{Castro}},
  \bibinfo{author}{\bibfnamefont{H.}~\bibnamefont{Appel}},
  \bibinfo{author}{\bibfnamefont{M.}~\bibnamefont{Oliveira}},
  \bibinfo{author}{\bibfnamefont{C.~A.} \bibnamefont{Rozzi}},
  \bibinfo{author}{\bibfnamefont{X.}~\bibnamefont{Andrade}},
  \bibinfo{author}{\bibfnamefont{F.}~\bibnamefont{Lorenzen}},
  \bibinfo{author}{\bibfnamefont{M.~A.~L.} \bibnamefont{Marques}},
  \bibinfo{author}{\bibfnamefont{E.~K.~U.} \bibnamefont{Gross}},
  \bibnamefont{and} \bibinfo{author}{\bibfnamefont{A.}~\bibnamefont{Rubio}},
  \bibinfo{journal}{physica status solidi (b)} \textbf{\bibinfo{volume}{243}},
  \bibinfo{pages}{2465} (\bibinfo{year}{2006}).

\bibitem[{\citenamefont{Andrade et~al.}(2015)\citenamefont{Andrade, Strubbe,
  De~Giovannini, Larsen, Oliveira, Alberdi-Rodriguez, Varas, Theophilou,
  Helbig, Verstraete et~al.}}]{Andrade_octopus_2015}
\bibinfo{author}{\bibfnamefont{X.}~\bibnamefont{Andrade}},
  \bibinfo{author}{\bibfnamefont{D.}~\bibnamefont{Strubbe}},
  \bibinfo{author}{\bibfnamefont{U.}~\bibnamefont{De~Giovannini}},
  \bibinfo{author}{\bibfnamefont{A.~H.} \bibnamefont{Larsen}},
  \bibinfo{author}{\bibfnamefont{M.~J.~T.} \bibnamefont{Oliveira}},
  \bibinfo{author}{\bibfnamefont{J.}~\bibnamefont{Alberdi-Rodriguez}},
  \bibinfo{author}{\bibfnamefont{A.}~\bibnamefont{Varas}},
  \bibinfo{author}{\bibfnamefont{I.}~\bibnamefont{Theophilou}},
  \bibinfo{author}{\bibfnamefont{N.}~\bibnamefont{Helbig}},
  \bibinfo{author}{\bibfnamefont{M.~J.} \bibnamefont{Verstraete}},
  \bibnamefont{et~al.}, \bibinfo{journal}{Phys. Chem. Chem. Phys.}
  \textbf{\bibinfo{volume}{17}}, \bibinfo{pages}{31371} (\bibinfo{year}{2015}).

\bibitem[{\citenamefont{Schleife
  et~al.}(2015{\natexlab{b}})\citenamefont{Schleife, Kanai, and
  Correa}}]{Schleife2015}
\bibinfo{author}{\bibfnamefont{A.}~\bibnamefont{Schleife}},
  \bibinfo{author}{\bibfnamefont{Y.}~\bibnamefont{Kanai}}, \bibnamefont{and}
  \bibinfo{author}{\bibfnamefont{A.~A.} \bibnamefont{Correa}},
  \bibinfo{journal}{Phys. Rev. B} \textbf{\bibinfo{volume}{91}},
  \bibinfo{pages}{014306} (\bibinfo{year}{2015}{\natexlab{b}}).

\bibitem[{\citenamefont{Shukri et~al.}(2016)\citenamefont{Shukri, Bruneval, and
  Reining}}]{Shukri_2016_electronic-stopping}
\bibinfo{author}{\bibfnamefont{A.~A.} \bibnamefont{Shukri}},
  \bibinfo{author}{\bibfnamefont{F.}~\bibnamefont{Bruneval}}, \bibnamefont{and}
  \bibinfo{author}{\bibfnamefont{L.}~\bibnamefont{Reining}},
  \bibinfo{journal}{Phys. Rev. B} \textbf{\bibinfo{volume}{93}},
  \bibinfo{pages}{035128} (\bibinfo{year}{2016}).

\bibitem[{\citenamefont{Markin et~al.}(2009)\citenamefont{Markin, Primetzhofer,
  Spitz, and Bauer}}]{Markin2009}
\bibinfo{author}{\bibfnamefont{S.~N.} \bibnamefont{Markin}},
  \bibinfo{author}{\bibfnamefont{D.}~\bibnamefont{Primetzhofer}},
  \bibinfo{author}{\bibfnamefont{M.}~\bibnamefont{Spitz}}, \bibnamefont{and}
  \bibinfo{author}{\bibfnamefont{P.}~\bibnamefont{Bauer}},
  \bibinfo{journal}{Phys. Rev. B} \textbf{\bibinfo{volume}{80}},
  \bibinfo{pages}{205105} (\bibinfo{year}{2009}).

\bibitem[{\citenamefont{Mason et~al.}(2007)\citenamefont{Mason, le~Page, Race,
  Foulkes, Finnis, and Sutton}}]{Mason2007}
\bibinfo{author}{\bibfnamefont{D.~R.} \bibnamefont{Mason}},
  \bibinfo{author}{\bibfnamefont{J.}~\bibnamefont{le~Page}},
  \bibinfo{author}{\bibfnamefont{C.~P.} \bibnamefont{Race}},
  \bibinfo{author}{\bibfnamefont{W.~M.~C.} \bibnamefont{Foulkes}},
  \bibinfo{author}{\bibfnamefont{M.~W.} \bibnamefont{Finnis}},
  \bibnamefont{and} \bibinfo{author}{\bibfnamefont{A.~P.}
  \bibnamefont{Sutton}}, \bibinfo{journal}{Journal of Physics: Condensed
  Matter} \textbf{\bibinfo{volume}{19}}, \bibinfo{pages}{436209}
  (\bibinfo{year}{2007}).

\bibitem[{\citenamefont{Grotemeyer}(2016)}]{Grotemeyer_Dissertation}
\bibinfo{author}{\bibfnamefont{M.~K.} \bibnamefont{Grotemeyer}}, Ph.D. thesis,
  \bibinfo{school}{Universit\"at Kiel} (\bibinfo{year}{2016}).

\bibitem[{\citenamefont{D{'}Agosta and Di~Ventra}(2013)}]{Agosta2013}
\bibinfo{author}{\bibfnamefont{R.}~\bibnamefont{D{'}Agosta}} \bibnamefont{and}
  \bibinfo{author}{\bibfnamefont{M.}~\bibnamefont{Di~Ventra}},
  \bibinfo{journal}{Phys. Rev. B} \textbf{\bibinfo{volume}{87}},
  \bibinfo{pages}{155129} (\bibinfo{year}{2013}).

\bibitem[{\citenamefont{Ullrich}(2006)}]{Ullrich2006}
\bibinfo{author}{\bibfnamefont{C.~A.} \bibnamefont{Ullrich}},
  \bibinfo{journal}{J. Chem. Phys.} \textbf{\bibinfo{volume}{125}},
  \bibinfo{pages}{234108} (\bibinfo{year}{2006}).

\bibitem[{\citenamefont{Kapoor}(2016)}]{Kapoor2016}
\bibinfo{author}{\bibfnamefont{V.}~\bibnamefont{Kapoor}},
  \bibinfo{journal}{Phys. Rev. A} \textbf{\bibinfo{volume}{93}},
  \bibinfo{pages}{063408} (\bibinfo{year}{2016}).

\bibitem[{\citenamefont{Lorente et~al.}(1994)\citenamefont{Lorente, Monreal,
  and Alducin}}]{Lorente_1994_Auger_neutralization}
\bibinfo{author}{\bibfnamefont{N.}~\bibnamefont{Lorente}},
  \bibinfo{author}{\bibfnamefont{R.}~\bibnamefont{Monreal}}, \bibnamefont{and}
  \bibinfo{author}{\bibfnamefont{M.}~\bibnamefont{Alducin}},
  \bibinfo{journal}{Phys. Rev. A} \textbf{\bibinfo{volume}{49}},
  \bibinfo{pages}{4716} (\bibinfo{year}{1994}).

\bibitem[{\citenamefont{Monreal}(2014)}]{Monreal_Review_2014}
\bibinfo{author}{\bibfnamefont{R.~C.} \bibnamefont{Monreal}},
  \bibinfo{journal}{Prog. Surf. Sci.} \textbf{\bibinfo{volume}{89}},
  \bibinfo{pages}{80} (\bibinfo{year}{2014}).

\bibitem[{\citenamefont{Balzer et~al.}(2018)\citenamefont{Balzer, Rasmussen,
  Schlünzen, Joost, and Bonitz}}]{balzer_prl_18}
\bibinfo{author}{\bibfnamefont{K.}~\bibnamefont{Balzer}},
  \bibinfo{author}{\bibfnamefont{M.}~\bibnamefont{Rasmussen}},
  \bibinfo{author}{\bibfnamefont{N.}~\bibnamefont{Schlünzen}},
  \bibinfo{author}{\bibfnamefont{J.~P.} \bibnamefont{Joost}}, \bibnamefont{and}
  \bibinfo{author}{\bibfnamefont{M.}~\bibnamefont{Bonitz}},
  \bibinfo{journal}{arXiv:1801.05267}  (\bibinfo{year}{2018}),
  \bibinfo{note}{submitted for publication}.

\bibitem[{\citenamefont{Keldysh}(1965)}]{keldysh64}
\bibinfo{author}{\bibfnamefont{L.}~\bibnamefont{Keldysh}},
  \bibinfo{journal}{Soviet Phys. JETP} \textbf{\bibinfo{volume}{20}},
  \bibinfo{pages}{1018} (\bibinfo{year}{1965}),
  \bibinfo{note}{(Zh.~Eksp.~Teor.~Fiz.~\textbf{47}, 1515 (1964))}.

\bibitem[{\citenamefont{Kadanoff and Baym}(1962)}]{kadanoff-baym-book}
\bibinfo{author}{\bibfnamefont{L.}~\bibnamefont{Kadanoff}} \bibnamefont{and}
  \bibinfo{author}{\bibfnamefont{G.}~\bibnamefont{Baym}},
  \emph{\bibinfo{title}{Quantum Statistical Mechanics}}
  (\bibinfo{publisher}{Benjamin}, \bibinfo{address}{New York},
  \bibinfo{year}{1962}).

\bibitem[{\citenamefont{Bonitz and Kremp}(1996)}]{bonitz-etal.96pla}
\bibinfo{author}{\bibfnamefont{M.}~\bibnamefont{Bonitz}} \bibnamefont{and}
  \bibinfo{author}{\bibfnamefont{D.}~\bibnamefont{Kremp}},
  \bibinfo{journal}{Physics Letters A} \textbf{\bibinfo{volume}{212}},
  \bibinfo{pages}{83 } (\bibinfo{year}{1996}), ISSN \bibinfo{issn}{0375-9601},
  \urlprefix\url{http://www.sciencedirect.com/science/article/pii/0375960196000564}.

\bibitem[{\citenamefont{Bonitz et~al.}(1996)\citenamefont{Bonitz, Kremp, Scott,
  Binder, Kraeft, and K\"ohler}}]{bonitz-etal.96jpb}
\bibinfo{author}{\bibfnamefont{M.}~\bibnamefont{Bonitz}},
  \bibinfo{author}{\bibfnamefont{D.}~\bibnamefont{Kremp}},
  \bibinfo{author}{\bibfnamefont{D.~C.} \bibnamefont{Scott}},
  \bibinfo{author}{\bibfnamefont{R.}~\bibnamefont{Binder}},
  \bibinfo{author}{\bibfnamefont{W.~D.} \bibnamefont{Kraeft}},
  \bibnamefont{and} \bibinfo{author}{\bibfnamefont{H.~S.}
  \bibnamefont{K\"ohler}}, \bibinfo{journal}{J. Phys.: Cond. Matt.}
  \textbf{\bibinfo{volume}{8}}, \bibinfo{pages}{6057} (\bibinfo{year}{1996}),
  \urlprefix\url{http://stacks.iop.org/0953-8984/8/i=33/a=012}.

\bibitem[{\citenamefont{Bonitz}(1996)}]{bonitz96pla}
\bibinfo{author}{\bibfnamefont{M.}~\bibnamefont{Bonitz}},
  \bibinfo{journal}{Physics Letters A} \textbf{\bibinfo{volume}{221}},
  \bibinfo{pages}{85 } (\bibinfo{year}{1996}), ISSN \bibinfo{issn}{0375-9601},
  \urlprefix\url{http://www.sciencedirect.com/science/article/pii/0375960196005567}.

\bibitem[{\citenamefont{Kremp et~al.}(1997)\citenamefont{Kremp, Bonitz, Kraeft,
  and Schlanges}}]{kremp-etal.97ap}
\bibinfo{author}{\bibfnamefont{D.}~\bibnamefont{Kremp}},
  \bibinfo{author}{\bibfnamefont{M.}~\bibnamefont{Bonitz}},
  \bibinfo{author}{\bibfnamefont{W.}~\bibnamefont{Kraeft}}, \bibnamefont{and}
  \bibinfo{author}{\bibfnamefont{M.}~\bibnamefont{Schlanges}},
  \bibinfo{journal}{Annals of Physics} \textbf{\bibinfo{volume}{258}},
  \bibinfo{pages}{320 } (\bibinfo{year}{1997}), ISSN \bibinfo{issn}{0003-4916},
  \urlprefix\url{http://www.sciencedirect.com/science/article/pii/S0003491697957031}.

\bibitem[{\citenamefont{Danielewicz}(1984)}]{DANIELEWICZ_84_ap2}
\bibinfo{author}{\bibfnamefont{P.}~\bibnamefont{Danielewicz}},
  \bibinfo{journal}{Annals of Physics} \textbf{\bibinfo{volume}{152}},
  \bibinfo{pages}{305 } (\bibinfo{year}{1984}), ISSN \bibinfo{issn}{0003-4916},
  \urlprefix\url{http://www.sciencedirect.com/science/article/pii/0003491684900939}.

\bibitem[{\citenamefont{K\"ohler}(1995)}]{koehler_prc_95}
\bibinfo{author}{\bibfnamefont{H.~S.} \bibnamefont{K\"ohler}},
  \bibinfo{journal}{Phys. Rev. C} \textbf{\bibinfo{volume}{51}},
  \bibinfo{pages}{3232} (\bibinfo{year}{1995}),
  \urlprefix\url{https://link.aps.org/doi/10.1103/PhysRevC.51.3232}.

\bibitem[{\citenamefont{B\'anyai et~al.}(1995)\citenamefont{B\'anyai, Thoai,
  Reitsamer, Haug, Steinbach, Wehner, Wegener, Marschner, and
  Stolz}}]{banyai_prl_95}
\bibinfo{author}{\bibfnamefont{L.}~\bibnamefont{B\'anyai}},
  \bibinfo{author}{\bibfnamefont{D.~B.~T.} \bibnamefont{Thoai}},
  \bibinfo{author}{\bibfnamefont{E.}~\bibnamefont{Reitsamer}},
  \bibinfo{author}{\bibfnamefont{H.}~\bibnamefont{Haug}},
  \bibinfo{author}{\bibfnamefont{D.}~\bibnamefont{Steinbach}},
  \bibinfo{author}{\bibfnamefont{M.~U.} \bibnamefont{Wehner}},
  \bibinfo{author}{\bibfnamefont{M.}~\bibnamefont{Wegener}},
  \bibinfo{author}{\bibfnamefont{T.}~\bibnamefont{Marschner}},
  \bibnamefont{and} \bibinfo{author}{\bibfnamefont{W.}~\bibnamefont{Stolz}},
  \bibinfo{journal}{Phys. Rev. Lett.} \textbf{\bibinfo{volume}{75}},
  \bibinfo{pages}{2188} (\bibinfo{year}{1995}),
  \urlprefix\url{https://link.aps.org/doi/10.1103/PhysRevLett.75.2188}.

\bibitem[{\citenamefont{Kwong et~al.}(1998)\citenamefont{Kwong, Bonitz, Binder,
  and K\"ohler}}]{kwong-etal.98pss}
\bibinfo{author}{\bibfnamefont{N.}~\bibnamefont{Kwong}},
  \bibinfo{author}{\bibfnamefont{M.}~\bibnamefont{Bonitz}},
  \bibinfo{author}{\bibfnamefont{R.}~\bibnamefont{Binder}}, \bibnamefont{and}
  \bibinfo{author}{\bibfnamefont{H.}~\bibnamefont{K\"ohler}},
  \bibinfo{journal}{phys. stat. sol. (b)} \textbf{\bibinfo{volume}{{\bf 206}}},
  \bibinfo{pages}{197} (\bibinfo{year}{1998}).

\bibitem[{\citenamefont{Binder et~al.}(1997)\citenamefont{Binder, K{\"o}hler,
  and Bonitz}}]{binder-etal.97prb}
\bibinfo{author}{\bibfnamefont{R.}~\bibnamefont{Binder}},
  \bibinfo{author}{\bibfnamefont{S.}~\bibnamefont{K{\"o}hler}},
  \bibnamefont{and} \bibinfo{author}{\bibfnamefont{M.}~\bibnamefont{Bonitz}},
  \bibinfo{journal}{Phys. Rev. B} \textbf{\bibinfo{volume}{{\bf 55}}},
  \bibinfo{pages}{5110} (\bibinfo{year}{1997}).

\bibitem[{\citenamefont{Bonitz et~al.}(2007)\citenamefont{Bonitz, Balzer, and
  van Leeuwen}}]{bonitz_prb_7}
\bibinfo{author}{\bibfnamefont{M.}~\bibnamefont{Bonitz}},
  \bibinfo{author}{\bibfnamefont{K.}~\bibnamefont{Balzer}}, \bibnamefont{and}
  \bibinfo{author}{\bibfnamefont{R.}~\bibnamefont{van Leeuwen}},
  \bibinfo{journal}{Phys. Rev. B} \textbf{\bibinfo{volume}{76}},
  \bibinfo{pages}{045341} (\bibinfo{year}{2007}),
  \urlprefix\url{http://link.aps.org/doi/10.1103/PhysRevB.76.045341}.

\bibitem[{\citenamefont{Balzer et~al.}(2009)\citenamefont{Balzer, Bonitz, van
  Leeuwen, Stan, and Dahlen}}]{balzer_prb_9}
\bibinfo{author}{\bibfnamefont{K.}~\bibnamefont{Balzer}},
  \bibinfo{author}{\bibfnamefont{M.}~\bibnamefont{Bonitz}},
  \bibinfo{author}{\bibfnamefont{R.}~\bibnamefont{van Leeuwen}},
  \bibinfo{author}{\bibfnamefont{A.}~\bibnamefont{Stan}}, \bibnamefont{and}
  \bibinfo{author}{\bibfnamefont{N.~E.} \bibnamefont{Dahlen}},
  \bibinfo{journal}{Phys. Rev. B} \textbf{\bibinfo{volume}{79}},
  \bibinfo{pages}{245306} (\bibinfo{year}{2009}),
  \urlprefix\url{http://link.aps.org/doi/10.1103/PhysRevB.79.245306}.

\bibitem[{\citenamefont{Kremp et~al.}(1999)\citenamefont{Kremp, Bornath,
  Bonitz, and Schlanges}}]{kremp_99_pre}
\bibinfo{author}{\bibfnamefont{D.}~\bibnamefont{Kremp}},
  \bibinfo{author}{\bibfnamefont{T.}~\bibnamefont{Bornath}},
  \bibinfo{author}{\bibfnamefont{M.}~\bibnamefont{Bonitz}}, \bibnamefont{and}
  \bibinfo{author}{\bibfnamefont{M.}~\bibnamefont{Schlanges}},
  \bibinfo{journal}{Phys. Rev. E} \textbf{\bibinfo{volume}{{\bf 60}}},
  \bibinfo{pages}{4725} (\bibinfo{year}{1999}),
  \urlprefix\url{http://link.aps.org/doi/10.1103/PhysRevE.60.4725}.

\bibitem[{\citenamefont{Bonitz et~al.}(1999)\citenamefont{Bonitz, Bornath,
  Kremp, Schlanges, and Kraeft}}]{bonitz_99_cpp}
\bibinfo{author}{\bibfnamefont{M.}~\bibnamefont{Bonitz}},
  \bibinfo{author}{\bibfnamefont{T.}~\bibnamefont{Bornath}},
  \bibinfo{author}{\bibfnamefont{D.}~\bibnamefont{Kremp}},
  \bibinfo{author}{\bibfnamefont{M.}~\bibnamefont{Schlanges}},
  \bibnamefont{and} \bibinfo{author}{\bibfnamefont{W.~D.}
  \bibnamefont{Kraeft}}, \bibinfo{journal}{Contrib. Plasma Phys.}
  \textbf{\bibinfo{volume}{{\bf 39}}}, \bibinfo{pages}{329}
  (\bibinfo{year}{1999}), ISSN \bibinfo{issn}{1521-3986},
  \urlprefix\url{http://dx.doi.org/10.1002/ctpp.2150390407}.

\bibitem[{\citenamefont{Stefanucci and van
  Leeuwen}(2013)}]{stefanucci_cambridge_2013}
\bibinfo{author}{\bibfnamefont{G.}~\bibnamefont{Stefanucci}} \bibnamefont{and}
  \bibinfo{author}{\bibfnamefont{R.}~\bibnamefont{van Leeuwen}},
  \emph{\bibinfo{title}{Nonequilibrium Many-Body Theory of Quantum Systems}}
  (\bibinfo{publisher}{Cambridge University Press},
  \bibinfo{address}{Cambridge}, \bibinfo{year}{2013}).

\bibitem[{\citenamefont{Balzer et~al.}(2010{\natexlab{a}})\citenamefont{Balzer,
  Bauch, and Bonitz}}]{balzer_pra_10}
\bibinfo{author}{\bibfnamefont{K.}~\bibnamefont{Balzer}},
  \bibinfo{author}{\bibfnamefont{S.}~\bibnamefont{Bauch}}, \bibnamefont{and}
  \bibinfo{author}{\bibfnamefont{M.}~\bibnamefont{Bonitz}},
  \bibinfo{journal}{Phys. Rev. A} \textbf{\bibinfo{volume}{81}},
  \bibinfo{pages}{022510} (\bibinfo{year}{2010}{\natexlab{a}}),
  \urlprefix\url{http://link.aps.org/doi/10.1103/PhysRevA.81.022510}.

\bibitem[{\citenamefont{Balzer et~al.}(2010{\natexlab{b}})\citenamefont{Balzer,
  Bauch, and Bonitz}}]{balzer_pra_10_2}
\bibinfo{author}{\bibfnamefont{K.}~\bibnamefont{Balzer}},
  \bibinfo{author}{\bibfnamefont{S.}~\bibnamefont{Bauch}}, \bibnamefont{and}
  \bibinfo{author}{\bibfnamefont{M.}~\bibnamefont{Bonitz}},
  \bibinfo{journal}{Phys. Rev. A} \textbf{\bibinfo{volume}{82}},
  \bibinfo{pages}{033427} (\bibinfo{year}{2010}{\natexlab{b}}),
  \urlprefix\url{http://link.aps.org/doi/10.1103/PhysRevA.82.033427}.

\bibitem[{\citenamefont{Verdozzi C. Wacker~A. and M.}(2016)}]{verdozzi_jpcs16}
\bibinfo{author}{\bibfnamefont{A.~C.-O.} \bibnamefont{Verdozzi C. Wacker~A.}}
  \bibnamefont{and} \bibinfo{author}{\bibfnamefont{B.}~\bibnamefont{M.}},
  \bibinfo{journal}{Journal of Physics: Conference Series}
  \textbf{\bibinfo{volume}{696}}, \bibinfo{pages}{011001}
  (\bibinfo{year}{2016}),
  \urlprefix\url{http://stacks.iop.org/1742-6596/696/i=1/a=011001}.

\bibitem[{\citenamefont{Schl\"unzen et~al.}(2016)\citenamefont{Schl\"unzen,
  Hermanns, Bonitz, and Verdozzi}}]{schluenzen_prb16}
\bibinfo{author}{\bibfnamefont{N.}~\bibnamefont{Schl\"unzen}},
  \bibinfo{author}{\bibfnamefont{S.}~\bibnamefont{Hermanns}},
  \bibinfo{author}{\bibfnamefont{M.}~\bibnamefont{Bonitz}}, \bibnamefont{and}
  \bibinfo{author}{\bibfnamefont{C.}~\bibnamefont{Verdozzi}},
  \bibinfo{journal}{Phys. Rev. B} \textbf{\bibinfo{volume}{93}},
  \bibinfo{pages}{035107} (\bibinfo{year}{2016}),
  \urlprefix\url{http://link.aps.org/doi/10.1103/PhysRevB.93.035107}.

\bibitem[{\citenamefont{Bonitz et~al.}(2018{\natexlab{b}})\citenamefont{Bonitz,
  Scharnke, and Schlünzen}}]{bonitz_cpp18}
\bibinfo{author}{\bibfnamefont{M.}~\bibnamefont{Bonitz}},
  \bibinfo{author}{\bibfnamefont{M.}~\bibnamefont{Scharnke}}, \bibnamefont{and}
  \bibinfo{author}{\bibfnamefont{N.}~\bibnamefont{Schlünzen}},
  \bibinfo{journal}{Contributions to Plasma Physics}
  \textbf{\bibinfo{volume}{58}} (\bibinfo{year}{2018}{\natexlab{b}}),
  \eprint{https://onlinelibrary.wiley.com/doi/pdf/10.1002/ctpp.201700052},
  \urlprefix\url{https://onlinelibrary.wiley.com/doi/abs/10.1002/ctpp.201700052}.

\bibitem[{tri()}]{trim}
\bibinfo{note}{TRIM and SRIM code packages, www.srim.org}.

\bibitem[{\citenamefont{Heese}(2017)}]{heese_bsc_17}
\bibinfo{author}{\bibfnamefont{S.}~\bibnamefont{Heese}},
  \emph{\bibinfo{title}{Dielectric function of graphene with yambo}}
  (\bibinfo{year}{2017}), \bibinfo{note}{bachelor thesis, Kiel University,
  unpublished}.

\bibitem[{\citenamefont{Pamperin
  et~al.}(2015{\natexlab{c}})\citenamefont{Pamperin, Bronold, and
  Fehske}}]{pamperin_2015_many}
\bibinfo{author}{\bibfnamefont{M.}~\bibnamefont{Pamperin}},
  \bibinfo{author}{\bibfnamefont{F.~X.} \bibnamefont{Bronold}},
  \bibnamefont{and} \bibinfo{author}{\bibfnamefont{H.}~\bibnamefont{Fehske}},
  \bibinfo{journal}{Phys. Rev. B} \textbf{\bibinfo{volume}{91}},
  \bibinfo{pages}{035440} (\bibinfo{year}{2015}{\natexlab{c}}).

\bibitem[{\citenamefont{Brenig}(1979)}]{brenig_zpb79}
\bibinfo{author}{\bibfnamefont{W.}~\bibnamefont{Brenig}},
  \bibinfo{journal}{Zeitschrift f{\"u}r Physik B Condensed Matter}
  \textbf{\bibinfo{volume}{36}}, \bibinfo{pages}{81} (\bibinfo{year}{1979}),
  ISSN \bibinfo{issn}{1431-584X},
  \urlprefix\url{https://doi.org/10.1007/BF01333956}.

\bibitem[{\citenamefont{Bonitz et~al.}(2012)\citenamefont{Bonitz, Rosenthal,
  Fujioka, Zaporojtchenko, Faupel, and Kersten}}]{bonitz_cpp12}
\bibinfo{author}{\bibfnamefont{M.}~\bibnamefont{Bonitz}},
  \bibinfo{author}{\bibfnamefont{L.}~\bibnamefont{Rosenthal}},
  \bibinfo{author}{\bibfnamefont{K.}~\bibnamefont{Fujioka}},
  \bibinfo{author}{\bibfnamefont{V.}~\bibnamefont{Zaporojtchenko}},
  \bibinfo{author}{\bibfnamefont{F.}~\bibnamefont{Faupel}}, \bibnamefont{and}
  \bibinfo{author}{\bibfnamefont{H.}~\bibnamefont{Kersten}},
  \bibinfo{journal}{Contributions to Plasma Physics}
  \textbf{\bibinfo{volume}{52}}, \bibinfo{pages}{890} (\bibinfo{year}{2012}),
  \eprint{https://onlinelibrary.wiley.com/doi/pdf/10.1002/ctpp.201200038},
  \urlprefix\url{https://onlinelibrary.wiley.com/doi/abs/10.1002/ctpp.201200038}.

\bibitem[{\citenamefont{Brenig and Pehlke}(2008)}]{brenig_2008_reaction}
\bibinfo{author}{\bibfnamefont{W.}~\bibnamefont{Brenig}} \bibnamefont{and}
  \bibinfo{author}{\bibfnamefont{E.}~\bibnamefont{Pehlke}},
  \bibinfo{journal}{Prog. Surf. Sci.} \textbf{\bibinfo{volume}{83}},
  \bibinfo{pages}{263} (\bibinfo{year}{2008}).

\bibitem[{\citenamefont{Bronold and Fehske}(2017)}]{BF17}
\bibinfo{author}{\bibfnamefont{F.~X.} \bibnamefont{Bronold}} \bibnamefont{and}
  \bibinfo{author}{\bibfnamefont{H.}~\bibnamefont{Fehske}},
  \bibinfo{journal}{J. Phys. D: Appl. Phys} \textbf{\bibinfo{volume}{50}},
  \bibinfo{pages}{294003} (\bibinfo{year}{2017}).

\bibitem[{\citenamefont{Langmuir and Mott-Smith}(1924)}]{LM24}
\bibinfo{author}{\bibfnamefont{I.}~\bibnamefont{Langmuir}} \bibnamefont{and}
  \bibinfo{author}{\bibfnamefont{H.}~\bibnamefont{Mott-Smith}},
  \bibinfo{journal}{Gen. Electr. Rev.} \textbf{\bibinfo{volume}{27}},
  \bibinfo{pages}{449} (\bibinfo{year}{1924}).

\bibitem[{\citenamefont{Robertson}(2013)}]{Robertson13}
\bibinfo{author}{\bibfnamefont{S.}~\bibnamefont{Robertson}},
  \bibinfo{journal}{Plasma Phys. Control. Fusion}
  \textbf{\bibinfo{volume}{55}}, \bibinfo{pages}{093001}
  (\bibinfo{year}{2013}).

\bibitem[{\citenamefont{Brinkmann}(2009)}]{Brinkmann09}
\bibinfo{author}{\bibfnamefont{R.~P.} \bibnamefont{Brinkmann}},
  \bibinfo{journal}{J. Phys. D: Appl. Phys.} \textbf{\bibinfo{volume}{42}},
  \bibinfo{pages}{194009} (\bibinfo{year}{2009}).

\bibitem[{\citenamefont{Franklin}(2003)}]{Franklin03}
\bibinfo{author}{\bibfnamefont{R.~N.} \bibnamefont{Franklin}},
  \bibinfo{journal}{J. Phys. D: Appl. Phys.} \textbf{\bibinfo{volume}{36}},
  \bibinfo{pages}{R309} (\bibinfo{year}{2003}).

\bibitem[{\citenamefont{Riemann}(1991)}]{Riemann91}
\bibinfo{author}{\bibfnamefont{K.-U.} \bibnamefont{Riemann}},
  \bibinfo{journal}{J. Phys. D: Appl. Phys.} \textbf{\bibinfo{volume}{24}},
  \bibinfo{pages}{493} (\bibinfo{year}{1991}).

\bibitem[{\citenamefont{Schwager and Birdsall}(1990)}]{SB90}
\bibinfo{author}{\bibfnamefont{L.~A.} \bibnamefont{Schwager}} \bibnamefont{and}
  \bibinfo{author}{\bibfnamefont{C.~K.} \bibnamefont{Birdsall}},
  \bibinfo{journal}{Phys. Fluids B} \textbf{\bibinfo{volume}{2}},
  \bibinfo{pages}{1057} (\bibinfo{year}{1990}).

\bibitem[{\citenamefont{Campanell and Umansky}(2016)}]{CU16}
\bibinfo{author}{\bibfnamefont{M.~D.} \bibnamefont{Campanell}}
  \bibnamefont{and} \bibinfo{author}{\bibfnamefont{M.~V.}
  \bibnamefont{Umansky}}, \bibinfo{journal}{Phys. Rev. Lett.}
  \textbf{\bibinfo{volume}{116}}, \bibinfo{pages}{085003}
  (\bibinfo{year}{2016}).

\bibitem[{\citenamefont{Langendorf and Walker}(2015)}]{LW15}
\bibinfo{author}{\bibfnamefont{S.}~\bibnamefont{Langendorf}} \bibnamefont{and}
  \bibinfo{author}{\bibfnamefont{M.}~\bibnamefont{Walker}},
  \bibinfo{journal}{Phys. Plasma} \textbf{\bibinfo{volume}{22}},
  \bibinfo{pages}{033515} (\bibinfo{year}{2015}).

\bibitem[{\citenamefont{Sheehan
  et~al.}(2013{\natexlab{b}})\citenamefont{Sheehan, Hershkowitz, Kaganovich,
  Wang, Raitses, Barnat, Weatherford, and Sydorenko}}]{SHK13}
\bibinfo{author}{\bibfnamefont{J.~P.} \bibnamefont{Sheehan}},
  \bibinfo{author}{\bibfnamefont{N.}~\bibnamefont{Hershkowitz}},
  \bibinfo{author}{\bibfnamefont{I.~D.} \bibnamefont{Kaganovich}},
  \bibinfo{author}{\bibfnamefont{H.}~\bibnamefont{Wang}},
  \bibinfo{author}{\bibfnamefont{Y.}~\bibnamefont{Raitses}},
  \bibinfo{author}{\bibfnamefont{E.~V.} \bibnamefont{Barnat}},
  \bibinfo{author}{\bibfnamefont{B.~R.} \bibnamefont{Weatherford}},
  \bibnamefont{and}
  \bibinfo{author}{\bibfnamefont{D.}~\bibnamefont{Sydorenko}},
  \bibinfo{journal}{Phys. Rev. Lett.} \textbf{\bibinfo{volume}{111}},
  \bibinfo{pages}{075002} (\bibinfo{year}{2013}{\natexlab{b}}).

\bibitem[{\citenamefont{Sydorenko et~al.}(2009)\citenamefont{Sydorenko,
  Kaganovich, Raitses, and Smolyakov}}]{SKR09}
\bibinfo{author}{\bibfnamefont{D.}~\bibnamefont{Sydorenko}},
  \bibinfo{author}{\bibfnamefont{I.~D.} \bibnamefont{Kaganovich}},
  \bibinfo{author}{\bibfnamefont{Y.}~\bibnamefont{Raitses}}, \bibnamefont{and}
  \bibinfo{author}{\bibfnamefont{A.}~\bibnamefont{Smolyakov}},
  \bibinfo{journal}{Phys. Rev. Lett.} \textbf{\bibinfo{volume}{103}},
  \bibinfo{pages}{145004} (\bibinfo{year}{2009}).

\bibitem[{\citenamefont{Taccogna et~al.}(2004)\citenamefont{Taccogna, Longo,
  and Capitelli}}]{TLC04}
\bibinfo{author}{\bibfnamefont{F.}~\bibnamefont{Taccogna}},
  \bibinfo{author}{\bibfnamefont{S.}~\bibnamefont{Longo}}, \bibnamefont{and}
  \bibinfo{author}{\bibfnamefont{M.}~\bibnamefont{Capitelli}},
  \bibinfo{journal}{Phys. Plasma} \textbf{\bibinfo{volume}{11}},
  \bibinfo{pages}{1220} (\bibinfo{year}{2004}).

\bibitem[{\citenamefont{Hu and Ziering}(1966)}]{HZ66}
\bibinfo{author}{\bibfnamefont{P.~N.} \bibnamefont{Hu}} \bibnamefont{and}
  \bibinfo{author}{\bibfnamefont{S.}~\bibnamefont{Ziering}},
  \bibinfo{journal}{Phys. Fluids} \textbf{\bibinfo{volume}{9}},
  \bibinfo{pages}{2168} (\bibinfo{year}{1966}).

\bibitem[{\citenamefont{Franklin}(1976)}]{Franklin76}
\bibinfo{author}{\bibfnamefont{R.~N.} \bibnamefont{Franklin}},
  \emph{\bibinfo{title}{Plasma phenomena in gas discharges}}
  (\bibinfo{publisher}{Clarendon Press}, \bibinfo{address}{Oxford},
  \bibinfo{year}{1976}).

\bibitem[{\citenamefont{Becker et~al.}(2010)\citenamefont{Becker, Grubert, and
  Loffhagen}}]{BGL10}
\bibinfo{author}{\bibfnamefont{M.~M.} \bibnamefont{Becker}},
  \bibinfo{author}{\bibfnamefont{G.~K.} \bibnamefont{Grubert}},
  \bibnamefont{and}
  \bibinfo{author}{\bibfnamefont{D.}~\bibnamefont{Loffhagen}},
  \bibinfo{journal}{Eur. Phys. J. Appl. Phys.} \textbf{\bibinfo{volume}{51}},
  \bibinfo{pages}{11001} (\bibinfo{year}{2010}).

\bibitem[{\citenamefont{Kushner}(2004)}]{Kushner04}
\bibinfo{author}{\bibfnamefont{M.~J.} \bibnamefont{Kushner}},
  \bibinfo{journal}{J. Appl. Phys.} \textbf{\bibinfo{volume}{95}},
  \bibinfo{pages}{846} (\bibinfo{year}{2004}).

\bibitem[{\citenamefont{Golubovskii et~al.}(2002)\citenamefont{Golubovskii,
  Maiorov, Behnke, and Behnke}}]{GMB02}
\bibinfo{author}{\bibfnamefont{Y.~B.} \bibnamefont{Golubovskii}},
  \bibinfo{author}{\bibfnamefont{V.~A.} \bibnamefont{Maiorov}},
  \bibinfo{author}{\bibfnamefont{J.}~\bibnamefont{Behnke}}, \bibnamefont{and}
  \bibinfo{author}{\bibfnamefont{J.~F.} \bibnamefont{Behnke}},
  \bibinfo{journal}{J. Phys. D: Appl. Phys.} \textbf{\bibinfo{volume}{35}},
  \bibinfo{pages}{751} (\bibinfo{year}{2002}).

\bibitem[{\citenamefont{Dussart et~al.}(2010)\citenamefont{Dussart, Overzet,
  Lefaucheux, Dufour, Kulsreshath, Mandra, Tillocher, Aubry, Dozias, Ranson
  et~al.}}]{DOL10}
\bibinfo{author}{\bibfnamefont{R.}~\bibnamefont{Dussart}},
  \bibinfo{author}{\bibfnamefont{L.~J.} \bibnamefont{Overzet}},
  \bibinfo{author}{\bibfnamefont{P.}~\bibnamefont{Lefaucheux}},
  \bibinfo{author}{\bibfnamefont{T.}~\bibnamefont{Dufour}},
  \bibinfo{author}{\bibfnamefont{M.}~\bibnamefont{Kulsreshath}},
  \bibinfo{author}{\bibfnamefont{M.~A.} \bibnamefont{Mandra}},
  \bibinfo{author}{\bibfnamefont{T.}~\bibnamefont{Tillocher}},
  \bibinfo{author}{\bibfnamefont{O.}~\bibnamefont{Aubry}},
  \bibinfo{author}{\bibfnamefont{S.}~\bibnamefont{Dozias}},
  \bibinfo{author}{\bibfnamefont{P.}~\bibnamefont{Ranson}},
  \bibnamefont{et~al.}, \bibinfo{journal}{Eur. Phys. J. D}
  \textbf{\bibinfo{volume}{60}}, \bibinfo{pages}{601} (\bibinfo{year}{2010}).

\bibitem[{\citenamefont{Kulsreshath et~al.}(2012)\citenamefont{Kulsreshath,
  Schwaederle, Overzet, Lefaucheux, Ladroue, Tillocher, Aubry, Woytasik,
  Schelcher, and Dussart}}]{KSO12}
\bibinfo{author}{\bibfnamefont{M.~K.} \bibnamefont{Kulsreshath}},
  \bibinfo{author}{\bibfnamefont{L.}~\bibnamefont{Schwaederle}},
  \bibinfo{author}{\bibfnamefont{L.~J.} \bibnamefont{Overzet}},
  \bibinfo{author}{\bibfnamefont{P.}~\bibnamefont{Lefaucheux}},
  \bibinfo{author}{\bibfnamefont{J.}~\bibnamefont{Ladroue}},
  \bibinfo{author}{\bibfnamefont{T.}~\bibnamefont{Tillocher}},
  \bibinfo{author}{\bibfnamefont{O.}~\bibnamefont{Aubry}},
  \bibinfo{author}{\bibfnamefont{M.}~\bibnamefont{Woytasik}},
  \bibinfo{author}{\bibfnamefont{G.}~\bibnamefont{Schelcher}},
  \bibnamefont{and} \bibinfo{author}{\bibfnamefont{R.}~\bibnamefont{Dussart}},
  \bibinfo{journal}{J. Phys. D: Appl. Phys.} \textbf{\bibinfo{volume}{45}},
  \bibinfo{pages}{285202} (\bibinfo{year}{2012}).

\bibitem[{\citenamefont{Eden et~al.}(2013)\citenamefont{Eden, Park, Cho, Kim,
  Houlahan, Li, Kim, Kim, Lee, Kim et~al.}}]{EPC13}
\bibinfo{author}{\bibfnamefont{J.~G.} \bibnamefont{Eden}},
  \bibinfo{author}{\bibfnamefont{S.-J.} \bibnamefont{Park}},
  \bibinfo{author}{\bibfnamefont{J.~H.} \bibnamefont{Cho}},
  \bibinfo{author}{\bibfnamefont{M.~H.} \bibnamefont{Kim}},
  \bibinfo{author}{\bibfnamefont{T.~J.} \bibnamefont{Houlahan}},
  \bibinfo{author}{\bibfnamefont{B.}~\bibnamefont{Li}},
  \bibinfo{author}{\bibfnamefont{E.~S.} \bibnamefont{Kim}},
  \bibinfo{author}{\bibfnamefont{T.~L.} \bibnamefont{Kim}},
  \bibinfo{author}{\bibfnamefont{S.~K.} \bibnamefont{Lee}},
  \bibinfo{author}{\bibfnamefont{K.~S.} \bibnamefont{Kim}},
  \bibnamefont{et~al.}, \bibinfo{journal}{IEEE Trans. Plasma Sci.}
  \textbf{\bibinfo{volume}{41}}, \bibinfo{pages}{661} (\bibinfo{year}{2013}).

\bibitem[{\citenamefont{{P. A. Tchertchian, C. J. Wagner, T. J. Houlahan Jr.,
  B. Li, D. J. Sievers, and J. G. Eden}}(2011)}]{TWH11}
\bibinfo{author}{\bibnamefont{{P. A. Tchertchian, C. J. Wagner, T. J. Houlahan
  Jr., B. Li, D. J. Sievers, and J. G. Eden}}}, \bibinfo{journal}{Contr. Plasma
  Phys.} \textbf{\bibinfo{volume}{51}}, \bibinfo{pages}{889}
  (\bibinfo{year}{2011}).

\bibitem[{\citenamefont{Ostrom and Eden}(2005)}]{OE05}
\bibinfo{author}{\bibfnamefont{N.~P.} \bibnamefont{Ostrom}} \bibnamefont{and}
  \bibinfo{author}{\bibfnamefont{J.~G.} \bibnamefont{Eden}},
  \bibinfo{journal}{Appl. Phys. Lett.} \textbf{\bibinfo{volume}{87}},
  \bibinfo{pages}{141101} (\bibinfo{year}{2005}).

\bibitem[{\citenamefont{Sternovsky}(2005)}]{Sternovsky05}
\bibinfo{author}{\bibfnamefont{Z.}~\bibnamefont{Sternovsky}},
  \bibinfo{journal}{Plasma Sources Sci. Technol.}
  \textbf{\bibinfo{volume}{14}}, \bibinfo{pages}{32} (\bibinfo{year}{2005}).

\bibitem[{\citenamefont{Riemann}(2003)}]{Riemann03}
\bibinfo{author}{\bibfnamefont{K.-U.} \bibnamefont{Riemann}},
  \bibinfo{journal}{J. Phys. D: Appl. Phys.} \textbf{\bibinfo{volume}{36}},
  \bibinfo{pages}{2811} (\bibinfo{year}{2003}).

\bibitem[{\citenamefont{Sheridan and Goree}(1991)}]{SG91}
\bibinfo{author}{\bibfnamefont{T.~E.} \bibnamefont{Sheridan}} \bibnamefont{and}
  \bibinfo{author}{\bibfnamefont{J.}~\bibnamefont{Goree}},
  \bibinfo{journal}{Phys. Fluids B} \textbf{\bibinfo{volume}{3}},
  \bibinfo{pages}{2796} (\bibinfo{year}{1991}).

\bibitem[{\citenamefont{Tsankov and Czarnetzki}(2017)}]{TC17}
\bibinfo{author}{\bibfnamefont{T.~V.} \bibnamefont{Tsankov}} \bibnamefont{and}
  \bibinfo{author}{\bibfnamefont{U.}~\bibnamefont{Czarnetzki}},
  \bibinfo{journal}{Plasma Sources Sci. Technol.}
  \textbf{\bibinfo{volume}{26}}, \bibinfo{pages}{055003}
  (\bibinfo{year}{2017}).

\bibitem[{\citenamefont{Lacroix et~al.}(2014)\citenamefont{Lacroix, Hermanns,
  Hinz, and Bonitz}}]{lacroix_prb14}
\bibinfo{author}{\bibfnamefont{D.}~\bibnamefont{Lacroix}},
  \bibinfo{author}{\bibfnamefont{S.}~\bibnamefont{Hermanns}},
  \bibinfo{author}{\bibfnamefont{C.~M.} \bibnamefont{Hinz}}, \bibnamefont{and}
  \bibinfo{author}{\bibfnamefont{M.}~\bibnamefont{Bonitz}},
  \bibinfo{journal}{Phys. Rev. B} \textbf{\bibinfo{volume}{90}},
  \bibinfo{pages}{125112} (\bibinfo{year}{2014}),
  \urlprefix\url{http://link.aps.org/doi/10.1103/PhysRevB.90.125112}.

\bibitem[{\citenamefont{Hopjan et~al.}(2016)\citenamefont{Hopjan, Karlsson,
  Ydman, Verdozzi, and Almbladh}}]{hopian_prl_16}
\bibinfo{author}{\bibfnamefont{M.}~\bibnamefont{Hopjan}},
  \bibinfo{author}{\bibfnamefont{D.}~\bibnamefont{Karlsson}},
  \bibinfo{author}{\bibfnamefont{S.}~\bibnamefont{Ydman}},
  \bibinfo{author}{\bibfnamefont{C.}~\bibnamefont{Verdozzi}}, \bibnamefont{and}
  \bibinfo{author}{\bibfnamefont{C.-O.} \bibnamefont{Almbladh}},
  \bibinfo{journal}{Phys. Rev. Lett.} \textbf{\bibinfo{volume}{116}},
  \bibinfo{pages}{236402} (\bibinfo{year}{2016}),
  \urlprefix\url{https://link.aps.org/doi/10.1103/PhysRevLett.116.236402}.

\end{thebibliography}
\bibliographystyle{apsrev}

\end{document}